\documentclass[fleqn,10pt]{wlscirep}
\usepackage[utf8]{inputenc}
\usepackage[T1]{fontenc}
\usepackage{lineno}
\usepackage[normalem]{ulem}
\usepackage{xcolor}
\usepackage{color, soul}

\newcommand{\kms}{\mbox{km s$^{-1}$}}

\newcommand\farcs{\mbox{$.\!\!^{\prime\prime}$}}%

\title{
    Bowshocks driven by the pole-on molecular jet of outbursting protostar SVS 13
}

\author[1,*]{Guillermo Bl{\'a}zquez-Calero}
\author[1]{Guillem Anglada}
\author[2, 3]{Sylvie Cabrit}
\author[1]{Mayra Osorio}
\author[4,\textdagger]{Alejandro C. Raga}
\author[5, 6]{Gary A. Fuller}
\author[1]{Jos{\'e} F. G{\'o}mez}
\author[7, \textdagger]{Robert Estalella}
\author[8]{Ana K. Diaz-Rodriguez}
\author[9, 10]{Jos{\'e} M. Torrelles}
\author[11]{Luis F. Rodr{\'i}guez}
\author[12]{Enrique Mac{\'i}as}
\author[13]{Itziar de Gregorio-Monsalvo}
\author[14]{S. Thomas Megeath}
\author[11]{Luis Zapata}
\author[15,16]{Paul T. P. Ho}

\affil[1]{Instituto de Astrof{\'i}sica de Andaluc{\'i}a, CSIC, Glorieta de la Astronom{\'i}a s/n, E-18008 Granada, Spain}
\affil[2]{Observatoire de Paris, PSL University, Sorbonne Universit{\'e}, CNRS, LERMA, F-75014, Paris, France}
\affil[3]{Univ. Grenoble Alpes, CNRS, IPAG, 38000 Grenoble, France}
\affil[4]{Instituto de Ciencias Nucleares, Universidad Nacional Aut{\'o}noma de M{\'e}xico, Apartado Postal 70-543, 04510 Ciudad de M\'exico, Mexico}
\affil[5]{Jodrell Bank Centre for Astrophysics, Department of Physics and Astronomy, The University of Manchester, Oxford Road, Manchester M13 9PL, UK}
\affil[6]{I. Physikalisches Institut, University of Cologne, Z\"ulpicher Str. 77, 50937 K\"oln, Germany}
\affil[7]{Departament de F{\'i}sica Qu\`antica i Astrof{\'i}sica, Institut de Ci{\`e}ncies del Cosmos, Universitat de Barcelona, IEEC-UB, Mart{\'i} i Franqu{\`e}s, 1, E-08028 Barcelona, Spain}
\affil[8]{UK ALMA Regional Centre Node, Jodrell Bank Centre for Astrophysics, Department of Physics and Astronomy, The University of Manchester, Oxford Road, Manchester M13 9PL, UK}
\affil[9]{Institut de Ci{\`e}ncies de l'Espai (ICE, CSIC), Can Magrans s/n, E-08193 Cerdanyola del Vall{\`e}s, Catalonia, Spain}
\affil[10]{Institut d'Estudis Espacials de Catalunya (IEEC), E-08034 Barcelona, Catalonia, Spain}
\affil[11]{Instituto de Radioastronom{\'i}a y Astrof{\'i}sica, Universidad Nacional Aut{\'o}noma de M{\'e}xico, P.O. Box 3-72, 58090, Morelia, Michoac{\'a}n, Mexico}
\affil[12]{ESO Garching, Karl-Schwarzschild-Str. 2, 85748, Garching bei M\"unchen, Germany}
\affil[13]{European Southern Observatory, Alonso de C{\'o}rdova 3107, Casilla 19, Vitacura, Santiago, Chile}
\affil[14]{Ritter Astrophysical Research Center, Department of Physics and Astronomy, University of Toledo, 2801 West Bancroft Street, Toledo, OH 43606, USA}
\affil[15]{Academia Sinica Institute of Astronomy and Astrophysics, P.O. Box 23-141, Taipei 106, Taiwan}
\affil[16]{East Asian Observatory, Hilo 96720, HI, USA}
\affil[*]{gblazquez@iaa.es}
\affil[ ]{\textsuperscript{\textdagger}Deceased.}

\begin{abstract}

    Outflows play a key role in the star and planet formation processes. Some outflows show discrete clumps of cold molecular gas moving at extremely high velocities (EHVs) of $\sim$100 km~s$^{-1}$, known as ``molecular bullets'', that are likely closely associated with their primary driving agent. Here we present ALMA CO(J=3-2) observations of a bright EHV molecular bullet that reveal its morphology in detail down to scales of 30 au and its kinematic structure across the entire intermediate velocity range ($\sim$30-100~km~s$^{-1}$). These provide important new insights into how outflows transfer mass and momentum to the surrounding medium. The observed channel maps display several sequences of ring-like features whose velocity increases and size decreases with projected distance from the driving source, each sequence tracing a thin, bow-shaped shell culminating on-axis in a bright EHV head. The shape, kinematics, and mass of each shell all agree remarkably well with the simplest textbook models of momentum-conserving bowshocks produced by a time-variable EHV jet. The dynamical timescale between consecutive shells is of a few decades, with the latest ejection event coinciding with the protostar optical/IR outburst observed in $\sim$1990. The very strong evidence for bowshock-driven entrainment induced by jet variability revealed by this work suggests that accretion bursts, and therefore variations in the disk snowlines, should occur on decade timescales, which could substantially impact grain growth and planet formation.

\end{abstract}
\begin{document}

\flushbottom
\maketitle
\newpage
\thispagestyle{empty}

\section*{Introduction}

The star formation process is intrinsically associated with a powerful ejection activity that is believed to play a crucial role in regulating the final mass of the central object, the star formation efficiency on larger scales\cite{frank2014, krumholz2019}, and the accretion and planet formation processes in protoplanetary disks\cite{manara2023,pascucci2023}. Highly-collimated ejections with typical velocities of $\sim$100~km~s$^{-1}$, are launched from the close vicinity of the young stellar object (YSO), constituting the so-called ``extremely high-velocity'' (EHV)\cite{bachiller1996} component of the outflows, also known as jets. Jets are often observed at optical, infrared, and cm radio wavelengths through the emission of their hot ($>$500~K) ionized atomic and molecular constituents\cite{bally2016, anglada2018, lee2020, ray2021, ray2023, federman2024}. These EHV jets do not present a smooth brightness distribution but usually consist of sequences of closely spaced compact brightness enhancements, referred to as knots, typically separated by dynamical timescales (i.e., the ratio between separation and velocity) of only a few years to several decades \cite{reipurthbally2001, anglada2007, anglada2018, reipurth2023, federman2024}.

In some of the youngest protostars, a cold ($\sim$50-100 K) molecular constituent of the EHV jets is detected in low-excitation CO and SiO transitions. These cold molecular jets were discovered as discrete peaks well defined both spatially and in velocity, and dubbed ``molecular bullets'' \cite{bachiller1990}. Bullets usually appear aligned in sequences typically spaced at dynamical time intervals of up to $\sim$800~yr (refs. \cite{bachiller1991, hatchell1999, zapata2005, lefloch2007, lee2009, santiago-garcia2009, hirano2010, podio2016, tychoniec2019, schutzer2022, dutta2024}) and down to a few decades close to the source \cite{hirano2010, moraghan2016, podio2021, yoshida2021, lee2020}. High angular resolution images reveal that bullets are resolved into chains of bright knots\cite{hirano2010}, so the cold molecular jets show a knotty structure similar to that observed in the hot optical, infrared and radio continuum jets. 

An ubiquitous manifestation of the ejection activity associated with star formation are the so-called (cold) ``molecular outflows'', with bipolar morphology and moderate collimation, which are observed in CO and other low-excitation molecular transitions\cite{rodriguez1982, lada1985, bachiller1996, konigl2000, bally2016}. These outflows, with typical velocities from a few km~s$^{-1}$ to $\sim$20 km~s$^{-1}$, constitute the ``standard high-velocity'' (SHV)\cite{lada1985, bachiller1996, konigl2000} component (sometimes referred to as the slow/low velocity component) of outflows. These SHV molecular outflows exhibit a relatively smooth morphology over large spatial scales ($\sim$0.1~pc), corresponding to dynamical timescales of $\sim$10$^4$~yr (refs.~\cite{lada1985, levreault1988}), are more massive than the EHV cold molecular jets, and have traditionally been interpreted as a shell of swept-up ambient gas.

Over the past three decades, numerous models have been developed to explain the origin of outflows from YSOs, the relationship between their distinct components, their interaction with the surrounding ambient cloud, and the nature of jet knots and bullets \cite{ragacabrit1993, tafalla2017, rabenanahary2022, lishu1996, wang2019, shang2023, pudritznorman1983, yvart2016, devalon2022}. The cold molecular constituent of the outflows (typically traced by simple molecular species such as CO) offers a unique opportunity to study the full range of velocities of the outflows, from the EHV regime associated with protostellar ejection, to the SHV regime approaching ambient cloud speeds. Such studies have now become feasible with the unsurpassed high spatial and spectral resolution provided by large, very sensitive mm interferometers, such as the Atacama Large Millimeter Array (ALMA). This velocity coverage provides crucial insights into the potential interactions between fast-moving EHV material ejected from the inner protostellar environment and the surrounding ambient cloud.
A promising new approach for exploring these interactions involves the very high spatial and spectral resolution study of the morphology and kinematics of the 
much fainter molecular line emission in the velocity regime between the SHV and EHV components, hereafter referred to as ``intermediate high-velocities'' (IHV).
Features suggestive of thin, shell-like substructures in the IHV regime have begun to emerge in recent ALMA observations \cite{fernandez-lopez2020, zhang2019, devalon2022, liu2025, lee2022, louvet2018}. However, their imaging and interpretation has been limited so far by their intrinsic faintness, narrow velocity ranges, unfavorable outflow orientations (e.g., side-on views, where kinematics along the flow axis cannot be accurately deprojected), or insufficient velocity resolution (e.g., when multiple channels are averaged to improve signal-to-noise).
Here we present ALMA observations of a near pole-on outflow with a combination of high sensitivity and high angular resolution, providing a detailed view of the morphology and kinematics of the faint IHV component with exquisite velocity sampling, all the  way up to the EHV regime.

The target of our study is the near pole-on outflow from the protostar SVS~13 (refs.~\cite{strom1976, bachillercernicharo1990, masson1990, bachiller2000, chen2016, lefevre2017}) in the NGC~1333 star-forming region (distance = 300 pc)\cite{ortiz-leon2018, gaia2023}. SVS~13 is a binary at cm and mm wavelengths, with a projected separation of 90 au between its two components, designated VLA4A and VLA4B \cite{anglada2000, anglada2004, diaz-rodriguez2022}. Only the eastern component (VLA4B), which recently underwent a photometric outburst in the optical and near-infrared \cite{eisloeffel1991}, is visible at these shorter wavelengths \cite{hodapp2014}. SVS~13 is thought to drive the outstanding chain of Herbig-Haro objects HH~7-11 (refs.~\cite{strom1976,hartigan2019}) and an associated SHV bipolar CO outflow that extends over $\sim$0.4~pc in the northwest-southeast direction\cite{liseau1988, plunkett2013} (see Extended Data Fig.~1). Three bright blueshifted EHV molecular bullets (designated 1 through 3, in order of increasing distance from SVS 13) and one redshifted bullet have been identified within $\sim$$20''$ from SVS 13 in interferometric maps at $\sim$0.5$''$-2$''$ resolution, reaching line-of-sight (LOS) velocities $>$100 km~s$^{-1}$ relative to the ambient cloud, with signs of morphological substructure and velocity gradients within the bullets\cite{bachiller2000, chen2016, lefevre2017}. On smaller scales in the near-infrared, a 0.2$''$-long highly collimated blueshifted [Fe~II] microjet \cite{hodapp2014, gardner2016} arising from the eastern, visible component (VLA~4B) of SVS~13 is observed. This jet appears to pierce inside the first of a series of three arcuate features traced by the H$_2$ $v=1-0$ S(1) line extending over $\sim$3$''$ to the southeast\cite{hodapp2014} (see also Extended Data Fig.~1).

In this paper we perform a high-sensitivity CO study of the SVS~13 Bullet 1 reaching an angular resolution of $\sim$$0.1''$ ($\sim$30 au) and a spectral resolution $\lesssim$0.5~km~s$^{-1}$ (see Methods). The faint IHV component of the outflow is imaged with minimal projection effects on the velocity across its entire IHV regime ($\sim$30$-$100~km~s$^{-1}$). Combined with mass estimates derived from the CO dataset, a thorough analysis reveals the dynamical connection between individual EHV knots and their surrounding faint IHV material, providing new insight into how EHV ejecta transfer momentum to their environment. Our results further complement previous near-infrared studies\cite{eisloeffel1991,hodapp2014}, opening a new pathway to relate EHV molecular knots to accretion variability in protoplanetary disks, and to assess the impact of such variability on grain growth and planet formation\cite{cieza2016, houge2023}.

\section*{Results} \label{sec:results}

\subsection*{Morphology and kinematic substructure of Bullet 1 at 30~au resolution}

We present results from two CO($J$=3-2) datasets, one at high angular resolution ($\sim$0.1$''$), covering velocities relative to the cloud up to approximately $-$102~km~s$^{-1}$, and the other at lower angular resolution ($\sim0.5''$), covering up to $-$129~km~s$^{-1}$ (see `ALMA and ACA observations and data processing' section in Methods).  In Fig.~\ref{fig:mom} we show the CO integrated and peak intensity images of Bullet 1, showing at $\sim$$0.1''$ = 30~au resolution its elongated morphology (PA $\simeq$ 160$^\circ$) and rich substructure, with apparent cavities and arcuate emission enhancements. A similar elongated morphology is seen (at lower resolution) in SO (see Methods and Extended Data Fig.~2). As first noticed from lower angular resolution data \cite{lefevre2017}, there is a close spatial correlation of the CO emission enhancements with the arc-shaped H$_2$ features \cite{hodapp2014}. Our results further show that (after proper motion correction), for the second and third H$_2$ arcs (where our $\sim$0.1''\ CO data are available), the CO emission peaks slightly upstream of the H$_2$ emission. 

The velocity centroid map (third column in Fig.~\ref{fig:mom}) shows a clear longitudinal LOS velocity gradient, with the absolute velocity increasing with distance from SVS~13, confirming previous lower resolution results\cite{lefevre2017, chen2016}. The velocity pattern is roughly symmetric with respect to the axis of Bullet 1 (see also ref. \cite{lefevre2017}), in contrast with early claims for transverse rotation gradients\cite{chen2016}. The main reason for the discrepancy is the poor angular resolution ($\sim 3''$) of ref.~\cite{chen2016} observations, which does not allow the bullet emission to be completely isolated, resulting in blending with lower-velocity western emission unrelated to the bullet, which mimics a transverse velocity gradient.

A more revealing and striking view of the kinematics is obtained by looking at individual line channels in the CO image cube (see Fig.~\ref{fig:channels}, Extended Data Fig.\ 3, Supplementary Figs.\ 1 and 2, and Supplementary Videos 1 and 2). The channel images (specially the data of $\sim$$0.1''$ = 30 au resolution) show several sequences of rings whose positions and sizes change smoothly as a function of the LOS velocity, getting smaller and moving away from the source as velocity increases (in absolute value). The rings are narrow, until they abruptly transform into a bright filled circle that we call the ``head''. We call each of these sequences of rings (and its head) a ``family''. The rings are detected at LOS velocities from approximately $-$9 to $-$125~km~s$^{-1}$ relative to the cloud, have radii in the range $0.2''$ to $2.5''$ (60 to 750 au) and extend southeast up to $\sim$$5''$ ($\sim$1500 au), in projection, from SVS~13. In general, the brightest part of a ring is that closest in projection to SVS~13 and, in some cases, the rings appear incomplete, as arcs.

 The smooth progression of these rings with LOS velocity can also be well seen in the position-velocity (PV) cuts along the flow axis in Fig.~\ref{fig:mom}. There, each sequence of rings appears as two faint continuous ``arms'' starting around $-30$~km~s$^{-1}$ and gradually converging into a single bright filled EHV head at the highest LOS velocity. The presence of two arms occurs because in each velocity channel, the empty ring of a given sequence intersects the central axis of the PV cut at two positions. This smooth and coherent behavior in both space and velocity shows that each ring family actually traces a thin bow-shaped hollow shell opening up toward SVS~13, and converging to an EHV knot at its apex, with a large velocity gradient along its surface. The width of the CO rings in channel maps sets an upper limit to the shell thickness $<$30~au.

For SVS~13, it has been possible to image coherent CO ring sequences at high spatial and spectral resolution
($\sim$30~au and 0.5~km~s$^{-1}$) over the whole IHV regime, from $\sim$30~km~s$^{-1}$ (in absolute value relative to the cloud) all the way up to the EHV regime ($\sim$100~km~s$^{-1}$). CO ring-like features in channel maps were previously imaged in the near pole-on outflow DO Tau\cite{fernandez-lopez2020}, but they were only detected over a relatively narrow velocity range and at low velocities ($\lesssim$20~km~s$^{-1}$ relative to the cloud), much smaller than those of the EHV jet ($\sim$100~km~s$^{-1}$)\cite{simon2016}. Another example is the HH46/47 CO outflow, where elliptical features have been identified\cite{zhang2019}, but at LOS velocities $\lesssim 50$~km~s$^{-1}$ (ref.~\cite{zhang2019}), considerably lower than those of the EHV jet ($\sim200$~km~s$^{-1}$)\cite{hartigan1993,birney2024}.

 For a better quantitative characterization of the observed ringed emission, we performed elliptical fits to more than 400 rings identified in the observed CO channel images (see Methods). The fitted ellipses are superposed on each channel map in Supplementary Figures 1 and 2, and in Supplementary Video 3, and are plotted in the top panels of Fig.~\ref{fig:ellipses}, where six families of rings (labeled I to VI) are identified. The parameters of all the elliptical fits are listed in Supplementary Data 1. In the bottom panels of Fig.\ \ref{fig:ellipses}, we plot the radius and LOS velocity of the six families of rings as a function of the projected distance of the ring's center to SVS~13. These panels show that each family of rings has a truly differentiated behavior, with the separation into families appearing naturally. The panels clearly show the tendency, within each family, of increasing speed and narrowing radius away from SVS~13, with the EHV head being the most distant, compact, and highest speed structure.

 In the bottom left panel of Fig.~\ref{fig:ellipses}, there is also a clear trend for more distant ring families to reach a larger maximum radius. This trend (indicated with a gray line in the figure) suggests that all hollow shells subtend a similar maximum half-angle of $\sim22^\circ$ after deprojection by an inclination angle with respect to the LOS $i=25^\circ$ (see Methods). The bright filled rings tracing the EHV heads of each family, on the other hand, subtend a half-angle of only 3$^\circ$-5$^\circ$ after deprojection, typical of knots observed in other EHV cold molecular jets on similar scales\cite{tafalla2017,podio2021}.  Given the timescale spacing of a few decades between consecutive families (see below), we identify the EHV heads of our ring families with the knotty substructure frequently detected in EHV jets\cite{lee2020,podio2021,jhan2022}, while the empty rings trace a much fainter bow-shaped IHV shell. Hints of faint bow-shaped shells with an EHV knot at their apex have recently been imaged in CO or SiO in a few flows\cite{lee2022, liu2025}. However, limitations in sensitivity and/or unfavorable outflow orientation hampered the characterization of the morphology and kinematics of the shells. The sequences of CO rings reported here in SVS~13 are the first to reveal the bow (dominant) velocity component parallel to the flow axis, with both high spatial and spectral resolution, and no strong projection effects.

\subsection*{Ejection origin, time variability, and relation to the optical outburst}

Remarkably, as shown in the top panels of Fig.~\ref{fig:ellipses}, within each of the families I, II and III, the centers of all the rings are very well aligned and point to the proximity of VLA 4B as their origin (see also Methods). A fit to the ring centers for families II and III gives a PA=$160^\circ$, coinciding with the PA of the overall elongation of Bullet 1 (see above), and similar to the PA defined on a larger scale by the chain of CO Bullets 1-3 ($\sim$155$^\circ$) \cite{bachiller2000,chen2016}. For families IV, V, and VI the data are poorer but still consistent with the same values of the PA. In contrast, Family I is aligned along a smaller PA=$135^\circ$, close to that of the [FeII] microjet (PA=145$^\circ$)\cite{hodapp2014}.

Assuming that the heads of the families are moving approximately at constant speed away from VLA~4B, and at an inclination angle $i=25^\circ\pm2^\circ$ with respect to the LOS  (except for Family I, for which we assume $i=22^\circ\pm3^\circ$) (see Methods), we can estimate their dynamical times from the ratio of their deprojected positions and mean velocities. We obtain dynamical times of $25\pm5$, $56\pm6$, $91\pm9$, $100\pm9$, and $134\pm15$~yr for families I-IV and VI, respectively (see Extended Data Fig. 4). For Family V, the head is not identified in our data and we obtain an upper limit of 115~yr from the positions and velocities of the rings. The typical separation in dynamical times between consecutive families is a few tens of years.

Interestingly, the epoch of ejection corresponding to the dynamical time of the head of Family I (the most recent ejection) is 1992$\pm$5, consistent with the epoch of the strong optical/infrared outburst of the visible component of SVS~13 (VLA~4B), which occurred in 1990$\pm1$ (ref.~\cite{eisloeffel1991}), indicating a close association between the optical/infrared outburst and the EHV molecular knot\cite{fischer2023}. This coincidence in time suggests that EHV knots in SVS~13 trace actual episodic ejection events.

It is intriguing that the ejection direction differs between the most recent Family I and the older ones (see Fig.~\ref{fig:ellipses}). To illustrate the apparent direction changes in the SVS~13 outflow, we plot in Extended Data Fig.~4 the PA relative to VLA~4B, measured at the head of each family, as a function of the estimated dynamical time. For comparison, we also show rough PAs (in general, derived from lower resolution studies from the literature) and timescales (up to $\sim$ 2500~yr) for other high-velocity features likely ejected from SVS~13 (refs.~\cite{bachiller2000,chen2016,hodapp2014,hartigan2019}; see values in Supplementary Table 1). We note a rapid jump in PA in only 30 yr, from $\sim$160$^\circ$ to $\sim$140$^\circ$,  between Family II and the most recent ejection (associated with Family I and the optical/infrared outburst), and a similar rapid jump between Bullet~2 and HH~11. This is in contrast to the previous stability of the PA at $\sim$160$^\circ$ over longer periods of $\sim$80 yr (families II to VI, H$_2$ arcs 2-3, and Bullets 2-3). This unusual pattern of PA changes suggests either twin jets from a yet unresolved close binary in VLA~4B (ref.~\cite{lefevre2017}) or abrupt disk reorientations during some outbursts. This complex issue will be further discussed in a dedicated future work, and does not affect the result of the present paper.

\subsection*{Bowshock model interpretation}%\label{sec:bowshock}

In the following, we show that the observed morphology, kinematics, and mass of the ringed CO emission described above can all be naturally reproduced by the standard model of a momentum-conserving bowshock created by a variable jet propagating in a dense medium. Under this interpretation, each sequence of rings in the channel maps corresponds to a thin bowshock viewed near pole-on, sliced up into iso-LOS velocity sections that are almost perpendicular to its axis (see Fig.~\ref{fig:geometry}). The largest diameter rings then correspond to the bow wings, far behind the present bowshock apex, and are therefore naturally closer to the source. These larger rings are also naturally slower, due to them having intercepted more ambient, stationary gas, in a momentum-conserving scenario.   

The formation process of a jet bowshock is summarized in Extended Data Fig.~5. In a jet with supersonic velocity variability, a two-shock ``internal working surface'' (IWS) forms within the jet beam, where fast ejecta catch up with slower moving material. The over-pressured shocked jet material inside the working surface is then ejected sideways from the jet beam, where it is swept back into a curved bowshock by the ambient gas (which acts as a ``head-wind'' in the reference system co-moving with the working surface). In the case of a uniform ambient medium, the shape and mean velocity field of the bowshock can be computed analytically by assuming a thin shell of fully mixed jet + ambient material with pure momentum conservation (i.e., neglecting the effect of pressure gradients)\cite{masson1993,ostriker2001}. We will use here a more general prescription \cite{tabone2018}, which allows for an arbitrary propagation speed, and for a non-static environment (see Methods). We also present new analytical expressions for the predicted shell surface density and mass, that we use below to check the self-consistency of the bowshock model for SVS~13 (see ``Derivation of the ambient density and jet mass flux from the bowshock mass'' in Methods).

We find that, with only six free parameters (see Methods) this analytical momentum-conserving bowshock model accounts remarkably well and in a natural way for the observed progression of emission morphology with LOS velocity in a given family of rings. Fig.~\ref{fig:modeleli} shows a comparison of the bowshock model results with the parameters of the elliptical fits to all the rings in families II, III, and VI. We obtain (see Extended Data Table 1) inclination angles of $i \simeq 22^\circ$ (compatible with the independent observational constraints described in the section ``On the inclination angle'' in Methods), sideways ejection velocities of $v_0\simeq 20$~km~s$^{-1}$, propagation speeds with $v_{\rm jet}\simeq110$-$135$~km~s$^{-1}$ and a negligible velocity of the external medium into which the jet is traveling ($v_{\rm amb} < 5$-$10$~km~s$^{-1}$ for the families where this parameter is better constrained). The characteristic scale of the bowshock ($L_0$) ranges from 0.4$''$ to 1.7$''$, increasing with the distance of the IWS from VLA~4B ($z_{\rm ws}$) that ranges from 4.6$''$ to 15$''$. 

In Fig.~\ref{fig:modelpv} we show a sample comparison of observed and modeled spectral channel images for families II and III, which have the best observational data (see also Supplementary Video 4 for the whole sequence of channels). The images in Fig.~\ref{fig:modelpv} further show that the asymmetry of the intensity of the rings, with the side closest in projection to SVS~13 being brighter, is also naturally explained as a result of optical depth and beam filling factor effects (see Methods and Extended Data Fig.~6). Finally, the bottom panels of Fig.~\ref{fig:modelpv} show, for families II and III, a comparison of the observed and modeled zeroth- and first-order moment maps, and PV cuts along the flow axis, showing again remarkable overall agreement. 

The agreement in both Figs.~\ref{fig:modeleli} and \ref{fig:modelpv} is surprisingly good given the idealized assumptions in the momentum-conserving bowshock model, which are the simplest possible (thin shell, negligible pressure gradients, uniform ambient medium, full mixing between jet and ambient material). 

In the framework of our bowshock model for the narrow CO rings, the bright EHV filled rings at the head of each family would trace the "internal jet working surface" of over-pressured shocked jet material (dark-grey rectangle in Extended Data Fig.~5), before it interacts with ambient gas to form the wider bowshock wings of mixed jet + ambient material that would correspond to the fainter empty rings. The poorer behavior of the model at the head of each family (discontinuity in the PV plots of Fig.~\ref{fig:modeleli}, small size and absence of filled-in emission in channel maps, and narrow head in the PV cuts shown in Fig.~\ref{fig:modelpv}) can be attributed to the fact that the emitting properties of the working surface, in particular its finite size, are not explicitly included in the model. Also, the EHV filled rings show a distinctive behavior in the plots of the LOS velocity as a function of the projected distance of the ring center to SVS~13 (see triangle symbols in bottom left panel of Fig.~\ref{fig:ellipses}), with their LOS velocity (in absolute value) decreasing with distance to the star, which is opposite to the trend displayed by the empty rings. As noted for the knots in the EHV molecular jet of IRAS~04166+2706 (ref.~\cite{tafalla2017}), this trend could result from the kinematic contribution of the material ejected sideways from the jet working surface, and would require additional specific modeling.

Finally, using the estimated mass of each observed shell, with values of a few $10^{-4}~M_\odot$ (see ``Determination of the mass'' in Methods and Extended Data Table 1) and the bowshock model parameters, we can infer the density in the ambient medium, $\rho_{\rm amb}$, and the mass-flux of jet material ejected sideways from the working surface, $\dot{M}_0$ (see ``Derivation of the ambient density and jet mass flux'' in Methods and Extended Data Table 1). Our values of $\rho_{\rm amb}$ decrease with distance to the source, from 3.9 $\times 10^{-19}$~g~cm$^{-3}$ (Family II) to 7.2$\times 10^{-21}$~g~cm$^{-3}$ (Family VI), as expected for an embedded object, and agree with independent estimates ($\rho_{\rm amb}$=3.3$\times 10^{-20}$~g~cm$^{-3}$ on a $\sim 3''$ scale) from NH$_3$ observations toward SVS~13 (see Methods). In contrast, the values of $\dot{M}_0$ for families II, III, and VI remain similar, with $\dot{M}_0=0.7$-$1.1\times 10^{-6}~M_\odot~\mathrm{yr}^{-1}$. These values of $\dot{M}_0$ are consistent with jet mass-fluxes for SVS~13, independently obtained from far-infrared observations of shock tracers \cite{sperling2020}, and with a canonical ratio $\sim10\%$ of the mass accretion rate \cite{cabrit2007,lee2020}, that we estimate as $\dot{M}_{\rm acc}$=$10^{-5}~M_\odot~\mathrm{yr}^{-1}$ in SVS~13 (see Methods). This excellent overall consistency reinforces the attractiveness and plausibility of the jet-driven bowshock interpretation for the shells.

The jet-driven bowshock interpretation would also naturally explain why the detected rings in SVS~13 subtend a maximum angle with respect to the axis ($\sim22^\circ$ after deprojection, gray line in bottom left panels of Fig.~3). Recent numerical simulations of time variable jets in an ambient density gradient\cite{rabenanahary2022} or a slow surrounding disk wind\cite{tabone2018} show that, beyond some angle from the flow axis, the bowshock wings get piled-up and ``pinched'' by external pressure gradients into conical walls opening away from the source, with a shear-like flow parallel to the walls. When viewed close to pole-on, this conical shear-flow would project  in low-velocity channel maps as smooth and extended circular patches, that would not be well imaged in our high-resolution ALMA observations.

\section*{Discussion}

We have found that the shape and kinematics of the observed CO rings, including their apparent acceleration pattern, with the LOS velocity increasing with projected distance to SVS~13, are in strikingly good agreement with the simplest possible models for momentum-conserving jet-driven bowshocks propagating into a dense and slow moving medium ($\lesssim10$~km~s$^{-1}$). The bright and compact EHV knots at the tip of each bowshock would then trace the IWSs where jet material is being shocked and ejected sideways from the narrow jet beam, as previously proposed for EHV knots in IRAS~04166+2706 (ref. \cite{tafalla2017}). The jet-driven bowshock model naturally reproduces all the observables, including the shell shape, kinematics, and mass, with the simplest possible assumptions and with a reasonable jet mass-flux of 10\% the accretion rate, without the need for any ad-hoc parametrization.

These findings set new constraints on existing ejection models and have several important broad implications. First, since bowshocks of the kind modeled here are formed by sideways ejection of shocked material, the fast EHV component would trace a genuinely narrow jet beam, rather than an axial density enhancement within a wider-angle fast wind (where potential sideways motion would be quenched by neighboring streamlines); this argues in favor of alternative, less well-studied MHD ejection configurations where fast material is launched into intrinsically narrow cones\cite{koenigl2011, ferreira2000, zanni2013}. The presence, outside the bowshocks, of a separate slower outer MHD disk wind at velocities of $\sim$10 km~s$^{-1}$, arising from radii $\gtrsim$1~au (as recently proposed in several conical outflows based on ALMA rotation signatures\cite{louvet2018, devalon2020, devalon2022}), cannot be discarded with the present data, however, as emission in this low velocity range is not well sampled in our ALMA images. 

Second, jet-driven bowshock shells are known to stop expanding sideways quite rapidly, at $\sim$10$^4$~yr (ref.~\cite{rabenanahary2022}), so the large-scale outflow feedback on the initial mass function (IMF) peak and star-formation efficiency (SFE) in the parent cloud may be considerably weaker for time-variable narrow jets than for the, often used in this context, wide-angle X-winds\cite{matzner1999,federrath2014,rohde2019,offner2014}. Third, our finding of jet-driven bowshocks associated with EHV knots in SVS~13 strongly supports the view that most EHV knots reflect genuine variations in ejection velocity\cite{raga1990,jhan2022}, rather than, for example, magnetic instabilities\cite{shang2023,liu2025} in a steady wind. Moreover, the close association observed between one EHV knot and a recent photometric outburst\cite{eisloeffel1991} provides evidence that links this velocity variability to accretion bursts. The interval between successive EHV knots further suggests that these accretion bursts would have duty cycles as short as a few decades (see also refs.~\cite{hodapp2014,jhan2022}); the resulting rapid variations in snowline and dust evaporation radius could impact grain growth and planet formation considerably\cite{cieza2016,houge2023}. More detailed modeling, including tailored numerical simulations, is needed to fully explore the theoretical implications of the data presented here on the ejection mechanism, and the origin of accretion bursts. 

From an observational perspective, our study of dynamical times in SVS~13 revealed a hint of periodicity on $\sim$35-year scale which implies that a new ejection, suitable for detailed multi-wavelength study, should occur in the next few years. Second epoch ALMA observations could also provide measurements of the proper motion and transverse expansion of the CO rings in channel maps, that could directly confirm and improve our model. Finally, it is important to validate the bowshock interpretation in a broader outflow sample. We anticipate that sufficiently deep and well-resolved ALMA CO observations of bright outflows, viewed at moderate ($<$40$^\circ$) inclinations to the LOS, should identify thin CO rings across the entire IHV regime ($\sim$30-100~km~s$^{-1}$) similar to those described here in SVS~13. Combined with mass constraints and detailed modelling, the approach adopted here in SVS~13 opens up a new way to probe the interaction of protostellar jets with their environment, helping to address the long-standing problem of the driving agent of protostellar outflows and the role these outflows play in the formation of stars and planets.

\clearpage
\begin{figure}[p!]
\begin{center}
\includegraphics[width=\textwidth]{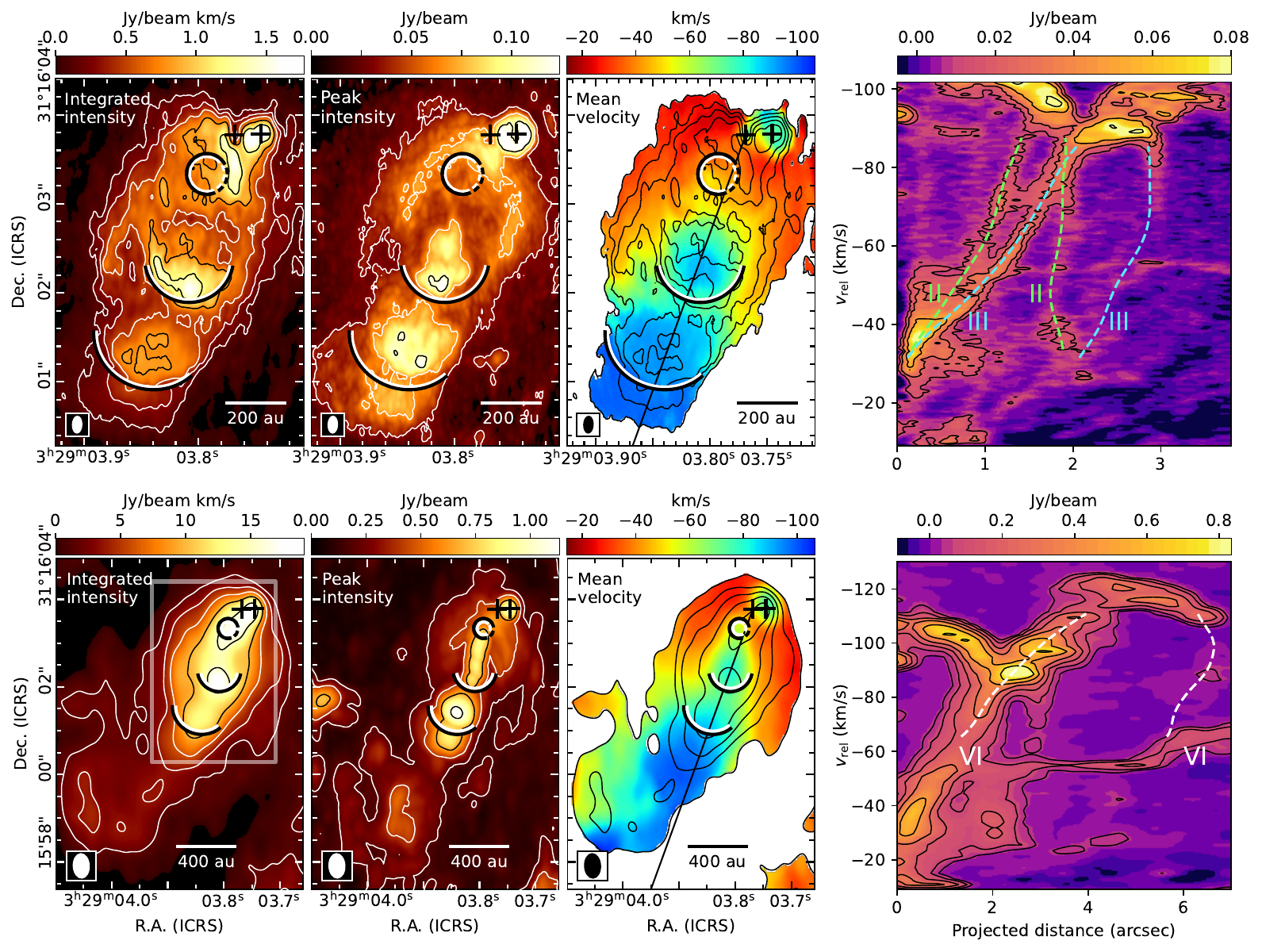}
\end{center}
\end{figure}

\setcounter{figure}{0}
\begin{figure*}[p!]
\begin{center}
\caption{
{\bf ALMA images and position-velocity diagrams of the blueshifted Bullet 1.}  Primary beam corrected ALMA images of the velocity-integrated intensity, peak intensity of the spectral cube, mean velocity, and position-velocity diagrams of the observed CO($J$=3-2) emission. The higher (lower) angular resolution results are shown in the top (bottom) panels. Synthesized beams are shown at the bottom left corners of the images. Plus signs indicate the positions of the protostars, VLA~4A (west) and VLA~4B (east), of the SVS~13 binary \cite{anglada2000, anglada2004, diaz-rodriguez2022}. 
The H$_2$ arcuate features \cite{hodapp2014} are plotted as black and white arcs, taking into account their observed proper motion and the $\sim 0.07''$ offset between the radio and optical/infrared positions of the star \cite{diaz-rodriguez2022, gaia2023}.
All velocities are relative to VLA~4B, ranging from $-$9.3 to $-$102.7 (top panels) and $-$126.4~km~s$^{-1}$ (bottom panels). These ranges exclude the CO disk emission but not contamination (stronger toward VLA 4A) from disk emission of other molecular transitions, with higher rest frequencies, falling within the frequency range of the blueshifted CO emission of the bullet. Note also that the low-resolution images include emission from the CO clumps 1 and 2 (see Extended Data Fig.~1 and Supplementary Table 1), near the southeast corner. The position-velocity diagrams have been obtained along the longitudinal axis of the bullet (PA = 160$^\circ$, indicated by the black line in the panels of the third column), with origin in VLA~4B. 
Dashed lines in these diagrams indicate the emission of the families of rings II, III, and VI (defined in Fig.~\ref{fig:ellipses}). 
Contours are: 3, 6, 10, 16, and 25 times $0.05$ (top) and $0.61$ (bottom) Jy~beam$^{-1}$~km~s$^{-1}$ (first and third columns); 3, 6, 10, and 16 times $0.011$ (top) and $0.066$ (bottom) Jy~beam$^{-1}$ (second column); 5, 10, 20, 35, 60 and 90 times 0.0025 (top) and 0.008 (bottom) Jy~beam$^{-1}$ (fourth column). See Methods for further details.
\label{fig:mom}}%Figure 1
\end{center}
 \end{figure*}

\clearpage
\begin{figure}[p!]
\begin{center}
\includegraphics[width=1.0\textwidth]{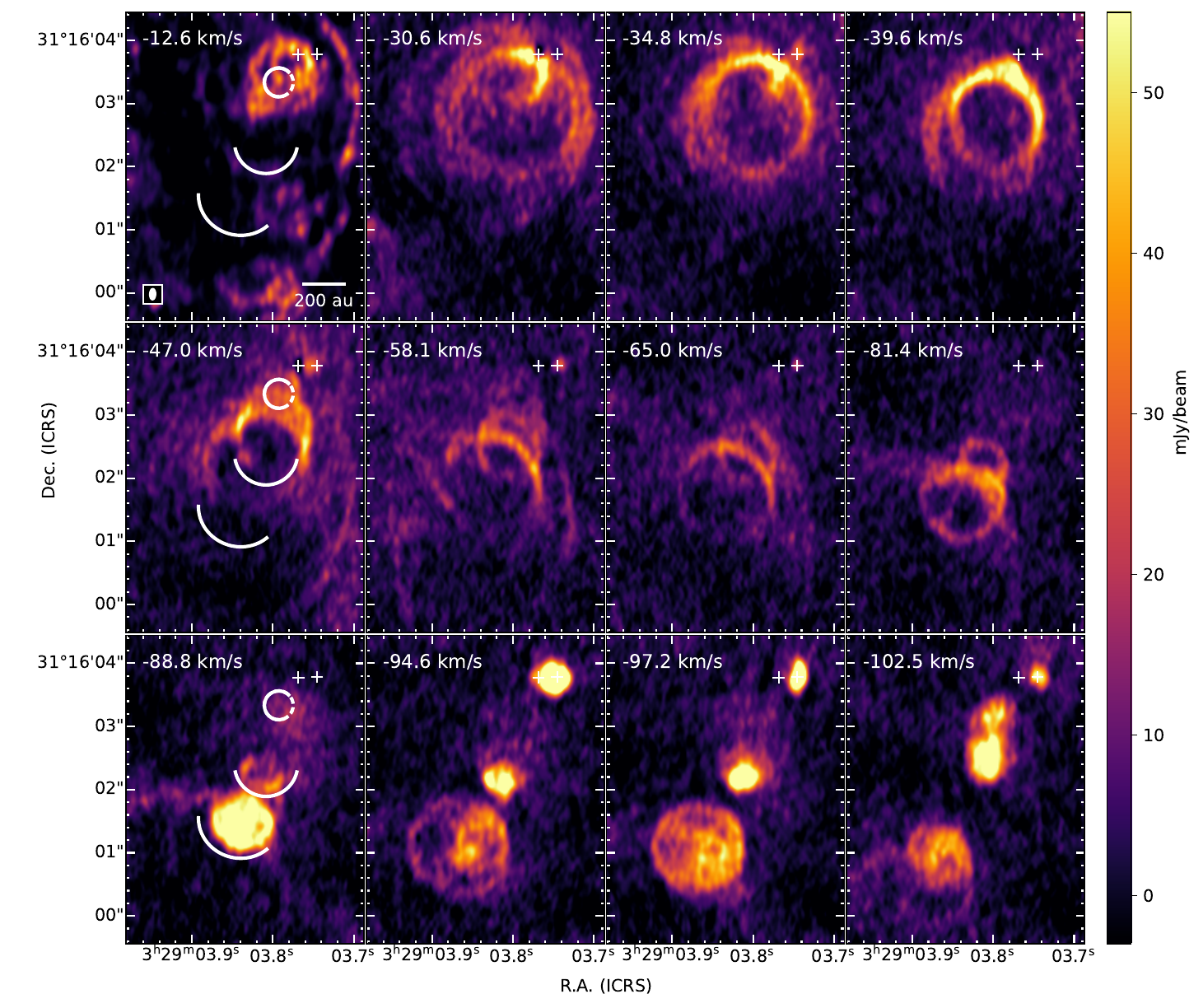}\\
\caption{
    {\bf Observed CO($J$=3-2) spectral channel images at high angular resolution.} A sample of high angular resolution channel maps of the CO($J$=3-2) emission observed by ALMA, which illustrate the ringed kinematic structure in Bullet~1. The synthesized beam is $0.173'' \times 0.091''$ (PA = $-2.2^\circ$), and is plotted as an ellipse in the bottom left corner of the first image. The full set of observed channel maps is shown in Supplementary Fig.~1 and Supplementary Video~1. The positions of the two protostars of the SVS~13 binary are indicated by plus signs. Note that several velocity channels are contaminated by emission from different molecular transitions, with higher rest frequency than CO, coming from the disks associated with VLA~4A and VLA~4B \cite{diaz-rodriguez2022}. This disk emission is stronger toward VLA 4A. The LOS velocity relative to VLA~4B is shown in the top left corner of each image. The velocity width of each of the channels shown in the figure is 0.53~km~s$^{-1}$, which corresponds to five native channels. The r.m.s noise of the images is 2.6 mJy beam$^{-1}$. The images have not been corrected by the primary beam response. The H$_2$ arcuate features \cite{hodapp2014} are plotted in the first column as white arcs. Lower angular resolution channel maps, covering a wider range of velocities, are shown in Extended Data Fig.~3, Supplementary Fig.~2 and Supplementary Video~2. 
\label{fig:channels}}%Figure 2
\end{center}
 \end{figure}

\clearpage
\begin{figure*}[t] 
\begin{center}
\includegraphics[width=1\textwidth]{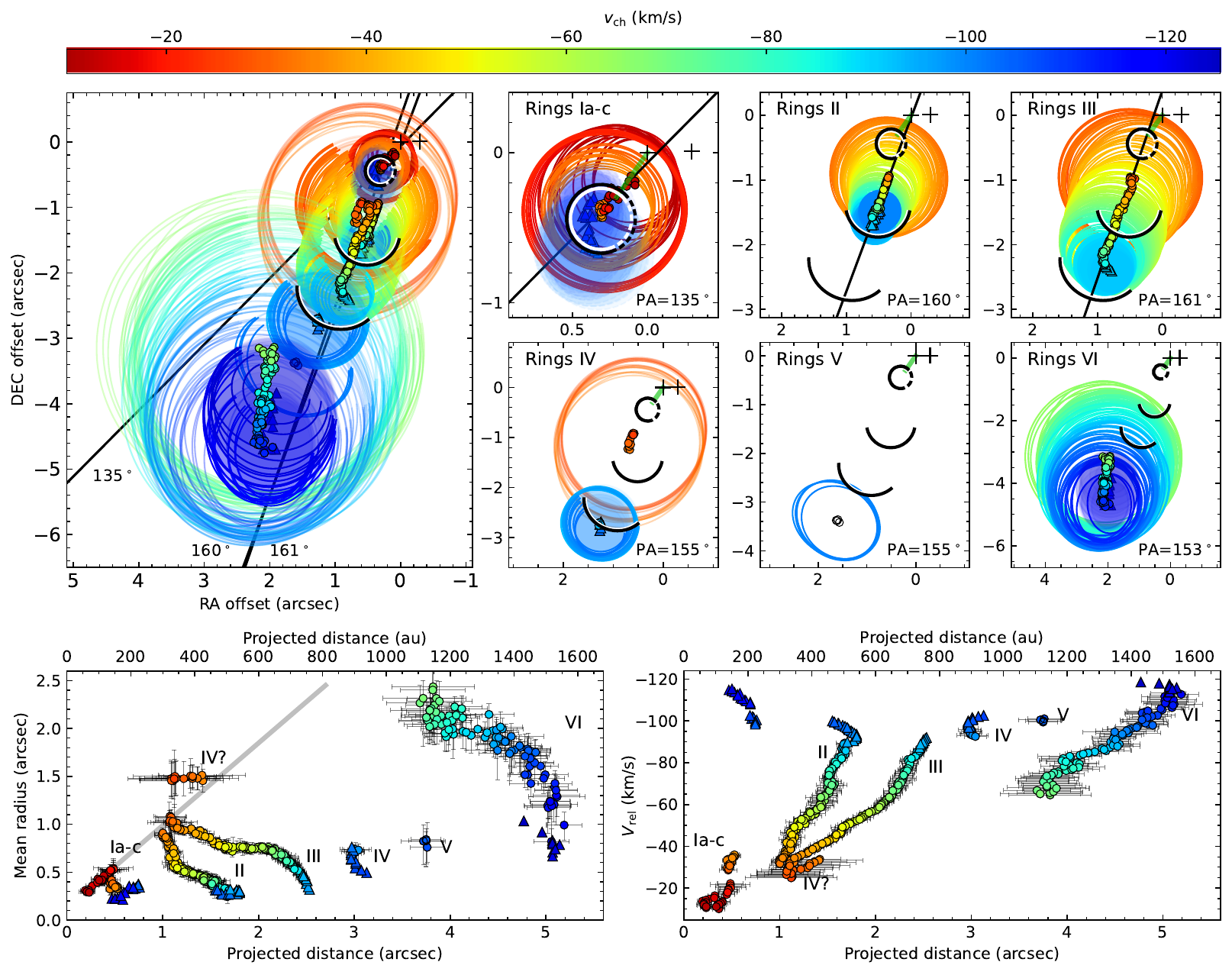}
\end{center}
\end{figure*}

\setcounter{figure}{2}
\begin{figure*}[th!]
\clearpage
\noindent
\caption{
    {\bf Elliptical fits to the ringed emission of the CO($J$=3-2) channel maps.} Top: Color-coded plots of elliptical fits (see Methods) to the observed CO emission. Rings are represented by the ellipses that best-fit their emission, and filled emission regions by shaded areas, with colors indicating the LOS velocity relative to VLA~4B of the channel map. Their centers are indicated by color dots (rings) and triangles (filled emission). The rings tend to group in sequences that we call ``families'', with their centers well aligned, their velocity increasing (in absolute value) and their size decreasing with distance from the stars (indicated with black plus signs), and ending in a region of filled emission (the ``head''). Separate plots for each family are shown in the right panels. Linear regression fits (plotted as black lines for families I to III) have been performed (for families IV to VI their origin is set at the position of VLA~4B). The [FeII] microjet \cite{hodapp2014} from VLA~4B is represented by a green segment. Black and white arcs represent the H$_2$ arc-like features \cite{hodapp2014}. Offsets are relative to the position of VLA 4B. Bottom: Plots of the mean radius and the LOS velocity of the rings, as a function of the projected distance of their centers to VLA~4B. Error bars represent the estimated uncertainties, adopted as the deviations with respect to the fitted value that cause the intensity of the fitted ring to decrease by 1$\times$r.m.s of the channel map (see Methods). The plots illustrate the trends (except for the heads) for the radius to decrease and the velocity to increase with distance, with the different families clearly differentiated in the plots. The gray line in the left panel traces the suggested outer edge of the families, forming an angle of $\sim$$44^\circ$ from the axis ($\sim$$22^\circ$ after deprojection by an inclination angle of $25^\circ$).    
\label{fig:ellipses}
 }
\end{figure*}

\clearpage
\begin{figure*}[th!]
\begin{center}
\includegraphics[width=1.0\textwidth]{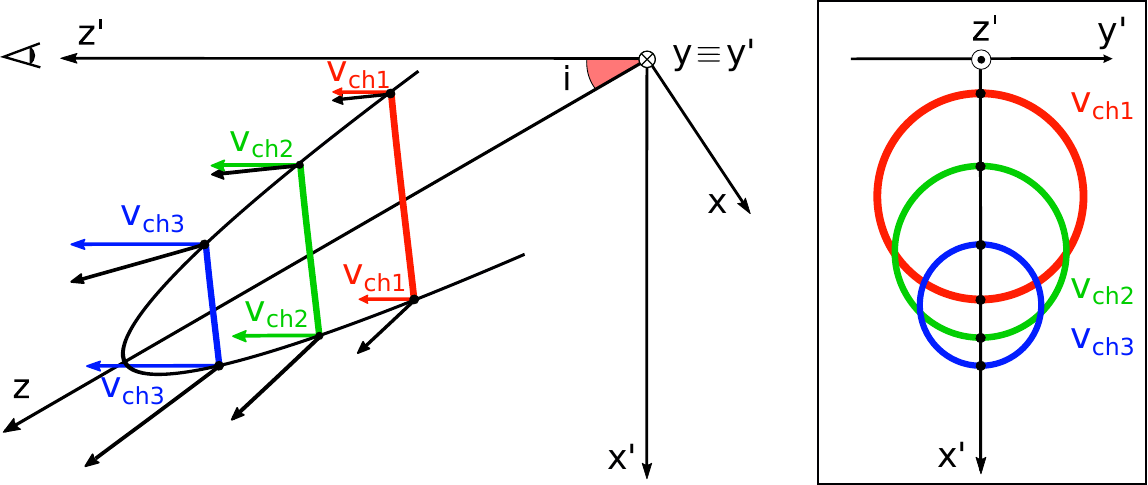}\\
\caption{
{\bf Decomposition of an approaching bowshock into velocity channel images.} The $x'y'$ plane is the plane of the sky, the $z'$ axis is the LOS, and the $z$ axis is the symmetry axis of the bowshock, oriented at an inclination angle $i$ with respect to the LOS. Left: Side view illustrating the velocity gradient of the blueshifted bowshock, where the magnitude of the velocities increases as we approach the bowshock head. Black arrows represent the velocity field of the bowshock shell, while colored arrows represent their projection onto the LOS. The points in the bowshock with LOS velocities $v_{\rm ch1}$, $v_{\rm ch2}$, and $v_{\rm ch3}$ are also shown in colors. Right: Projection onto the plane of the sky of the points in the bowshock with LOS velocities $v_{\rm ch1}$, $v_{\rm ch2}$, and $v_{\rm ch3}$. Each colored curve corresponds to the emission observed in a channel map. Note that the LOS velocities ($v_{\rm ch}$) are defined as positive when the motion is away from the observer (redshifted), but in the coordinate system used, positive $v_{z'}$ velocities are directed toward the observer (blueshifted); therefore, $v_{\rm ch}=-v_{z'}$.
}
\label{fig:geometry}
\end{center}
\end{figure*}

\clearpage
\begin{figure}[p!]
\begin{center}
\includegraphics[width=1.0\textwidth]{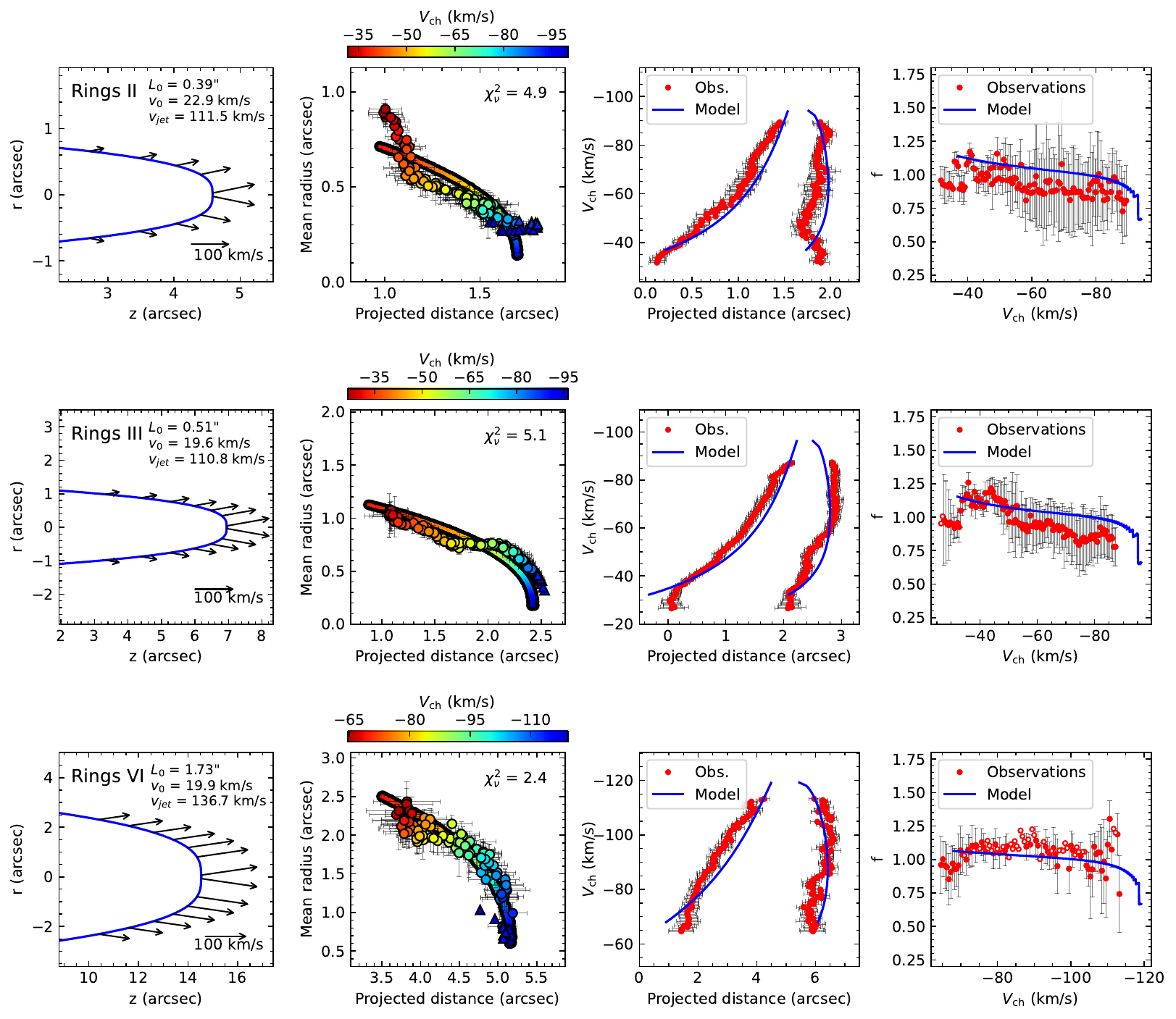}
\caption{
{\bf Bowshock model fitting to the elliptical fits of the rings.} Panels correspond to families II, III and VI, respectively, from top to bottom. The bowshock model fitting was performed through a least-square minimization to the data (see Methods). The best-fit model parameters are presented in Extended Data Table 1. First column: Shape and velocity field (black arrows) of the bowshock model, where $z$ is the symmetry axis (whose origin is VLA 4B) and $r$ is the cylindrical radius. The main parameters of the model, the characteristic scale ($L_0$), the velocity at which the jet material is initially ejected sideways by the working surface ($v_0$), and the velocity of the internal working surface along the $z$-axis ($v_{\rm jet}$), are listed in the top right corner of the panels. The fitted inclination angles are $i=20$-$25^\circ$, and the fitted velocity of the medium into which the jet is traveling is $v_{\rm amb}\simeq0$~km~s$^{-1}$ (see Methods and Extended Data Table~1). Second column: Observed (dots indicate rings and triangles indicate filled emission of the head) and bowshock model (continuum line) mean radius of the elliptical rings in channel maps as a function of the projected distance to VLA 4B. Velocities are indicated in a color scale. The reduced chi-square of the fitting is shown on the right top corner. Third column: Observed (red dots) and model (blue line) position-velocity diagram for the rings (excluding the heads). Fourth column: Observed (red dots) and model (blue line) elongation factor of the rings ($f$), taken as the ratio of the ring axes along the longitudinal and transverse directions. Error bars indicate the estimated uncertainties of the data points as defined in Fig.~\ref{fig:ellipses}, empty dots indicate that the axial ratio uncertainty could not be obtained (see ``Ellipse fitting'' in Methods). All velocities are LOS velocities relative to the velocity of VLA~4B.
}
 \label{fig:modeleli}. 

\end{center}
 \end{figure}

\clearpage
\begin{figure*}[t] 
\begin{center}
\includegraphics[width=0.95\textwidth]{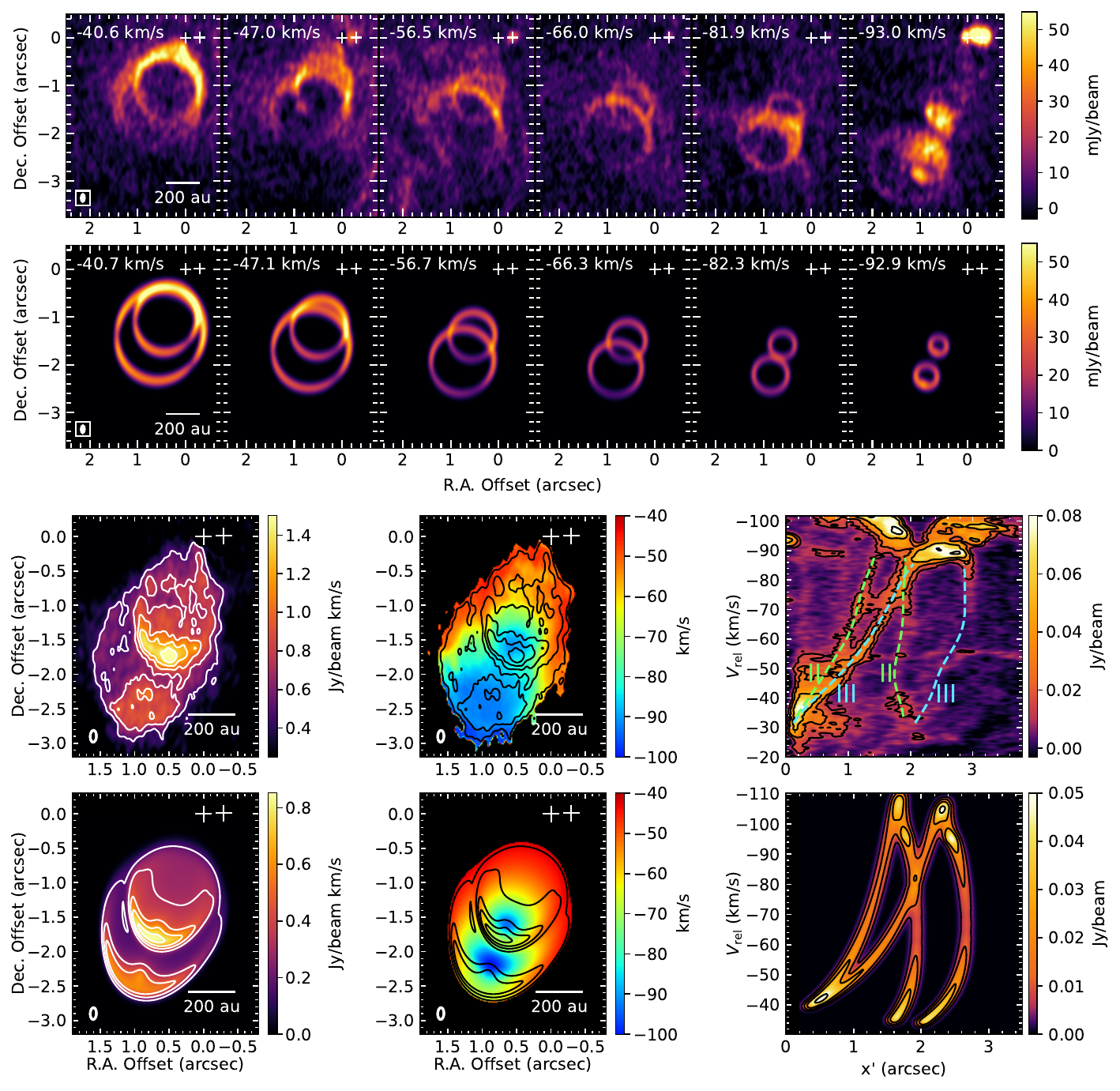}
\caption{
{\bf Comparison of the bowshock model with the observations.} Comparison of observational and model results for ring families II and III. The model parameters are given in Fig.~\ref{fig:modeleli} and Extended Data Table 1. An intrinsic velocity dispersion $v_T$=2~km~s$^{-1}$ has been assumed in the modeling (see Methods). The channel width is $\Delta v_{\rm ch}$=0.53~km~s$^{-1}$ in both model and observations. Offsets are relative to VLA~4B. Top rows: Comparison of a set of observed and bowshock model spectral channel images (first and second rows, respectively). The LOS velocity relative to VLA 4B is indicated in the images. Bottom rows: Comparison of observational and model results (third and fourth rows, respectively) for the integrated intensity and mean velocity images, and position-velocity diagrams. The LOS velocity range in model calculations is $-$41.6 to $-$112.0~km~s$^{-1}$, corresponding approximately to the range used to construct the observed images ($-$41.6 to $-$100.8~km~s$^{-1}$, where the emission of families II and III is dominant, although there is still some contamination from Family I within $\sim$$0.5''$ southeast from VLA~4B). The disk emission has been masked in the observed images; yet, the observed emission is stronger than that of the model, since the observations likely include some emission from other features that is difficult to avoid. Contours of the integrated intensity are 4, 6, 8, 10, and 12 times 0.12 Jy~beam$^{-1}$~km~s$^{-1}$ for the observations, and 2, 4, 6, 8, and 10 times 0.08 Jy~beam$^{-1}$~km~s$^{-1}$ for the model images. Contours of the position-velocity diagrams (PA = 160$^\circ$, width = $0.1''$) are 5, 10, 20, and 35 times 2.5 mJy~beam$^{-1}$ for the observations, and 2, 5, 10, and 20 times 2.5 mJy~beam$^{-1}$ for the model. Dashed lines in the observed position-velocity diagram indicate the emission from families II and III.
}
\label{fig:modelpv}
\end{center}
\end{figure*}

\clearpage
\vfill\eject

\section*{Methods}

\subsection*{ALMA and ACA observations and data processing.}\label{sec:observations}

The observations were carried out in Band 7 (0.9 mm), during Cycle 3 (program 2015.1.01229.S; PI: G. Anglada) and Cycle 4 (program 2016.1.101305.S; PI: G. Anglada). Here, we present a summary of the observations focused on the CO($J$=3-2) data, and we refer to ref.~\cite{diaz-rodriguez2022} for a more detailed description of the spectral line and continuum setup.

The phase center of all the observations was set at RA(ICRS) = $03^\mathrm{h}29^\mathrm{m}03.75^\mathrm{s}$, Dec(ICRS) = $31^\circ16' 04.00''$, within $\sim$$0.3''$ of the position of VLA4B.

Cycle 3 ALMA 12-m array observations were carried out during two runs on 2016 September 9 and 10, using 42 antennas with baselines in the range of 15-3225~m, providing an angular resolution of $\sim$$0.1''$ and a maximum recoverable scale of $\sim$$12''$ (strict upper limit, obtained from the shortest baseline) and of $\sim$$3''$ (good quality imaging, obtained from the $5^{\rm th}$ percentile shortest baseline). The native channel width was 122~kHz, corresponding to 0.106~km~s$^{-1}$ at 345.795990~GHz, the assumed rest frequency of the CO($J$=3-2) transition. The total bandwidth was 234.375~MHz, covering an LSR velocity range from $-$93.32 to 109.47~km~s$^{-1}$.

Cycle 4 ALMA 12-m array data were obtained on 2016 November 24, using 43 antennas with baselines in the range of 15-704~m, providing an angular resolution of $\sim$0.4$''$ and a maximum recoverable scale of $\sim$$12''$ (from the shortest baseline) and of $\sim$$3''$ (from the $5^{\rm th}$ percentile baseline). The native channel width was 244~kHz, corresponding to 0.211~km~s$^{-1}$, and the total bandwidth was 937.500~MHz, covering a broader LSR velocity range than in Cycle 3, from $-$398.34 to 414.25~km~s$^{-1}$.

In both cycles, the calibrators J0237+2848, J0238+1636, and J0336+3218 were used for the ALMA 12-m array bandpass, absolute flux, and complex gain calibration, respectively.

Cycle 4 observations with the 7-m Atacama Compact Array (ACA) were carried out on 2016 November 15, and 2017 July 7 and 26, with an angular resolution of $\sim$4$''$ and a maximum recoverable scale of $\sim$$35''$. The native channel width was 122~kHz (0.106~km~s$^{-1}$) with a total bandwidth of $250$~MHz, covering an LSR velocity range from $-$100.00 to 116.64~km~s$^{-1}$. The bandpass calibrator was J0510+1800 on November 15 and July 7, and J0522$-$3627 on July 26. The absolute flux calibrator was Ceres on November 15, and Uranus for the remaining observations. The complex gain calibrator was J0319+4130 for all observations. 

The observations were pipeline calibrated using CASA versions 4.7.01 and 4.7.2, for Cycle 3 and 4, respectively. Additionally, the 12-m array Cycle 3 data (1.6~h on-source with $\sim 0.1''$ resolution) were phase self-calibrated. Self-calibration was unsuccessful for the Cycle 4 data (only 4.15~min on-source with $\sim 0.4''$ resolution). All the imaging was performed with CASA 5.7.2, using the task TCLEAN. Further details on the calibration and image processing are given in ref.~\cite{diaz-rodriguez2022}. 

The images presented in this paper were made using Briggs weighting with robust 0.5 and have not been corrected by the primary beam response, except when indicated. High angular resolution CO($J$=3-2) images of Bullet 1 were made from the Cycle 3 ALMA 12-m array data, obtaining a synthesized beam of $0.173'' \times 0.091''$ (PA = $-2.2^\circ$), while lower angular resolution images, with a synthesized beam of $0.527'' \times 0.333''$ (PA = $2.7^\circ$), were obtained from the lower angular resolution Cycle 4 ALMA 12-m array data. In general, the LOS velocities appearing in the paper are relative to that of the source VLA~4B ($V_{\rm LSR}$ = +9.3 km~s$^{-1}$)\cite{diaz-rodriguez2022}, with the high-angular resolution data reaching blueshifted velocities up to $-$102.7~km~s$^{-1}$, while the lower angular resolution data extend over the full velocity range of Bullet 1. For most of the analysis carried out in this paper we have used high angular resolution channel maps with a velocity width of 0.53~km~s$^{-1}$, corresponding to the average of five native channels, and lower angular resolution channel maps with a velocity width of 2.12~km~s$^{-1}$, corresponding to the average of 10 native channels.

We obtained primary beam corrected images of the velocity-integrated intensity (zeroth-order moment of the spectral cube), of the peak intensity of the spectral cubes, and of the mean velocity field (first-order moment, or intensity-weighted average velocity) of the blueshifted Bullet 1. Both high and low angular resolution images have been obtained, using the available high-angular resolution data, with LOS velocities from $-$9.3 to $-$102.7~km~s$^{-1}$, and the lower angular resolution data extending over a wider velocity range, from $-$9.3 to $-$126.4~km~s$^{-1}$, in order to include the highest blueshifted velocities of the bullet. For the high angular resolution data a clipping of 14 mJy beam$^{-1}$ (4-$\sigma$ before primary beam correction) has been applied to the individual channel maps (with the native channel width of 0.106~km~s$^{-1}$) before constructing the moments. These images are presented in Fig.~\ref{fig:mom}. We note that the selected velocity ranges (both in the high- and lower-angular resolution moments) exclude the CO emission from the disk; however, the moments appear contaminated near the VLA~4A and VLA~4B positions by disk emission from other molecular transitions (with rest frequencies higher than CO) that fall within the frequency range covered by the blueshifted CO emission from Bullet 1. This emission excess is particularly evident toward VLA 4A, where the disk molecular emission is stronger \cite{diaz-rodriguez2022}. We also note that the low angular resolution moments include emission from the CO clumps 1 and 2, located near the southeast corner of the images (see Extended Data Fig.~1 and Supplementary Table 1). Contamination from these clumps, which are independent of Bullet 1, make the images appear asymmetric. 

PV diagrams (with slice widths of $0.1''$ and $0.4''$ for the high and lower angular resolution, respectively) were obtained along the longitudinal axis of Bullet~1 (PA = 160$^\circ$) and with origin at VLA~4B. These diagrams are shown in Fig.~\ref{fig:mom}.

Because of the small primary beam of the ALMA 12-m array (18$''$), only Bullet~1 has been imaged with these data. The larger size of the ACA primary beam (32$''$) allows us to detect Bullet~2 in CO($J$=3-2), but at low angular resolution $\sim$4$''$ (see Supplementary Fig.~3).

\subsection*{Identification of the observed line emission as CO($J$=3-2) at extremely high velocities.}

It is known that there is line emission associated with the SVS~13 disks corresponding to transitions of other molecular species, with different rest frequencies\cite{diaz-rodriguez2022}. For several of these transitions the difference in frequency with respect to the CO($J$=3-2) transition is similar to the Doppler shift corresponding to the range of velocities observed in the bullet. Therefore, in principle, it could occur that the emission that we attribute to EHV blueshifted CO in the bullet is indeed emission from other molecular transitions with higher rest frequencies arising from material moving at much lower velocities. We argue that this is not the case. Since CO is the most abundant detectable molecule, if the emission observed in some channels corresponds to other molecules, then we would expect to detect a CO ``replica'' of this emission in the velocity channels corresponding to the difference in frequency with respect to CO, which is not observed.

In addition, we identified SO($8_8$-$7_7$) emission in the bullet from a spectral window dedicated to continuum observations, with a low spectral resolution of 27.2~km~s$^{-1}$. Considering the rest frequency of this transition, 344.310612~GHz, the emission reaches velocities about $-$120~km~s$^{-1}$ (relative to the LSR velocity of VLA~4B), the same as the CO in the bullet. In Extended Data Fig.\ 2, we show the integrated SO emission in the bullet. With a spatial resolution of $0.54'' \times 0.36''$, we can distinguish the main morphological substructure we found in the CO channel maps. Thus, we confirm that the main substructure observed in the bullet is traced by different molecular transitions, and that the substructure reported in the CO does not arise from the overlapping of different molecular lines in the same channel maps. Moreover, the main kinematical features of Bullet~1 we have observed in CO(3-2) were already present, with the same velocities, in (lower resolution) CO(2-1) data, as well as in other molecular transitions, such as SiO(5-4) and SO($6_5$-$5_4$) lines \cite{lefevre2017}. This coincidence in velocity is obviously highly unlikely if it were just the result of contamination from other species.

\subsection*{Ellipse fitting.}\label{sec:ellfit}

The ellipse fitting to the rings in the channel maps has been carried out using an interactive program written in Python. This program finds the geometrical parameters (center coordinates, semimajor-axis, ellipticity, and PA) that maximize the mean intensity that falls within an elliptical annulus (we chose a width of a few pixels, $\sim$1/3 of the beam size). The program takes as an initial guess the geometrical parameters of an ellipse that visually fits a targeted ring, calls the function ``minimize'' from the Python's package {\sc SciPy} \cite{virtanen2020}, and returns the best-fit parameters. The fitting method L-BFGS-B was chosen, which allows bound-constrained minimizations. It is crucial to constrain the parameter-space to ensure that the fitting is performed to the targeted ring of a given family. In this manner, we avoid additional emission from a different ring family or other structure affecting the fit. We adopt as uncertainties on the fitted parameters the deviations with respect to their nominal values that cause the mean intensity within the corresponding elliptical annulus to decrease by one times the root mean square (r.m.s) noise of the channel map.  

We note that, when two rings of different families intersect in a given channel map, the inner region common to both rings often appears enclosed within the arcs (one from each ring) that connect the intersection points (see the middle row of panels in Fig.~\ref{fig:channels}); these connected arcs simulate a (fake) third small ring, which must be excluded from the fit. This means that the classification of rings into families has to be performed prior to the fitting, once it is clear which are the true rings that should be fitted. To better ensure the correct identification of rings in families, we perform the fitting on a channel by channel basis, using as initial parameters the solution of the previous channel fit. This usually gives good solutions, since the rings of a given family vary smoothly with velocity. We note that the elliptical fits of the heads of the families, which appear as filled emission, are visual estimates.

In Supplementary Figs.~1 and 2 we present the whole set of channel maps, with the elliptical fits superimposed on the rings (see also Supplementary Video 3, where we show how the elliptical fits change with the channel maps).

\subsection*{Description of the families of CO rings and comparison with the H$_{\mathbf 2}$ arcs.}

Using the ellipses fitted to the rings and filled-in features in the observed channel maps, a quantitative analysis of their positions, velocity gradients, and comparison with other outflow tracers can be performed. In particular, the good alignment of the centers of the ringed features can be quantified through total least-squares linear fitting (often referred as orthogonal regression), where the orthogonal distance of the measured points to the fitting line is minimized \cite{romero-munoz2014}. Families of rings II and III are the best defined and the fitting to the positions of the centers of the ellipses give lines at PA = $160.0^\circ\pm0.4^\circ$ for Family II, and PA = $160.9^\circ\pm0.4^\circ$ for Family III (see top panels in Fig.~\ref{fig:ellipses}). In both cases, the position of VLA 4B (the eastern component of the SVS~13 binary) is consistent with the fit, strongly suggesting that these two sequences of rings originate in VLA 4B. 

Some families appear incomplete (as ring families IV and V) and/or overlapping, as might be the case of Family I, which shows some anomalies such as channels with more than one ring and discontinuities in the sequence of rings. This family has some of the smallest rings, and is the one nearest to the origin, so their study would specially benefit of the highest available angular resolution data, but the high-velocity end of this family falls outside the velocity coverage of the high-resolution data, and the properties of the highest velocities have been inferred from the low-resolution data and are more uncertain. Indeed, it is unclear whether this is a single family or a sequence of several families, and thus we have labeled them as Ia, Ib, and Ic. A fit to the centers of the whole Family I gives a considerably smaller PA = 135.3$^\circ\pm2.7^\circ$, and an origin in VLA 4A cannot be completely discarded. Indeed, this PA is closer to that of the HH objects (see Extended Data Fig.~4) and this coincidence could give clues regarding their origin. Nevertheless, the matching of the positions of the ring centers with the [FeII] microjet \cite{hodapp2014} (see top panels in Fig.~\ref{fig:ellipses}) favors an origin of rings I in VLA~4B.

Families IV and V have only a small number of rings, and a similar fitting to infer their origin is not possible, but we note that, if we set VLA~4B as their origin, the centers of the rings and head features define a PA of $\sim$155$^\circ$, which is close to that of families II and III. On the other hand, the head of Family VI shows a PA of 158$^\circ$ if VLA~4B is set as its origin, which is also similar to the fitted PA of families II and III. However, the centers of the rings of Family VI show a large dispersion and progressively depart from this direction to a smaller PA ($\sim153^\circ$). This could be because they are more distant and have suffered a more intense interaction with the ambient cloud. Nonetheless, the positions of the rings of this family have larger uncertaintes because they are weaker, they are obtained from the low resolution data, and additional emission that does not belong to this family could hamper the fitting. Indeed, as illustrated by the PV diagram in the lower right corner of Fig.~\ref{fig:mom}, the left arm of the emission plot shows a protrusion at LOS velocities (relative to VLA 4B) of $\sim -120$~km~s$^{-1}$, slightly higher in absolute value than those of the head of Family VI. This suggests that another source, possibly with an independent origin, perhaps resulting from a shock, may be appearing blended with the EHV emission of Family VI. To avoid possible contamination, we have conservatively excluded channels with LOS velocities $\le$ $-$118.6~km~s$^{-1}$ from our analysis of Family VI.

The families of CO rings seem to define the same pattern observed in the H$_2$ features \cite{hodapp2014}, reinforcing the association of CO and H$_2$ features; the jet and the first bubble define a PA similar to that of Family I, while the more distant arcs define a larger PA, similar to that of families II and III, which is close to that of the (larger scale) sequence of blueshifted Bullets 1-3 (PA$\simeq$155$^\circ$)\cite{chen2016, bachiller2000}. Note that all these PA are larger than that of the HH 7-11 system (120$^\circ$-133$^\circ$; see Extended Data Fig.~4), which is similar to that of the axis of the blueshifted cavity of the large-scale standard CO outflow \cite{plunkett2013}.

\subsection*{Coordinate systems.}

To analyze the observations, we define two reference frames. The first reference frame is associated with the shell of high-velocity CO emitting material, with its origin at the position of the YSO, the $z$ axis along the shell axis, the $y$ axis perpendicular to $z$ and on the plane of the sky, and the $x$ axis perpendicular to both $z$ and $y$ (Fig.~4). The shell axis has an inclination $i$ with respect to the LOS. The second, ``observer'' reference frame has the same origin, with the $z'$ axis along the LOS toward the observer, the $x'$ axis along the projection of the shell axis on the plane of the sky, and the $y'$ axis coinciding with the $y$ axis, so that the plane $x'y'$ is the plane of the sky. Thus, the two coordinate systems are related by a rotation of angle $i$ around the $y$(=$y'$)-axis, and the transformation between the two sets of coordinates is 
\begin{equation}\label{eq:coord}
\left\{\begin{array}{rcl}
x' &=& x \cos i + z \sin i, \\
y' &=& y, \\
z' &=& z \cos i - x \sin i.
\end{array}\right.
\end{equation}

Note that the velocity of a channel map, $v_\mathrm{ch}$, is the LOS velocity relative to the YSO as measured by the observer, and is defined positive when the motion is away from the observer (redshifted). However, the system of coordinates used has the $z'$ axis directed toward the observer and, thus, positive $v_{z'}$ velocities are directed toward the observer (blueshited). Therefore, there is a change of sign between $v_\mathrm{ch}$ and $v_{z'}$:
\begin{equation}
v_\mathrm{ch} = -v_{z'}.
\end{equation}

\subsection*{On the inclination angle.}

We consider three independent ways to obtain an estimate of the SVS 13 outflow inclination angle (angle between the LOS and the flow axis; i.e., $i=0$ for a pole-on flow). The high-angular resolution H$_2$ S(1) results\cite{hodapp2014} show that the farthermost H$_2$ arc (HC3) has a proper motion of $\sim$30~mas~yr$^{-1}$ ($\sim$43~km~s$^{-1}$ at 300 pc) relative to VLA~4B. This arc is positionally associated with the head of Family III, that has a LOS velocity of $-$90~km~s$^{-1}$ relative to the velocity of VLA~4B ($V_{\rm LSR}$ = +9.3~km~s$^{-1}$)\cite{diaz-rodriguez2022}. Thus, combining the proper motion of the HC3 arc and the LOS velocity of the Family III head, we obtain an inclination angle (angle between the LOS and the flow axis) $i=25^\circ\pm2^\circ$ for the head of Family III and HC3. Unfortunately, the proper motion of the second H$_2$ arc (HC2), which is associated with Family II, could not be measured by ref.~\cite{hodapp2014}, so we cannot obtain its inclination angle through the same procedure. Since families II, III, IV, V and VI all share a similar position angle of $\sim$160$^\circ$, it is reasonable to assume that all of them also have a similar value of the inclination angle ($i=25^\circ\pm 2^\circ$). Family I has a different PA, of 135$^\circ$, and might have a different inclination angle. Indeed, combining the observed IR proper motions and LOS velocities, a value of the inclination angle for the H$_2$ bubble (HC1) and the [FeII] jet was inferred by ref.~\cite{hodapp2014}, resulting $i=22^\circ\pm 4^\circ$ after updating the distance to 300 pc. Since HC1 is positionally associated with the head of Family I, the same inclination angle of $22^\circ\pm 4^\circ$ seems appropriate for this family.

Recently, the inclination and position angle of the disk of dust associated with VLA 4B has been measured\cite{reynolds2024} from a Gaussian $uv$ fit to the visibilities obtained from 1.3 mm ALMA observations with a very high angular resolution of 64 mas$\times$28 mas (19.2 au$\times$8.4 au). From these results, and assuming that the jet is perpendicular to the disk, we derive an inclination angle $i=20.3^\circ\pm 1.5^\circ$ and a position angle PA = $170^\circ\pm4^\circ$ for the jet from VLA 4B.

By fitting a bowshock model to families II, III, and VI (see ``Bowshock model'' section below in these Methods), we obtain for these families an independent estimate of the inclination angle of $i\simeq22^\circ$ (see Extended Data Table 1).

Thus, the IR proper motions (and LOS velocities of their positionally associated CO heads), the inclination angle of the disk of dust, and the fitting of a bowshock model, consistently suggest that the inclination angle of the SVS 13 EHV flow falls in the range $i$=20$^\circ$-25$^\circ$.

On the other hand, inclination angles of $i=24^\circ\pm 2^\circ$ for the Herbig-Haro object HH~11 and $i=25^\circ\pm 5^\circ$ for HH~7C (a substructure of HH~7) have been reported \cite{hartigan2019}.

\subsection*{Dynamical times of the heads of the families of rings.}

We have obtained the dynamical times of the heads, $t_{\rm head}$, of the families of rings (see Extended Data Fig.~4), assuming they have traveled with constant velocity in a straight direction along the jet axis, so
\begin{equation}\label{eq:thead}
	t_{\rm head}=\frac{x'_{\rm head}}{v'_{\rm head}\tan i},
\end{equation}
where $x'_{\rm head}$ is the projected distance (on the plane of the sky) from the source to the head, $v'_{\rm head}$ is its LOS velocity relative to VLA 4B, and $i$ is the inclination angle of the flow with respect to the LOS (we use $i=22^\circ\pm2^\circ$ for Family I and $i=25^\circ\pm2^\circ$ for the rest of the families, see ``On the inclination angle'' section in Methods). Both $x'_{\rm head}$ and $v'_{\rm head}$ are obtained by locating the maximum emission of each family of rings. For this, we first obtain several estimates of $x'_{\rm head}$ by locating the positions of the local maxima of emission near the jet axis in the peak intensity images (see second column of Fig~\ref{fig:mom}). Toward each of these positions we obtain spectra within a box with size of the order of the synthesized beam, and take $v'_{\rm head}$ as the LOS velocity of the emission peak. We then estimate the dynamical time of the head, $t_{\rm head}$, as the mean value of the dynamical times calculated using every measured pair of $x'_{\rm head}$ and $v'_{\rm head}$ values. The standard deviation of the dynamical times obtained in this way for a given head is a few years (as indicated in the error bars of Extended Data Fig.~4). 

Using this procedure, we obtain dynamical times $t_{\rm head}$ = 25$\pm$5, 56$\pm$6, 91$\pm$9, 100$\pm$9, $<$115, and 134$\pm$15~yr, for families I to VI, respectively. We note that we could not find a clear maximum associated with the head of Family V, which is expected to be located somewhere between the positions of the heads of families IV and VI, but in a velocity range only available in the lower resolution images. Hence, we give only an upper limit for the dynamical time of Family V, obtained by substituting $x'_{\rm head}$ and $v'_{\rm head}$ in equation (\ref{eq:thead}) by the central position and the LOS velocity relative to VLA 4B, respectively, of the rings of Family V observed in the highest velocity channel map of our high-angular resolution data.

\subsection*{Column density calculation from a rotational transition.}

The molecular column density can be obtained from the observation of a molecular transition following simple procedures \cite{estalella2008}. From the radiative transfer equation, the line intensity towards a given position in the sky, and for a given LOS velocity, $I_\nu(x',y',v_{z'})$, is given by 
\begin{equation}
	I_\nu (v_{z'}) = \left[B_\nu(T_\mathrm{ex}) - B_\nu(T_{\rm bg})\right] \left[1-e^{-\tau_\nu(v_{z'})}\right],
\label{eq:i1}
\end{equation}
where $B_\nu(T_\mathrm{ex})$ is the source function, assumed to be constant along the LOS, $B_\nu(T_{\rm bg})$ is the background intensity, and $\tau_\nu(v_{z'})$ is the LOS integrated optical depth as a function of the LOS velocity. For simplicity in the nomenclature, the explicit dependence on the position $(x',y')$ has been omitted in this and the following equations. 

From equation~(\ref{eq:i1}) we obtain:
\begin{equation}
\tau_v (v_{z'}) = - \ln{\left[1- \frac{I_\nu (v_{z'})}{B_\nu(T_\mathrm{ex}) - B_\nu(T_{\rm bg})}\right]}.
\label{eq:tau1}
\end{equation}

The optical depth of a transition between the rotational levels $J$$\rightarrow$$J$$-$1 is related to the column density of molecules in the $J$ level, as:
\begin{equation}
 \frac{d N_J(v_{z'})}{dv_{z'}} = \frac{8\pi\nu^3_{J,J-1}}{c^3 A_{J,J-1}} \left(e^{h\nu_{J,J-1}/k_B T_\mathrm{ex}}-1\right)^{-1} \tau_v (v_{z'}), 
\label{eq:dNJ}
\end{equation}
where $c$ is the speed of light, $h$ and $k_B$ are the Planck's and Boltzmann's constants, $A_{J,J-1} = (64/3) \pi^4  c^{-3} h^{-1} J\, (2J+1)^{-1} \mu_{\rm mol}^2 \nu^3_{J,J-1}$ is the Einstein spontaneous emission coefficient, $\mu_{\rm mol}$ is the dipole moment of the molecule$, \nu_{J,J-1}$ is the frequency of the transition, $d N_J(v_{z'})$ is the column density of molecules in the upper level $J$ per LOS velocity bin $d v_{z'}$, and $\tau_v (v_{z'})$ is given by equation~(\ref{eq:tau1}).

To obtain the total column density of the molecular species considered, we have to add the populations of all the rotational levels, $N_\mathrm{mol}(v_{z'})=\sum^\infty_{\mathrm{j=0}}N_j(v_{z'})$, that in local thermodynamic equilibrium is given by:
\begin{equation}
	N_\mathrm{mol}(v_{z'}) = \frac{N_J(v_{z'})}{2J+1}\, e^{(J+1)h\nu_{J,J-1}/2k_B T_\mathrm{ex}}\, Q(T_\mathrm{ex}),
\label{eq:nmol1}
\end{equation}
where $Q(T_\mathrm{ex})=\sum^\infty_{j=0}\left(2j +1\right) e^{-(j+1) h \nu_{j,j-1}/2k_B T_\mathrm{ex}}$ is the partition function. Therefore, using equations (\ref{eq:dNJ}) and (\ref{eq:nmol1}), we obtain: 
\begin{equation}
  d N_\mathrm{mol}(v_{z'}) = 
\frac{3 h Q(T_\mathrm{ex})}{8 \pi^3 J \mu_{\rm mol}^2}\,
\frac{e^{(J+1)h\nu_{J,J-1}/2k_B T_\mathrm{ex}}}{e^{h\nu_{J,J-1}/k_B T_\mathrm{ex}}-1}\,
\tau_v (v_{z'}) \, dv_{z'}, 
\label{eq:nmol2}
\end{equation}
where $\tau_v (v_{z'})$ is given by equation~(\ref{eq:tau1}). In practice, the observations are made with a finite angular resolution, so the measured intensities and, therefore, the derived column densities are indeed beam-averaged quantities. Note that beam dilution, because of a small beam filling factor, cannot be distinguished from a small optical depth. If the emission is truly optically thin a possible beam dilution does not affect the derived beam-averaged column density, but beam dilution would result in an underestimate of the column density if the emission is optically thick, since the opacity will be underestimated. In our analysis, we assume that the beam filling factor is $\sim$1 in the $0.1''$ resolution observations, which we think is a good approximation except perhaps in the lower part of the rings where the emission can have a large optical depth and a small beam filling factor (see below).

If the emission is optically thin, we are in Rayleigh-Jeans regime, the background is negligible ($T_\mathrm{ex} \gg T_\mathrm{bg}$), and the partition function is approximated as $Q(T_\mathrm{ex})\simeq 2 k_B J T_\mathrm{ex}/h \nu_{J,J-1}$, equation~(\ref{eq:nmol2}) simplifies as:
\begin{equation}
 d N_\mathrm{mol}(v_{z'}) \simeq \frac{3 c^2 k_B T_\mathrm{ex}}{8 \pi^3 h \nu_{J,J-1}^4 \mu_{\rm mol}^2}\, 
 I_\nu(v_{z'})\, e^{(J+1)h\nu_{J,J-1}/2k_B T_\mathrm{ex}}\, dv_{z'}.
\label{eq:nmolthin}
\end{equation}
Although in our observations the emission is mostly optically thin, we will use in our calculations the complete equation~(\ref{eq:nmol2}). 

Thus, from the CO ($J$=3$\rightarrow$2) observations, and using equation~(\ref{eq:nmol2}), we can obtain the beam-averaged CO column density, $N_{\rm CO}(x',y',v_{\rm ch},\Delta v_{\rm ch})$, associated with a given pixel with position $(x',y')$, in a channel map with central LOS velocity $v_{\rm ch}=-v_{z'}$, and channel width $\Delta v_{\rm ch}=|\Delta v_{z'}|$ as:
\begin{equation}
N_{\rm CO}(x',y',v_{\rm ch},\Delta v_{\rm ch}) = - \frac{h Q(T_\mathrm{ex})}{8 \pi^3 \mu_{\rm CO}^2}\, 
\frac{e^{2 h\nu_{32}/k_B T_\mathrm{ex}}}{e^{h\nu_{32}/k_B T_\mathrm{ex}}-1}\, \ln{\left[1- \frac{I_\nu({\rm CO}(3\mbox{-}2); x',y',v_{\rm ch})}{B_\nu(T_\mathrm{ex}) - B_\nu(T_{\rm bg})}\right]}\, \Delta v_{\rm ch}, 
\label{eq:NCO1}
\end{equation}
where $\mu_{\rm CO}$ is the dipole moment of the CO molecule, and $I_\nu({\rm CO}(3\mbox{-}2); x',y',v_{\rm ch})$ is the observed intensity of the CO ($J$=3$\rightarrow$2) transition toward the position considered.

\subsection*{Excitation temperature.} 

An estimate of the excitation temperature can be obtained from the observation of two rotational molecular transitions, $J$$\rightarrow$$J$$-$1 and $J'$$\rightarrow$$J'$$-$1. From Boltzmann's equation, and using equations (\ref{eq:tau1}) and (\ref{eq:dNJ}) to obtain the ratio of populations between the $J$ and $J'$ levels, we get:
\begin{equation}
 \frac{J'} {J} \, 
 \frac{e^{h\nu_{J',J'-1}/k_B T_\mathrm{ex}}-1} {e^{h\nu_{J,J-1}/k_B T_\mathrm{ex}}-1} \, 
 \frac{\ln{\left\{1- \frac{I_\nu(J\rightarrow J-1,\, v_{z'})}
{f_b \left[B_\nu(\nu_{J,J-1},\, T_\mathrm{ex}) - B_\nu(\nu_{J,J-1},\, T_{\rm bg})\right]} \right\}}}
 {\ln{\left\{1- \frac{I_\nu(J'\rightarrow J'-1,\, v_{z'})}
{f'_b \left[B_\nu(\nu_{J',J'-1}, \, T_\mathrm{ex})-B_\nu(\nu_{J',J'-1}, \, T_{\rm bg})\right]} \right\}}}
 = e^{-(E_J-E_{J'})/k_B T_\mathrm{ex}},
\label{eq:Jratio3}
\end{equation}
where $\nu_{J,J-1}$ and $\nu_{J',J'-1}$ are the frequencies, $I_\nu(J\rightarrow J-1,\, v_{z'})$ and $I_\nu(J'\rightarrow J'-1,\, v_{z'})$ the intensities, and $E_J$ and $E_{J'}$ the upper energy levels of the two transitions, $B_\nu(\nu,\, T)$ is the Planck function at frequency $\nu$ and temperature $T$, and $f_b$ and $f'_b$ the beam-filling factors at the native angular resolution of the $J$$\rightarrow$$J$$-$1 and $J'$$\rightarrow$$J'$$-$1 observations, respectively.

Assuming the solid angle of the source is the same in both transitions, equation (\ref{eq:Jratio3}) can be written in terms of the intensity of one of the transitions (e.g., the $J$$\rightarrow$$J$$-$1 transition) and the ratio $R_{J/J'} \equiv I'_\nu($$J$$\rightarrow$$J$$-$$1,v_{z'})/I_\nu($$J'$$\rightarrow$$J'$$-$$1,v_{z'}) = (f'_b/f_b)\, I_\nu($$J$$\rightarrow$$J$$-$$1,v_{z'})/I_\nu($$J'$$\rightarrow$$J'$$-$$1,v_{z'})$, where $I'_\nu($$J$$\rightarrow$$J$$-$$1,v_{z'})$ is the intensity of the $J$$\rightarrow$$J$$-$$1$ transition when convolved to the $J'$$\rightarrow$$J'$$-$$1$ beam, resulting:
\begin{equation}
 \frac{J'} {J} \, 
 \frac{e^{h\nu_{J',J'-1}/k_B T_\mathrm{ex}}-1} {e^{h\nu_{J,J-1}/k_B T_\mathrm{ex}}-1} \, 
 \frac{\ln{\left\{1- \frac{I_\nu(J\rightarrow J-1,\, v_{z'})}
{f_b \left[B_\nu(\nu_{J,J-1},\, T_\mathrm{ex})-B_\nu(\nu_{J,J-1},\, T_{\rm bg})\right]} \right\}}}
 {\ln{\left\{1- \frac{I_\nu(J\rightarrow J-1,\, v_{z'})}
{f_b R_{J/J'} \left[B_\nu(\nu_{J',J'-1}, \, T_\mathrm{ex})-B_\nu(\nu_{J',J'-1}, \, T_{\rm bg})\right]} \right\}}}
= e^{-h \nu_{J,J-1}/k_B T_\mathrm{ex}}.
\label{eq:Jratio4}
\end{equation}
 In this way, only the filling factor of the (unconvolved) $J$$\rightarrow$$J$$-$1 transition, $f_b$, is needed. This approach is particularly useful when this transition is observed at very high angular resolution, so $f_b \simeq 1$. This equation can be solved numerically to obtain $T_\mathrm{ex}$.

For the particular case of the $J$$\rightarrow$$J$$-$1 = 3$\rightarrow$2 and $J'$$\rightarrow$$J'$$-$1 = 2$\rightarrow$1 transitions we have:
\begin{equation}
 \frac{2} {3} \, 
 \frac{e^{h\nu_{21}/k_B T_\mathrm{ex}}-1} {e^{h\nu_{32}/k_B T_\mathrm{ex}}-1} \, 
 \frac{
\ln{\left\{1- \frac{I_\nu(3\rightarrow2, \, v_{z'})}{f_{32} 
\left[B_\nu(\nu_{32}, \, T_\mathrm{ex}) - B_\nu(\nu_{32}, \, T_{\rm bg})\right]} \right\}}}
{\ln{\left\{1- \frac{I_\nu(3\rightarrow2, \, v_{z'})}{f_{32}
R_{3/2} 
\left[B_\nu(\nu_{21}, \, T_\mathrm{ex}) - B_\nu(\nu_{21}, \, T_{\rm bg})\right]} \right\}}}
 = e^{-h \nu_{32}/k_B T_\mathrm{ex}}.
\label{eq:Jratio5}
\end{equation}

CO(2$\rightarrow$1) line observations of the SVS~13 bullets, with a synthesized beam of $2.8'' \times 2.6''$, were carried out with the Submillimeter Array (SMA)\cite{chen2016}. From the contour levels in the channel maps presented in ref.~\cite{chen2016}, we estimated the CO(2$\rightarrow$1) intensity at different positions and LOS velocities. After convolution of our CO(3$\rightarrow$2) data to match the SMA beam, we estimated the intensity ratio $R_{3/2}$ at these positions and velocities. Then, solving equation (\ref{eq:Jratio5}) we obtained $T_\mathrm{ex}$ using the $R_{3/2}$ ratios and the measured CO(3$\rightarrow$2) intensities toward the same positions in our high-angular resolution ALMA observations where, with a native resolution of $0.173''\times0.091''$, the emission fills well the beam, and for which we therefore assume a beam-filling factor $f_{32}\simeq1$. We obtained typical values of the ratio $R_{3/2} \gtrsim 4$, and CO(3$\rightarrow$2) intensities in our high-angular resolution ALMA observations of $I_\nu$(3$\rightarrow$2) $\simeq$ 1 Jy arcsec$^{-1}$, resulting in excitation temperatures $T_\mathrm{ex} \gtrsim 100$ K. We note that, unfortunately, for the parameter space corresponding to our data, the inferred $T_\mathrm{ex}$ is very sensitive to uncertainties in the intensity ratio $R_{3/2}$, especially if overestimated (e.g., an overestimate of 15\% in $R_{3/2}$ yields an overestimate of $T_\mathrm{ex}$ by a factor of 2.4). We note that these temperatures are considerably higher than the value of $T_\mathrm{ex}$ = 25~K previously obtained from much lower resolution ($20''$-$30''$) data\cite{masson1990}. This value was likely underestimated since our CO(3$\rightarrow$2) images show intensity peaks of 80 mJy beam$^{-1}$, corresponding to brightness temperatures of $\sim$60~K. Since the CO(3$\rightarrow$2) emission is not necessarily optically thick and/or does not completely fill the beam, this sets a lower limit of $\sim$60 K to the excitation temperature, in agreement with the value inferred from the ratio of CO(3$\rightarrow$2)/CO(2$\rightarrow$1) intensities. Therefore, we adopt $T_\mathrm{ex}$ = 100~K in our calculations.

\subsection*{Determination of the mass.} 

We can estimate the mass associated to each pixel $(x',y')$ in a given channel image of velocity $v_{\rm ch}$ as:
\begin{equation}
 \Delta M(x',y',v_{\rm ch}) = \mu_\mathrm{H_2} m_\mathrm{H} X_\mathrm{CO}^{-1} D^2 N_\mathrm{CO}(x',y',v_{\rm ch}) \Delta x' \Delta y', 
\label{eq:dm}
\end{equation}
where $X_\mathrm{CO}$ is the CO abundance relative to molecular hydrogen, assumed to be $8.5 \times 10^{-5}$ (ref.~\cite{frerking1982}), $\mu_\mathrm{H_2}$=2.8 is the mean molecular mass per hydrogen molecule, and takes into account helium and metals that are heavier but less abundant than H$_2$ (ref.~\cite{kauffmann2008}), $m_\mathrm{H}$ is the mass of the hydrogen atom, $D$ is the distance to the source, and $\Delta x'$ and $\Delta y'$ are the angular dimensions of a pixel in the image. $N_\mathrm{CO}(x',y',v_{\rm ch})$ is the column density of CO molecules that can be obtained from equation (\ref{eq:NCO1}) using $T_\mathrm{ex}$ = 100 K, as inferred in the above section. The mass of a given family is obtained by adding the masses of the pixels identified as belonging to that family (in case of intersection of rings in a pixel, the mass is equally split between the corresponding families). We do not expect the adopted value of $T_\mathrm{ex}$, or its possible gradients within the source, to have a large impact on the derived masses, since using a value of $T_\mathrm{ex}$=50 K (even smaller than the lower limit of 60 K set by the observed brightness temperatures; see previous section) the obtained masses drop by only 30\%.

Using equation (\ref{eq:dm}) we derived the masses of the families, obtaining 6.7$\times 10^{-5}$, 2.3$\times 10^{-4}$, 3.1$\times 10^{-4}$, 9.5$\times 10^{-5}$, and 3.2$\times 10^{-4}~M_\odot$, for families I, II, III, IV, and VI, respectively. For Family V, only a lower limit of 5.0$\times 10^{-7}~M_\odot$ is given, corresponding to the mass of the rings that could be identified in the high resolution data. The head, that should appear at higher velocities, only available in the low resolution data, could not be identified in these data. We present these values of the mass, along with other properties of features associated with the SVS~13 outflow in Supplementary Table 1.

\subsection*{Estimation of the mass accretion rate.}

Assuming that the bolometric luminosity, $L_\mathrm{bol}$, is mostly due to accretion, the mass-accretion rate can be calculated as \cite{hartmann2009}
\begin{equation}
	\dot{M}_\mathrm{acc} \simeq \frac{R_*L_\mathrm{acc}}{GM_*}, 
\end{equation}
where we adopt a mass of the VLA 4B protostar $M_*=0.6~M_\odot$ (ref.~\cite{diaz-rodriguez2022}), a radius $R_*$ $\simeq$ 3-5~$R_\odot$ (ref.~\cite{palla1993}), and an accretion luminosity $L_\mathrm{acc}\simeq L_\mathrm{bol}\simeq 50~L_\odot$ (ref.~\cite{enoch2009}), resulting in a mass accretion rate $\dot{M}_\mathrm{acc} = 0.8$-$1.3 \times 10^{-5}~M_\odot~\mathrm{yr}^{-1}$.

\subsection*{Momentum-conserving bowshock model.}\label{sec:model}

Here we summarize the foundations and properties of the analytic, momentum-conserving bowshock model of ref.~\cite{tabone2018}, that we compare with our data. This model was initially developed for the leading bowshock at the jet head\cite{masson1993, ostriker2001}. It was recently adapted by ref.~\cite{tabone2018} to the case of a bowshock generated by a jet IWS in a slower moving ambient medium. The IWS is a two-shock structure generated inside the jet by a velocity jump that propagates downstream at the mean jet speed (see ref.~\cite{raga1990}). As illustrated in Extended Data Fig.~5, overpressured shocked jet material is then ejected sideways from the IWS, where it intercepts slower-moving ambient material that sweeps it back into a curved bowshock. Considering the bowshock as a stationary, thin shell of well-mixed material, its shape and velocity field can then be derived self-consistently from mass and momentum balance. The corresponding equations (ref.~\cite{tabone2018}) are recalled below, and show that this model is quite constrained, with only six free adjustable parameters to fit a given family of rings. In addition, we provide in the following subsections new derivations of the predicted bowshock mass and surface density, not presented by ref.~\cite{tabone2018}. They enable us to infer the ambient density and jet mass-flux for each ring family fitted with this model, and to explain the observed ring brightness asymmetries.

Thus, we consider an IWS moving in the $z$-axis direction with velocity $v_{\rm jet}$, and ejecting material sideways at velocity $v_0$ and mass rate ${\dot M_0}$ into a uniform ambient medium of density $\rho_{\rm amb}$, which is also moving away from the outflow source at a velocity $v_{\rm amb}$ parallel to $v_{\rm jet}$. The working surface is located at distance $z_{\rm ws}$ from the source, and we consider a reference frame ($x^*, r$) co-moving with the working surface, where $x^*=z_{\rm ws}-z$. The bowshock wings are formed by the interaction of the jet material ejected sideways from the working surface with the streaming ambient material that impinges on it at a relative velocity $v_{\rm jet}-v_{\rm amb}$ (see Extended Data Fig.~5). Assuming that the bowshock forms a stationary, well-mixed thin shell, its radius at each position, $r_b(x^*)$, results from the mass and $(x^*,r)$-momentum conservation equations (equations (1)-(3) of ref.~\cite{tabone2018}, in the narrow jet limit where $r_{\rm jet}\rightarrow 0$ ):
\begin{eqnarray} 
	{\dot M} & = & {\dot M}_0+ {\dot M}_{\rm amb} = {\dot M}_0+\pi\, r_b^2\, \rho_{\rm amb}(v_{\rm jet}-v_{\rm amb})\,, \label{mcon} \\
	{\dot \Pi}_{x^*} & = & {\dot M}_{\rm amb} (v_{\rm jet}-v_{\rm amb}) = 
 \pi\, r_b^2\, \rho_{\rm amb}(v_{\rm jet}-v_{\rm amb})^2={\dot M}\, v_{x^*}\,, \label{xcon} \\
  {\dot \Pi}_r & = & {\dot M_0}\, v_0={\dot M}\, v_r\,,
  \label{rcon}
\end{eqnarray}
where ${\dot M}$, ${\dot \Pi}_{x^*}$, and ${\dot \Pi}_r$ are the total mass rate, $x^*$-momentum rate, and $r$-momentum rate (respectively) flowing along the thin shell up to a given value of $x^*$. The velocities $v_{x^*}$ and $v_r$ are the components of the well mixed material within the shell along the $x^*$- and $r$-axes (respectively), and 
\begin{equation}
{\dot M}_{\rm amb} \equiv \pi\, r_b^2\, \rho_{\rm amb}(v_{\rm jet}-v_{\rm amb})  \,
  \label{eq:mw}
\end{equation}
is the mass-rate of fresh ambient material swept into the bowshock, from its apex up to a given radius $r_b(x^*)$ (see Extended Data Fig.~5). We then combine the two $(x^*,r)$-momentum equations (equations (\ref{xcon}) and (\ref{rcon})) to obtain the differential equation for the bowshock shape:
\begin{equation}
	\frac{dr_b}{dx^*} \equiv \tan\alpha  = \frac{v_r}{v_{x^*}} = 
  \frac{{\dot \Pi}_r}{{\dot \Pi}_{x^*}}
= \frac{{\dot M}_0 \, v_0}
	{\pi\, r_b^2 \, \rho_{\rm amb}(v_{\rm jet}-v_{\rm amb})^2}\,,
    \label{drb}
\end{equation}
where $\alpha$ is the angle between the jet axis  and the local tangent to the bow surface at $x^*$ (see Extended Data Fig.~5). Equation (\ref{drb}) can be directly integrated to obtain the shape of the bowshock wings:
\begin{equation}
	r_b(x^*)=\left(L_0^2 \, x^*\right)^{1/3},   
  \label{rb}
\end{equation}
with
\begin{equation}
	L_0\equiv \sqrt{\frac{3{\dot M}_0 \, v_0}{\pi\, \rho_{\rm amb}(v_{\rm jet}-v_{\rm amb})^2}}
  \label{l0}
\end{equation}
being the characteristic scale\cite{tabone2018}, which controls the radius of curvature of the bowshock. Using the expression for $\tan \alpha$ in equation~(\ref{drb}), the two components of velocity at each position $x^*$ can also be obtained as:
\begin{eqnarray}
 	v_{x^*} & = & \frac{{\dot \Pi}_{x^*}}{\dot M} = 
  \frac{\pi\, r_b^2\, \rho_{\rm amb}(v_{\rm jet}-v_{\rm amb})^2}{\dot M_0+\pi\, r_b^2\, \rho_{\rm amb} (v_{\rm jet}-v_{\rm amb})} 
  = \frac{1}{\frac{1}{v_{\rm jet}-v_{\rm amb}} + \frac{1}{v_0} \tan \alpha} \,,
\label{vx} \\ 
  	v_r & = & \tan \alpha\, v_{x^*} \,.
   \label{vr}
\end{eqnarray}
where $\tan \alpha$ at position ${x^*}$ is inferred from equation (\ref{rb}) as
\begin{equation}
\tan\alpha \equiv \frac{dr_b}{dx^*} = \frac{1}{3} \left( \frac{L_0}{x^*}\right)^{2/3} 
= \frac{1}{3} \left( \frac{L_0}{r_b}\right)^{2}.
    \label{tangalpha}
\end{equation}

\subsubsection*{Bowshock model fit to ring families}

Equations (\ref{rb}), (\ref{vx}), (\ref{vr}), and (\ref{tangalpha}) show that the momentum-conserving jet-driven bowshock model is tightly constrained. In the assumed narrow jet regime ($r_{\rm jet}\rightarrow 0$, equivalent to $r_{\rm jet}\ll L_0$),  only three parameters determine the entire bowshock shape and kinematics in the comoving reference frame: (i) the size scale, $L_0$, of the bowshock (defined by equation (\ref{l0})); (ii) the velocity, $v_0$, at which the jet material is ejected sideways by the IWS; (iii) the relative velocity, $v_{\rm jet}$$-$$v_{\rm amb}$, between the jet working surface and the surrounding ambient medium. Three additional parameters determine how the bowshock appears in the observer's frame: (iv) the propagation speed, $v_{\rm jet}$, of the IWS with respect to the source (giving $v_z = v_{\rm jet}$$-$$v_{x^*}$); (v) the distance, $z_{\rm ws}$, of the working surface (= bowshock apex) from the source (giving $z = z_{\rm ws} - x^*$); (vi) the inclination angle, $i$, of the jet axis with respect to the LOS (determining the projected shape and position in the sky of the observed rings at each LOS velocity, see Fig.~\ref{fig:geometry}). 

Taking as free parameters  $L_0$, $v_0$, $v_{\rm jet}$, $z_{\rm ws}$, $i$, and $v_{\rm amb}$, and using the elliptical fits performed to the observed rings (see section ``Elliptical fits'' in Methods, and Supplementary Data 1), we obtained the best-fit models for the best defined families of rings (families II, III, and VI). To do this, we performed a least-square minimization to (i) the mean ring radius as a function of the projected distance to the source (second column of Fig.~\ref{fig:modeleli}), and to (ii) the positions of the two intersections of each ring with the $x'$-axis as a function of the LOS velocity (third column of Fig.~\ref{fig:modeleli}).

The best-fit model parameters are listed in Extended Data Table 1. We obtain, for all the three families, a similar value for the inclination angle, $i\simeq22^\circ$, and for the velocity of the ambient, $v_{\rm amb} \simeq 0$~km~s$^{-1}$ (with 1-$\sigma$ upper limits of 8, 5, and 20 km~s$^{-1}$ for families II, III, and VI, respectively). For the rest of the parameters, we obtain: $L_0=0.40''\pm0.02''$, $v_0=23\pm3$~km~s$^{-1}$, $v_{\rm jet}=112~\pm4$~km~s$^{-1}$, $z_{\rm ws}=4.6''\pm0.4''$ (Family II); $L_0=0.51''\pm0.03''$, $v_0=20\pm2$~km~s$^{-1}$, $v_{\rm jet}=111\pm4$~km~s$^{-1}$, $z_{\rm ws}=7.0''\pm 0.6''$ (Family III); and $L_0=1.73''\pm0.07''$, $v_0=20\pm3$~km~s$^{-1}$, $v_{\rm jet}=137\pm3$~km~s$^{-1}$, $z_{\rm ws}=15''\pm 1''$ (Family VI).

In addition to the success of the bowshock model in reproducing the observed shape of the shells (equation (\ref{rb})) and their kinematics (equations (\ref{vx}) and (\ref{vr})) (see Fig.~5), we note that the bounding envelope of the outer edges of the observed shells has a conical geometry (as suggested by the gray line in the bottom left panel of Fig.~\ref{fig:ellipses}) that is in agreement with what is expected in a bowshock scenario, as reproduced in numerical simulations of jet-driven shells\cite{rabenanahary2022}.

\subsubsection*{Derivation of the ambient density and jet mass flux from the bowshock mass.}

Below we show that once the free parameters $L_0$, $v_0$, and $v_{\rm jet}$$-$$v_{\rm amb}$ are determined from the bowshock shape and kinematics (as outlined above), the density of the ambient gas, $\rho_{\rm amb}$, can be estimated from the mass of the bowshock shell $M_f$ (i.e., the mass of a given family of rings), measured from the observations (see ``Determination of the mass'' in Methods). The mass rate of jet material initially ejected sideways by the working surface, $\dot{M}_0$, and the mass rate of ambient material being incorporated into the bowshock, $\dot{M}_{\rm amb}$ (see equation (\ref{eq:mw})), can also be derived. These inferred physical parameters provide a useful check of the plausibility of the bowshock interpretation.

In the narrow jet regime ($r_{\rm jet}\ll L_0$), the mass of the shell can be obtained by integrating the mass rate flowing along the shell, $\dot M=\dot{M}_0+\dot{M}_{\rm amb}$ (equation (\ref{mcon})), over the flow crossing time along the bowshock: 
\begin{equation}
\label{eq:intmass_mr}
	M_f=\int_{t_0}^{t_f}\dot{M}\, dt = \int_0^{r_{f}}\dot{M} \, \frac{dr_b}{v_r}, 
\end{equation}
where $r_f$ is the radius at the outer edge of the bowshock shell, $t_f$$-$$t_0$ the time it takes for a fluid parcel to flow from the tip of the bow to $r_f$, and $v_r=dr_b/dt$.\\  
From the conservation of $r$-momentum, given by equation (\ref{rcon}), we have 
\begin{equation}
	v_r = \frac{\dot{M}_0 \, v_0}{\dot{M}},
 \label{eq:vr}
\end{equation}
and from the definition of $\dot M_{\rm amb}$ (equation (\ref{eq:mw})) and $L_0$ (equation (\ref{l0})), the mass conservation (equation (\ref{mcon})) can be written as:
\begin{equation}\label{eq:mdotm0}
	\dot{M} = \dot{M}_0+\dot{M}_{\rm amb} = \dot{M}_0\left[1+\frac{3}{\gamma}\left(\frac{r_b}{L_0}\right)^2\right],
\end{equation}
with
\begin{equation}
	\gamma\equiv \frac{v_{\rm jet}-v_{\rm amb}}{v_0}\,.
  \label{gam}
\end{equation}
Substituting equations (\ref{eq:vr}) and (\ref{eq:mdotm0}) in equation (\ref{eq:intmass_mr}), we obtain 
\begin{equation}
M_f=\int_0^{r_{f}}\frac{\dot{M}_0}{v_0}\left[1+\frac{3}{\gamma}\left(\frac{r_b}{L_0}\right)^2\right]^2\, dr_b.
\end{equation}
 Using the change of variable $u=(3/\gamma)^{1/2} (r_b/L_0)$ and resolving the integral:
\begin{equation}
M_f=\frac{\dot{M}_0 \, L_0}{v_0}\left(\frac{\gamma}{3}\right)^{1/2}\int_0^{u_f}\left(1+u^2\right)^2\, du= \frac{\dot{M}_0 \, L_0}{v_0}\left(\frac{\gamma}{3}\right)^{1/2}\left(\frac{u_f^5}{5} + \frac{2u_f^3}{3} + u_f\right),
\end{equation}
where $u_f=(3/\gamma)^{1/2} (r_{f}/L_0)$. Using equation (\ref{l0}) to eliminate $\dot{M}_0$, and the definition of $\gamma$ in equation (\ref{gam}), we can calculate the density of the ambient medium, $\rho_{\rm amb}$, as a function of the mass of the shell, $M_f$, as:
\begin{equation}\label{eq:rhow_mr}
	\rho_{\rm amb} = \frac{M_f}{3\pi L^2_0 r_f} \left(\frac{\gamma}{3}\right)^{-2}\left(\frac{u^4_f}{5}+\frac{2 u^2_f}{3} + 1\right)^{-1}. 
\end{equation}
Note that, when $r_{f}\gg L_0$ ($u_f\gg 1$), $\rho_{\rm amb}$ can be approximated as:
\begin{equation}
	\rho_{\rm amb} \simeq \frac{M_f}{3\pi L^2_0 r_f} \left(\frac{\gamma}{3}\right)^{-2}\left(\frac{u^4_f}{5}\right)^{-1} = \frac{5}{3\pi} \frac{M_f\, L_0^2}{r_{f}^5}. 
\end{equation}
Once we have an estimate of $\rho_{\rm amb}$, we can calculate from equation (\ref{l0}) the mass rate of jet material ejected from the working surface, $\dot{M}_0$, as:
\begin{equation}\label{eq:mdot0_rhow}
	\dot{M}_0 = \pi\, L_0^2 \, \rho_{\rm amb} \frac{(v_{\rm jet}-v_{\rm amb})^2}{3\, v_0}=\frac{M_f v_0}{r_f}\left(\frac{u^4_f}{5} + \frac{2u^2_f}{3}+1\right)^{-1}.
\end{equation}
Note that, from equations (\ref{eq:mdot0_rhow}), (\ref{eq:mw}), and (\ref{tangalpha}), the ratio of the jet mass rate to the ambient mass rate up to a given radius $r_b$ is given by
\begin{equation}
	\frac{\dot{M}_0}{\dot{M}_{\rm amb}} = \frac{\gamma}{3}\left(\frac{L_0}{r_b}\right)^2 = \gamma\,  \tan\alpha,
 \label{m0mamb}
\end{equation}
so only the tip of the bowshock, where $r_b<L_0 \, (\gamma/3)^{1/2}$, is dominated by ejected jet material. 

Evaluating equations (\ref{eq:mw}) and (\ref{m0mamb}) at the maximum radius of the shell, $r_{f}$, we can obtain the mass rate of ambient material incorporated over the whole wings of the bowshock:
\begin{equation}\label{eq:mw_f}
	\dot{M}_{{\rm amb}, f}= \pi\, r_{f}^2 \, \rho_{\rm amb} (v_{\rm jet}-v_{\rm amb})
 = \dot{M}_0 \frac{3}{\gamma}\left(\frac{r_f}{L_0}\right)^2 \,.
\end{equation}

Taking $r_f=0.75''$, $1.09''$, and $2.5''$ (for families II, III, and VI, respectively), we obtain $\rho_{\rm amb}$, $\dot{M}_0$, and $\dot{M}_{{\rm amb}, f}$ from equations (\ref{eq:rhow_mr}), (\ref{eq:mdot0_rhow}), and (\ref{eq:mw_f}), using the fitted bowshock parameters and the observed mass of each family, $M_f$ (see Methods and Extended Data Table 1). We obtain ambient densities $\rho_{\rm amb}$ = 5.0$\times$10$^{-19}$, 1.7$\times$10$^{-19}$, and 9.8$\times$10$^{-21}$~g~cm$^{-3}$ for families II, III, and VI, respectively. Thus, we derive steadily decreasing values of $\rho_{\rm amb}$ with distance from the source (see values of $z_{\rm ws}$ in Extended Data Table~1), which is physically consistent with the source being an embedded object, located in the central, denser region of a molecular core. Furthermore, we note that the values of the ambient density obtained from our model are in agreement with the estimates of 3.3$\times$10$^{-20}$~g~cm$^{-3}$ inferred from NH$_3$ observations ($\sim 3''$ resolution) with the VLA (A.K. Diaz-Rodriguez and G. Anglada, personal communication).

While $\rho_{\rm amb}$ drops by a factor of $\sim$50 with distance, $L_0$ increases in such a way that equation (\ref{eq:mdot0_rhow}) yields values of the same order for the mass ejection rate from the working surface in all three families, $\dot{M}_0$ = 1.4$\times$10$^{-6}$, 9.6$\times$10$^{-7}$, and 9.3$\times$10$^{-7}~M_\odot~\mathrm{yr}^{-1}$ for families II, III, and VI, respectively. These values are consistent with a simple picture where the successive bowshocks are driven by similar, recurrent jet outburst episodes. 

Taking $\dot{M}_0$ as an estimate of the mass-loss rate of the blueshifted jet \cite{tabone2018}, we infer a (two-sided) mass loss rate $\dot{M}_\mathrm{jet}$ $\simeq 2\times10^{-6}~M_\odot~\mathrm{yr}^{-1}$. It is worth noting that, since the mass accretion rate inferred for VLA 4B is $\dot{M}_\mathrm{acc}$ = 0.8-1.3 $\times$ 10$^{-5}~M_\odot~\mathrm{yr}^{-1}$ (see Methods), we obtain a (two-sided) ejection to accretion ratio of $\dot{M}_\mathrm{jet}/\dot{M}_\mathrm{acc}\simeq 0.2$, fully consistent with the typical ratio in protostellar jets \cite{cabrit2007,lee2020}. Regarding the mass rate at which ambient material is currently being incorporated into the observed portion of the bowshock, we obtain $\dot{M}_{{\rm amb}, f}$ = 3.2$\times10^{-6}$, 2.5$\times10^{-6}$, and 9.2$\times10^{-7}~M_\odot~\mathrm{yr}^{-1}$ for families II, III, and VI, respectively. While $\dot{M}_{{\rm amb}, f} \simeq 2\,\dot{M}_0$ for families II and III, we obtain $\dot{M}_{{\rm amb}, f} \simeq \dot{M}_0$ for Family VI. This occurs because the ratio $({r_f}/{L_0})$ entering equation (\ref{eq:mw_f}) is smaller in Family VI, where the rings with largest radii are faint and difficult to identify. Therefore, we could be missing a substantial part of the mass in the bowshock wings, that are predominantly composed of ambient material.

In summary, the momentum-conserving jet bowshock models that best reproduce the  morphology and kinematics of each shell can also naturally reproduce their observed masses, given the ambient density and accretion rate in SVS~13-VLA~4B. This success lends strong support to the proposed interpretation.

\subsubsection*{Predicted bowshock surface density and CO intensity in the channel maps.}

Below we derive another analytical property of the bowshock model, not presented in ref.~\cite{tabone2018}, namely the shell surface density, $\sigma$, at each position. We note that $\dot{M} = 2 \pi\, r_b \, \sigma \, v_t$, where the tangential velocity along the thin shell is given by $v_t=v_{x^*}/\cos\alpha$, and use equations (\ref{drb}), (\ref{vx}), (\ref{vr}) to calculate the surface density of the shell as
\begin{equation}
	\sigma=\frac{1}{2} \, \rho_{\rm amb} \, r_b \,\cos\alpha\left(\gamma \, \tan\alpha + 1\right)^2 \,,
	\label{sig}
\end{equation}
where $r_b(x^*)$ is given by equation (\ref{rb}), the constant $\gamma$ was defined in equation (\ref{gam}), $\tan\alpha(x^*)$ is given by equation (\ref{tangalpha}), and 
\begin{equation}
	\cos\alpha= \left(1+\tan^2\alpha\right)^{-1/2} = \left(1+\frac{L_0^4}{9\, r_b^4}\right)^{-1/2}.
  \label{cal}
\end{equation}
We then have a full solution giving the shape (equation (\ref{rb})), velocity (equations (\ref{vx}) and (\ref{vr})) and surface density (equation (\ref{sig})) for a thin shell bowshock flow. 

Once we know the shape of the bowshock shell, with the velocity and surface density at all its points, we can obtain their projected positions on the plane of the sky and their LOS velocities. In this way, we can assign each point of the shell to a pixel in a given channel map of velocity width $\Delta v_{\rm ch}$, obtaining the mass and column density for each of the pixels of the model channels. In practice, a thermal+turbulent velocity dispersion, $v_T$, is expected to be present so that the surface density is spread over a range of velocities, which is taken into account when calculating the distribution in velocity channels. In our case, we find a 3D velocity dispersion $v_T\simeq 2$~km~s$^{-1}$, which corresponds to a LOS velocity dispersion of $\simeq 1.2$~km~s$^{-1}$. From the CO column density, obtained through its abundance relative to H$_2$, we can derive the CO optical depth and intensity predicted by the model using equations~(\ref{eq:nmol2}) and (\ref{eq:i1}), respectively. Finally, channel maps directly comparable with observations can be simulated by convolving the intensities with the appropriate beam.

\subsection*{Line opacity and ring brightness asymmetry.}\label{sec:opacity}

An outstanding feature of the observed rings is that, in general, they appear brighter in the side closer in projection to the origin. This is well reproduced by the bowshock model, and can be understood in terms of opacity and beam filling-factor effects due to the geometry of the bowshock relative to the observer, as illustrated in Extended Data Fig.~6.

As shown in the top row of this figure (left and central panels), because of the geometry and orientation of the SVS 13 bowshocks with respect to the observer, the ring emitting at a given LOS velocity (i.e., in a given channel map) appears spread over a wider region on the side closer in projection to the origin (smaller values of $x'$) than on the opposite side of the ring (larger values of $x'$), where the emission appears projected over a very narrow arcuate region. This projection effect results in a higher column density (and optical depth) towards the narrower side of the ring. 

The middle row of Extended Data Fig.~6 shows the predicted emission maps if the emission was optically thin all over the ring: with infinite resolution (left panel of middle row) the narrower side, of higher optical depth, would appear brighter. When convolved with a finite beam (central panel in middle row), however, the beam-integrated flux from each side of the ring would be the same, and the image of the ring would appear symmetric (except for a local effect due to the elongation of the beam)\cite{osorio2014}. The higher surface brightness in the narrow emitting region would be fully compensated by its smaller beam filling-factor.

In contrast, the bottom row of Extended Data Fig.~6 shows the predicted emission maps when the effects of finite optical depth, $\tau$, are taken into account. The intensity toward a given line of sight is lower by a factor of $[1-\exp{(-\tau)}]/\tau$ with respect to the optically thin regime. The decrease in intensity is more noticeable on the narrow side of the ring (where the opacity is higher), reducing the brightness contrast between the two sides at infinite resolution (left panel in bottom row). When the emission is convolved with a finite beam (see central panel in bottom row), an asymmetry is produced because the emission in the extended region (on the side closer to the origin) fills the beam better than on the side of the ring farthest from the origin, which is distributed in a very narrow region. The last column in the figure includes noise in the model images (middle and bottom)
for a more realistic comparison with the observations (top).

\section*{Data availability}
The ALMA raw data are available from the ALMA Science Archive (https://almascience.eso.org/aq/), using the project identifiers 2015.1.01229.S (for Cycle 3 data) and 2016.1.101305.S (for Cycle 4 data). The reduced data is available via Zenodo (ref. \cite{zenodo2025}) at https://zenodo.org/records/17249853.

\section*{Code availability}
This work has made extensive use of the open-source Python libraries NumPy \cite{harris2020}, SciPy \cite{virtanen2020}, Matplotlib \cite{hunter2007}, Photutils \cite{bradley2022}, Astropy \cite{astropy2013,astropy2018,astropy2022}, and {\sc LMFIT} \cite{newville2014}. The code used to compute the channel maps of the bowshock model is available at GitHub \cite{gblazquez2025}.

%\bibliography{biblio}

\section*{Acknowledgements}

We dedicate this work to the memory of our dear colleagues and friends, Alejandro (Alex) Raga and Robert Estalella. Their passion and brilliance were central to the theoretical foundations of this research and played a crucial role in shaping the article as it stands today. This manuscript owes much of its depth and rigour to their insight, dedication, and collaboration. Although they did not live to see its publication, their spirit and intellectual legacy live on in every page. We remain deeply grateful for the time we shared and the science we built together. We thank Benoit Tabone for his useful comments and suggestions on this work. G.A., G.B.-C., IdG-M., A.K.D.-R., G.A.F., J.F.G., M.O., acknowledge financial support from grants PID2020-114461GB-I00, PID2023-146295NB-I00, and CEX2021-001131-S, funded by MCIN/AEI/10.13039/501100011033. G.B.-C., G.A.F., and M.O. acknowlege financial support from Junta de Andalucia (Spain) grant P20-00880 (FEDER, EU). G.B-C acknowledges support from grant PRE2018-086111, funded by MCIN/AEI/ 10.13039/501100011033 and by `ESF Investing in your future', and thanks ESO Science Support Discretionary Fund for their finantial support under the 2024 SSDF 06 project. S.C. acknowledges support from Conseil Scientifique of Observatoire de Paris and from the Programme National de Physique et Chimie du Milieu Interstellaire (PCMI) of CNRS/INSU co-funded by CEA and CNES. A-K.D.R acknowledges support from STFC Grant ST/T001488/1. G.A.F. also acknowledges support from the Collaborative Research Centre 956, funded by the Deutsche Forschungsgemeinschaft (DFG) project ID 184018867, the DFG for funding through SFB 1601 ``Habitats of massive stars across cosmic time'' (sub-project B1), and the University of Cologne and its Global Faculty programme. R.E. acknowledges partial financial support from the grants PID2020-117710GB-I00 and CEX2019-000918-M funded by MCIN/ AEI /10.13039/501100011033. J.M.T. acknowledges support from the PID2023-146675NB grant funded by MCIN/AEI/10.13039/501100011033 and by the programme Unidad de Excelencia Mar\'{\i}a de Maeztu CEX2020-001058-M. L.F.R. acknowledges support from grant CBF-2025-I-2471 of SECIHTI, Mexico. L.A.Z. acknowledges financial support from CONACyT-280775, UNAM-PAPIIT IN110618, and IN112323 grants, M\'exico. This work makes use of the following ALMA data: ADS/JAO.ALMA\#2015.1.01229.S, ADS/JAO.ALMA\#2016.1.01305.S. ALMA is a partnership of ESO (representing its member states), NSF (USA) and NINS (Japan), together with NRC (Canada) and NSC and ASIAA (Taiwan) and KASI (Republic of Korea), in cooperation with the Republic of Chile. The Joint ALMA Observatory is operated by ESO, AUI/NRAO and NAOJ. This publication use images based on observations made with the NASA/ESA Hubble Space Telescope, and obtained from the Hubble Legacy Archive, which is a collaboration between the Space Telescope Science Institute (STScI/NASA), the Space Telescope European Coordinating Facility (ST-ECF/ESA) and the Canadian Astronomy Data Centre (CADC/NRC/CSA). 

\section*{Author Contributions}
G.B.-C. led the data reduction, analysis, modeling, interpretation, and writing of the paper. G.A, S.C., M.O, A.C.R., G.A.F, and R.E. contributed important inputs to the methodology, modeling, and interpretation of the results, and to the writing of paper sections. G.A., J.F.G., and A.K.D.-R contributed to the data reduction. G.A. led the ALMA observation proposal, conceived and prepared with contributions from M.O., J.F.G., A.K.D.-R., J.M.T., L.F.R., E.M., I.d.G.-M., and P.T.P.H. All authors participated in discussions of the results, and contributed to the paper preparation and revision.

\section*{Competing Interests}
The authors declare no competing interests.

\section*{Supplementary Material}

\subsection*{Supplementary Information}
Supplementary Figs.~1-3, Table 1, and captions for Videos 1-4 (pages 36-61 of this document).

\subsection*{Supplementary Video 1}
CO ($J$=3$\rightarrow$2) channel maps of Bullet 1 in SVS~13, in the line-of-sight velocity range (relative to VLA 4B) from $-$0.9 to $-$102.5~km~s$^{-1}$, observed with the ALMA 12-m array with an angular resolution of $0\farcs17 \times 0\farcs09$.
The LSR velocity of VLA~4B is $+$9.3~km~s$^{-1}$ (ref. \citen{diaz-rodriguez2022}). \href{https://static-content.springer.com/esm/art%3A10.1038%2Fs41550-025-02716-2/MediaObjects/41550_2025_2716_MOESM2_ESM.mp4}{Link to supplementary\_video\_1.mp4}

\subsection*{Supplementary Video 2}
CO ($J$=3$\rightarrow$2) channel maps of Bullet 1 in SVS~13, in the line-of-sight velocity range (relative to VLA~4B) from $-$9.0 to $-$129.7~km~s$^{-1}$, observed with the ALMA 12-m array with an angular resolution of $0\farcs53 \times 0\farcs33$. 
The LSR velocity of VLA~4B is $+$9.3~km~s$^{-1}$ (ref. \citen{diaz-rodriguez2022}). \href{https://static-content.springer.com/esm/art%3A10.1038%2Fs41550-025-02716-2/MediaObjects/41550_2025_2716_MOESM3_ESM.mp4}{Link to supplementary\_video\_2.mp4}

\subsection*{Supplementary Video 3}
Decomposition of the observed CO ($J$=3$\rightarrow$2) channel map emission of Bullet 1 in SVS~13 into elliptical rings. The line-of-sight velocity, relative to VLA~4B ($V_{\rm LSR}$ = +9.3~km~s$^{-1}$; ref. \citen{diaz-rodriguez2022}), ranges from $-$0.9 to $-$102.5~km~s$^{-1}$. \href{https://static-content.springer.com/esm/art%3A10.1038%2Fs41550-025-02716-2/MediaObjects/41550_2025_2716_MOESM4_ESM.mp4}{Link to supplementary\_video\_3.mp4}

\subsection*{Supplementary Video 4}
Comparison of the observed and bowshock model CO ($J$=3$\rightarrow$2) emission of the families of rings II and III in the line-of-sight velocity range (relative to VLA~4B) from $-$38.3 to $-$102.5~km~s$^{-1}$. \href{https://static-content.springer.com/esm/art%3A10.1038%2Fs41550-025-02716-2/MediaObjects/41550_2025_2716_MOESM5_ESM.mp4}{Link to supplementary\_video\_4.mp4}

\subsection*{Supplementary Data 1}
Ellipse fits to the observed rings in the channel maps. \href{https://static-content.springer.com/esm/art%3A10.1038%2Fs41550-025-02716-2/MediaObjects/41550_2025_2716_MOESM6_ESM.xlsx}{Link to supplementary\_data\_1.xlsx}

\newpage
%%%%%%%% EXTENDED DATA FIGURES %%%%%%%%%%%%%%

\newpage

\begin{figure*}[t!] 
{\centering
	\includegraphics[width=0.85\textwidth]{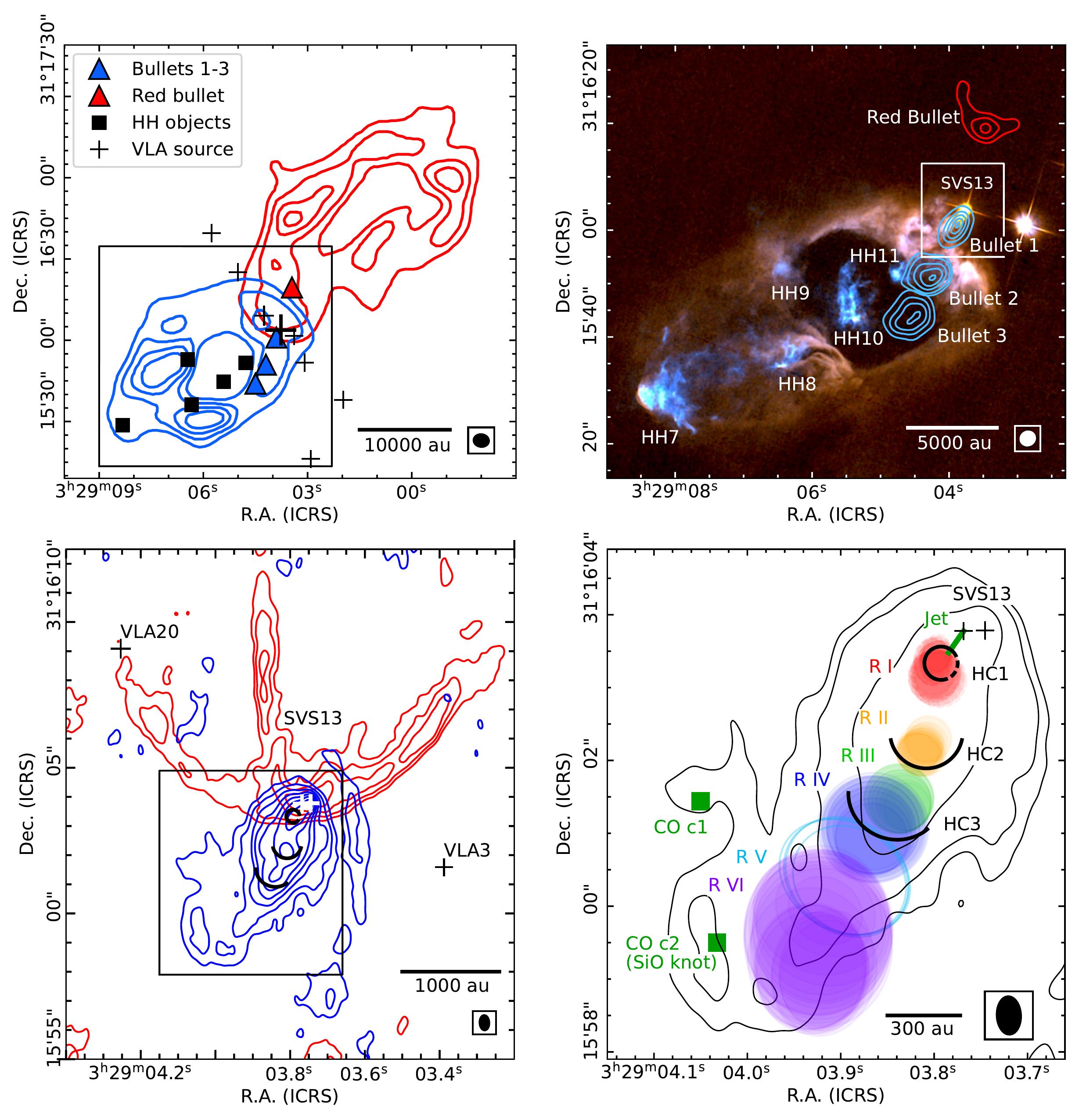}\\
}
\caption*{
{\bf Extended Data Fig.\ 1: Overview of the outflows in the vicinity of SVS 13.} Observed outflow features surrounding SVS~13 at different scales (see their observational properties in Supplementary Table 1). Large boxes indicate zoomed-in regions. The synthesized beams are shown in the bottom right corner of the panels. Velocities are relative to VLA~4B ($v_\mathrm{LSR}=+9.3$~km~s$^{-1}$)\cite{diaz-rodriguez2022}. Top left: CARMA CO(J=1-0) map of the large-scale SHV molecular outflow\cite{plunkett2013, stephens2017}. The emission has been integrated from -11.6 to -3.1 km~s$^{-1}$ (blue lobe) and from 0.7 to 8 km~s$^{-1}$ (red lobe). The location of the EHV molecular bullets \cite{bachiller2000, chen2016} (triangles), Herbig-Haro objects (squares), and radio sources proposed as YSOs\cite{rodriguez1999, diaz-rodriguez2022} (plus signs, where SVS 13 is the largest) is indicated. Top right: HST optical image (F606W filter) of the region, showing the outflow cavity, with the SMA map of the CO(2-1) emission of the EHV bullets\cite{chen2016} (with velocities from -161 to +129 km/s) overlapped in blue contours. Bottom left: Our ALMA CO(3-2) map (not corrected for the primary beam response) of the central region of the molecular outflow (beam=$0.527''\times0.333''$, PA=2.7 deg) is shown in contours. The emission has been integrated from -126.4 to -6.5~km~s$^{-1}$ (blue lobe) and from 6.5 to 60.8~km~s$^{-1}$ (red lobe). Contours are 3, 6, 10, 15, 21, 28, and 37 times 0.41 Jy~beam$^{-1}$~km~s$^{-1}$ (blue lobe) and 0.32 Jy~beam$^{-1}$~km~s$^{-1}$ (red lobe).The H$_2$ arcuate features60 are plotted as arcs. Bottom right: Close-up toward Bullet 1. Our ALMA CO(3-2) map of the blueshifted emission (beam=$0.537''\times0.333''$, PA=2.7 deg), integrated from -9.3 to -126.4 km~s$^{-1}$ (thus including the SHV, IHV, and EHV components), is shown in contours. Contours are 3, 6, 10, 16, and 25 times 0.61 Jy~beam$^{-1}$~km~s$^{-1}$. The colored filled areas represent the heads of the detected families of rings (see Fig. 3). The green squares represent the location of two CO clumps (where c2 coincides in position and velocity with the SiO knot reported by ref. \citen{lefevre2017}).
}
\label{fig:intro}
\end{figure*}

\newpage
\begin{figure*}[t!] 
{\centering
\includegraphics[width=\textwidth]{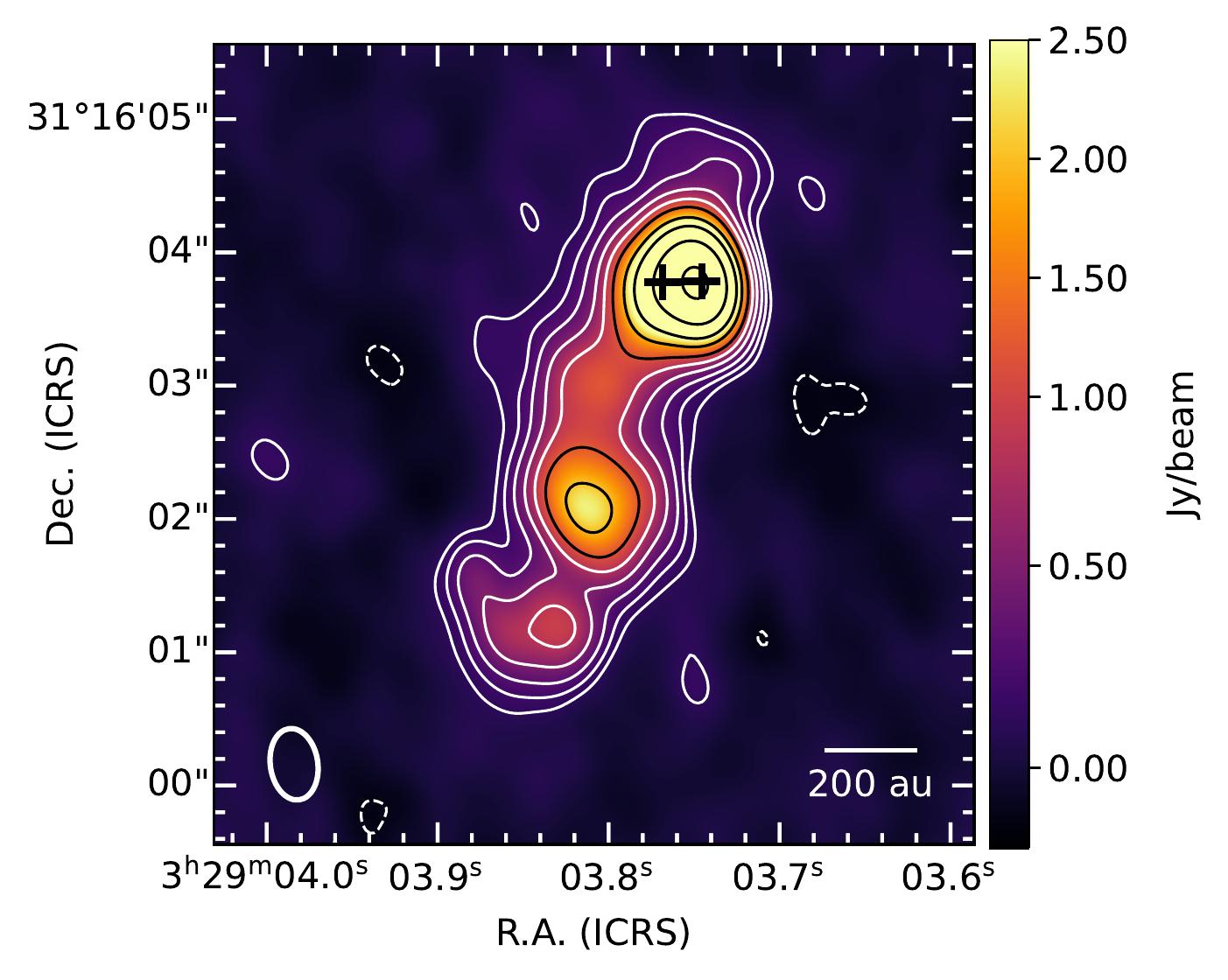}\\
}
\caption*{
{\bf Extended Data Fig.\ 2: Observed SO image of Bullet 1.} 
Image of the velocity-integrated intensity of the SO(8$_8$-7$_7$) line observed with the ALMA 12-m array using a low spectral resolution spectral window dedicated to continuum observation, with a channel spacing of 13.6~\kms. The emission has been integrated in the LOS velocity range from $-$9.8 to $-$118.6~\kms\ relative to the velocity of VLA 4B ($+$9.3~\kms)\cite{diaz-rodriguez2022}. The positions of the two protostars of the SVS~13 binary \cite{anglada2000} are indicated by plus signs. The synthesized beam, shown in the bottom left corner, is $0\farcs54 \times 0\farcs36$ (PA = $8.37^\circ$). Contours are $-$3, 3, 5, 8, 13, 20, 30, 50, 80, 140, 260 times 0.04 Jy~beam$^{-1}~\kms$. The image has not been corrected by the primary beam response.
}
\end{figure*}

\begin{figure*}[t!] 
{\centering
\includegraphics[width=0.92\textwidth]{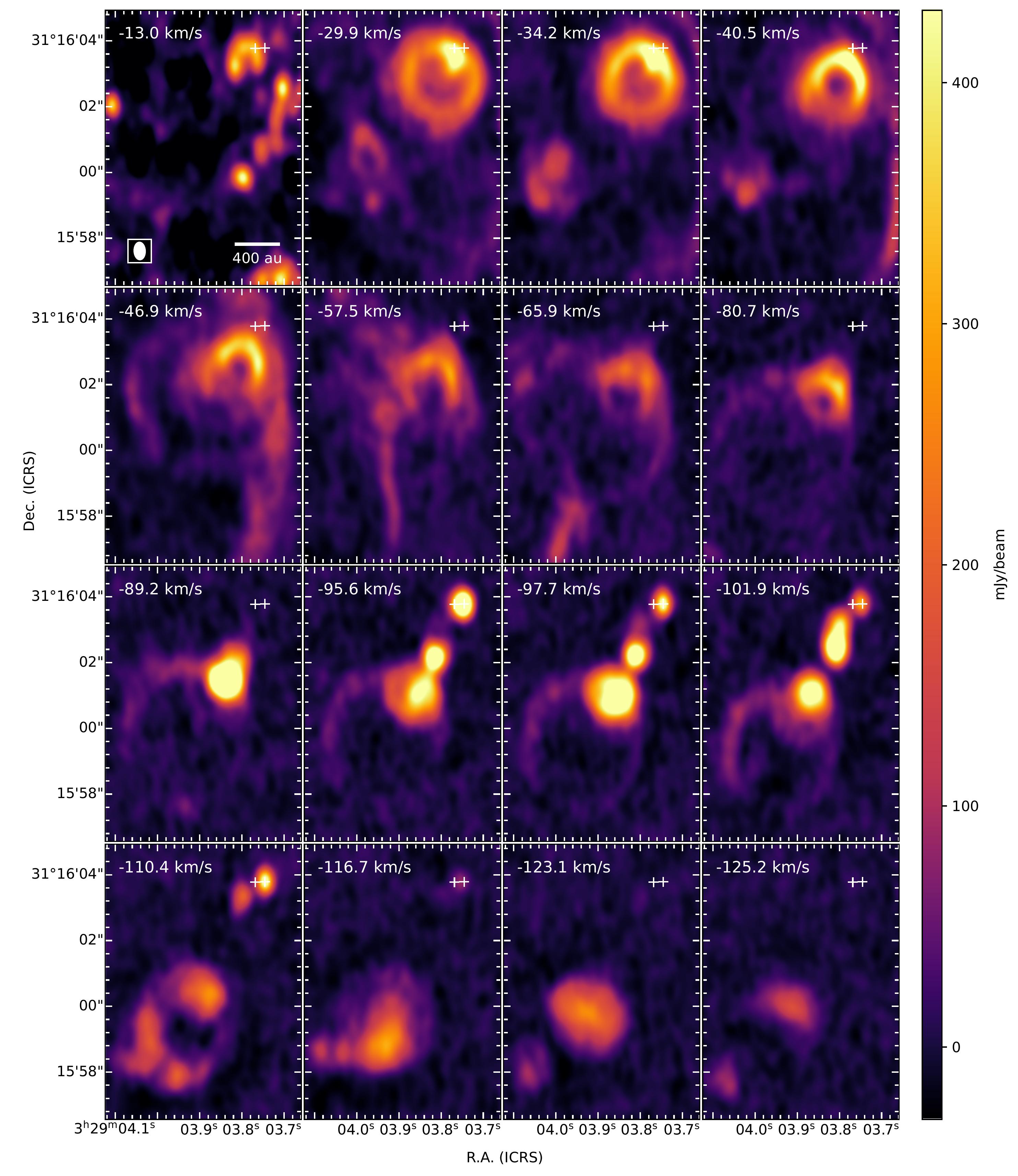}\\
}
\caption*{
{\bf Extended Data Fig.\ 3: Observed CO spectral channel images of Bullet 1 at low angular resolution.} A sample of spectral channel images of the CO($J$=3-2) emission observed by ALMA with a synthesized beam of
$0\farcs554 \times 0\farcs352$ (PA = 3.1$^\circ$), where natural weighting has been used.
The obtained data cover a range of velocities wider than the high angular resolution data shown in Fig.~\ref{fig:channels}. The positions of the two protostars of the SVS~13 binary \cite{anglada2000} are indicated by plus signs. The LOS velocity, relative to the velocity of VLA~4B ($V_{\rm LSR}$ = +9.3 \kms)\cite{diaz-rodriguez2022}, is shown in the top left corner of each image. The width of each of the spectral channels shown in the figure is 2.12 \kms, which corresponds to the average of 10 native channels. The r.m.s of the images is 8 mJy beam$^{-1}$. Images have not been corrected by the primary beam response. The synthesized beam is plotted as an ellipse in the bottom left corner of the first image.
}
 \label{fig:lowres}
\end{figure*}

\newpage
\begin{figure*}[t!]
{\centering
\includegraphics[width=0.75\textwidth]{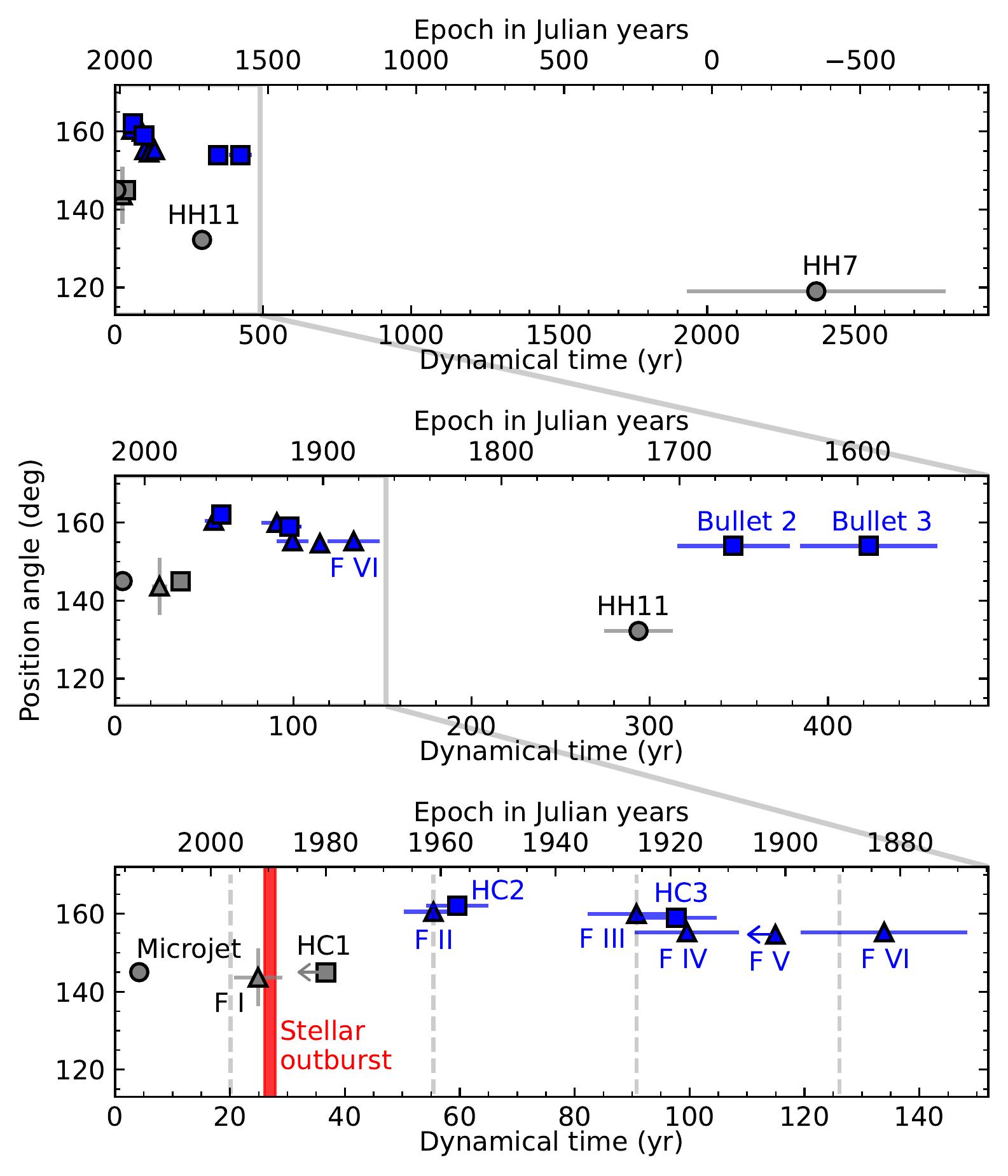}\\
}
\caption*{
{\bf Extended Data Fig.~4: Position angles and dynamical times of the outflow features associated with SVS 13.}  Position angle as a function of the dynamical time for the heads of the families of rings (plotted as triangles and labeled F~I to F~VI; see Methods), as well as for other features associated with the SVS~13 outflow (we plot atomic features as circles, and molecular as rectangles), such as HH objects\cite{hartigan2019,khanzadyan2003} (plotted as squares and labeled HH7, HH11), [FeII] jet (triangle labeled as Microjet), and H$_2$ arcs\cite{hodapp2014} (squares labeled as HC1, HC2, HC3). Colors indicate features with position angles around 160 deg (in blue) and in the range of 120-140 deg (in gray). The data for the families of rings (F~I-VI) are presented as mean values, and error bars represent the standard deviation (see 'Dynamical times of the heads of the families of rings' in Methods). For the rest of the objects we represent the values and uncertainties as reported in the literature\cite{khanzadyan2003, hartigan2019, hodapp2014, chen2016}. We assumed inclination angles from 22 to 25 deg. See Methods and Supplementary Table 1 for details on the calculations. Epoch is given in Julian years, with epoch 2000 corresponding to the standard definition of Julian epoch J2000.0. Dynamical times correspond to the date of our high-angular resolution ALMA observations (epoch 2016.69).
}
\label{fig:dyntimes}
\end{figure*}

\newpage
\begin{figure*}[t!]
{\centering
\includegraphics[width=0.75\textwidth]{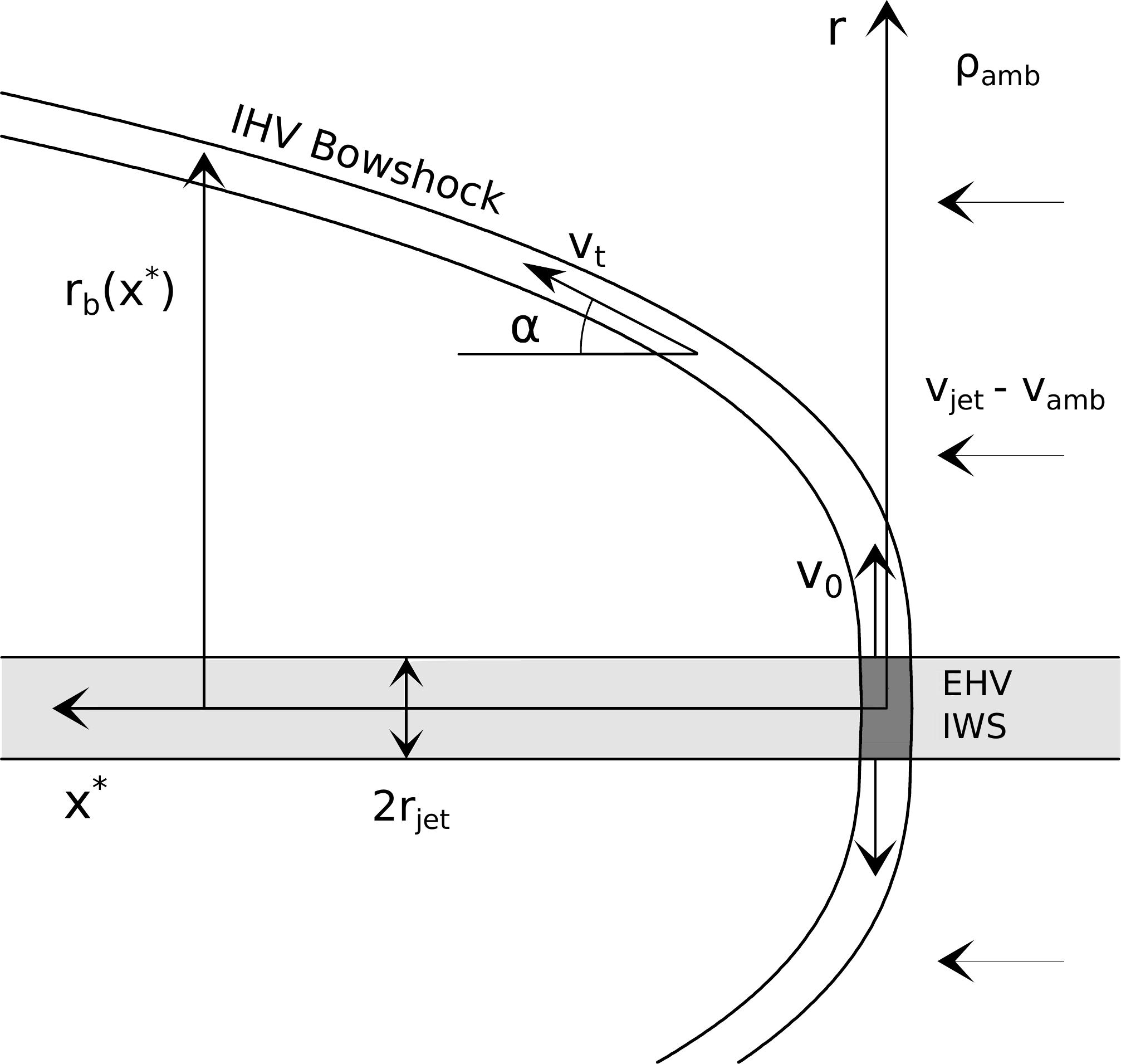}\\
}
{\bf Extended Data Fig.\ 5: Schematic diagram of the thin shell bowshock model.} 
The bowshock is seen in a reference system moving at the velocity $v_{\rm jet}$ of the working surface. 
The cylindrical jet beam has a diameter $2\,r_{\rm jet}$) and the impinging ambient gas moves to the left at a velocity $v_{\rm jet}-v_{\rm amb}$). We show a cylindrical coordinate system $(x^*,r)$, where $r$ is the cylindrical radius and $x^*$ the distance measured from the head of the working surface towards the outflow source. The working surface ejects material sideways at a velocity $v_0$ (which is approximately equal to the post-cooling region sound speed of $\sim 10$~km~s$^{-1}$). This sideways ejection interacts with the impinging ambient gas, forming a thin shell bowshock that has a well defined locus $r_b(x^*)$, and locally has a slope $\tan\alpha=dr_b/dx^*$.
\label{a1}
\end{figure*}

\newpage
\begin{figure*}[b!]
 {\centering
\includegraphics[width=0.9\textwidth]{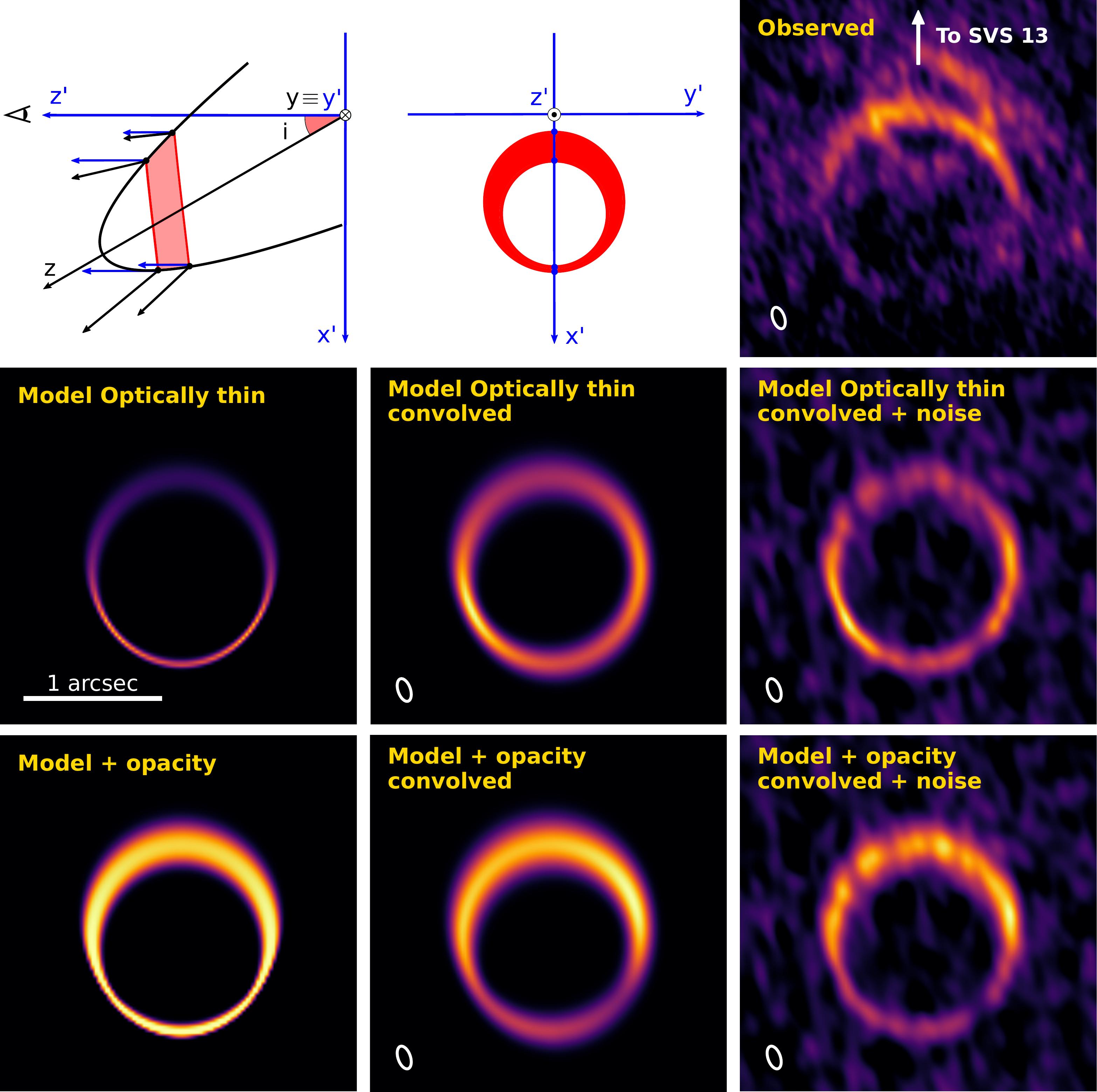}\\
}
\caption*{
{\bf Extended Data Fig.\ 6: Effect of the optical depth on the observed emission of the rings.} Top left: Geometry of the bowshock shell, where black vectors show velocities of the shell, blue vectors their LOS projections, and the shaded band outlines the spatial extent of the emitting region in the velocity range of a given channel map. Top center: Sketch of the corresponding channel image. Due to projection effects, the observed ring emission looks wider on the side closest to the star, while the far side appears narrower. Top right: Observed channel image (rotated so that the top faces SVS~13) illustrating the asymmetry of the rings, with the side closest in projection to the star appearing brighter and wider than the opposite side. Middle row: At low optical depths, in the side closest to the star, the emission appears spread over a wider range of radii and with lower intensity than on the far side, where it appears narrower but with higher intensity (left). When convolved with an elongated beam (center), the intensity becomes almost uniform throughout the ring, except near the positions where the beam axis is tangent to the ring (a well-known beam filling factor effect\cite{osorio2014}). When noise is added (right), no substantial asymmetries are detected. Bottom row: When the optical depth is high enough (left), the side closer to the star becomes almost as bright, but more extended, than the far side, where any increase in intensity is hindered by opacity saturation. When convolved with the beam (center), the intensity drops in the  far (narrower) side of the ring, because of the smaller beam filling factor, producing an asymmetry similar to that observed. A model image with noise (right panel) shows that a qualitative agreement with observations (top panel) can be achieved. In the modeling, the assumed velocity dispersion is 2~km~s$^{-1}$, the inclination angle is 20 deg, and the channel width is 0.53~km~s$^{-1}$.
}
\label{fig:bullet2aca}
\end{figure*}

\newpage
\clearpage 
\begin{figure*}[t!]
{\centering
	\includegraphics[width=\textwidth]{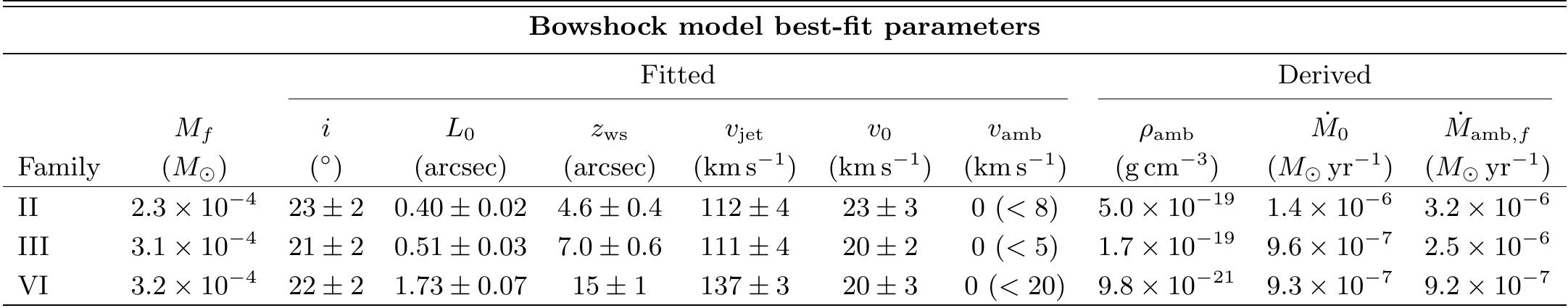}\\
}

{\bf Extended Data Table\ 1: Bowshock fitting results for Family II, III, and VI.} 
%$M_{\rm rings}$
$M_f$ is the mass of the shell measured from the observations,
$i$ is the inclination angle between the shell symmetry axis and the line of sight,
$L_0$ is the characteristic scale of the bowshock shell, % (equation (\ref{rb})),
$z_{\rm ws}$ is the position of the internal working surface relative to VLA~4B,
$v_{\rm jet}$ is the velocity of the working surface,
$v_0$ is the initial velocity at which jet material is ejected sideways, %Fits are better considering $v_{\rm amb}$ close to 0.
$v_{\rm amb}$ is the velocity of the ambient (upper limits in parenthesis are the 1-$\sigma$ errors of the fit),
$\rho_{\rm amb}$ is the density of the medium into which the jet is traveling, %(estimated from equation (\ref{eq:rhow_mr})), 
$\dot{M}_0$ is the mass-rate at which the jet material is ejected sideways from the internal working surface, %(estimated from equation (\ref{eq:mdot0_rhow})), 
and $\dot{M}_{{\rm amb}, f}$ is the mass-rate of ambient material that is being incorporated into the bowshock shell up to its outer edge radius.
A distance of 300~pc has been adopted \cite{ortiz-leon2018, gaia2023}.
\label{opacity}
\end{figure*}

\clearpage
\newpage

\begin{center}
    {\Huge\bf Supplementary Information}

\vspace*{\fill}
\noindent
{\bf \large Contents}
\vspace{2em}

\noindent
Supplementary Figure 1 \dotfill 2 \\
Supplementary Figure 2 \dotfill 12 \\
Supplementary Figure 3 \dotfill 20 \\
Supplementary Table 1 \dotfill 21 \\
Supplementary Video 1 (caption) \dotfill 22 \\
Supplementary Video 2 (caption) \dotfill 23 \\
Supplementary Video 3 (caption) \dotfill 24 \\
Supplementary Video 4 (caption) \dotfill 25 \\
References \dotfill 26 \\
\vspace*{\fill}
\end{center}

\begin{figure*}[t!] 
{\centering
	\includegraphics[width=\textwidth]{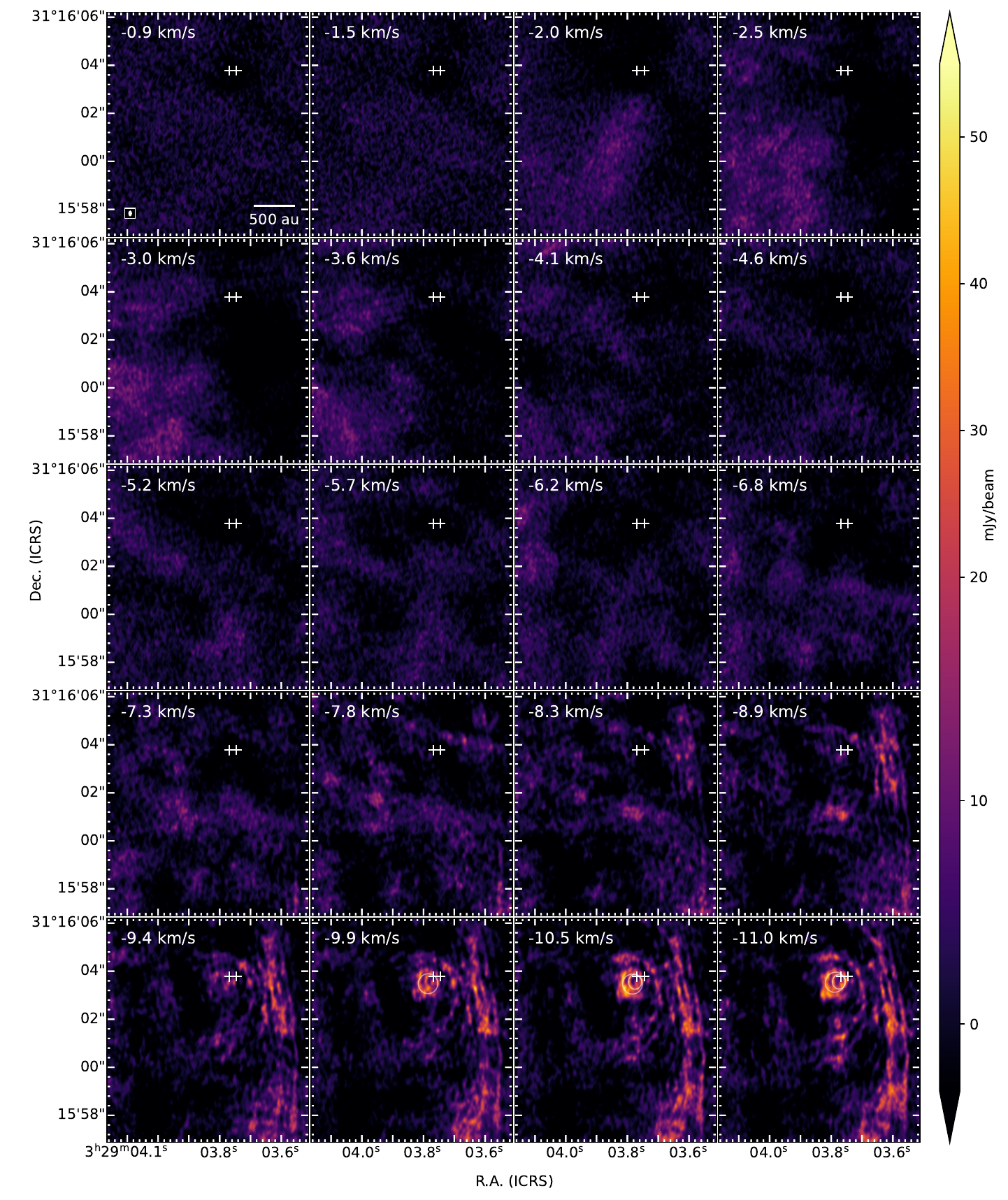}\\
}
{
\footnotesize
{\bf Supplementary Figure 1: Full set of CO (${\mathbf J}$=3$\rightarrow$2) spectral channel images of Bullet 1 in SVS~13 observed at high angular resolution.} Channel maps observed with the ALMA 12-m array with a synthesized beam of $0.17'' \times 0.09''$ (PA = $-2.2^\circ$). The line-of-sight velocity relative to the velocity of VLA~4B ($V_{\rm LSR}$ = +9.3~km~s$^{-1}$; ref. \citen{diaz-rodriguez2022_sup}) ranges from $-$0.9 to $-$102.5~km~s$^{-1}$ , and is shown in the top left corner of each image. The positions of the two protostars, VLA~4A (west) and VLA~4B (east), of the SVS~13 binary \cite{anglada2000_sup} are indicated by plus signs. The white ellipses are the elliptical fits to the families of Rings Ia-c, II, III, IV and V. The channel width is 0.53~km~s$^{-1}$ and the r.m.s. of the images (uncorrected by the primary beam response) is 2.6 mJy beam$^{-1}$. The beam is plotted as an ellipse in the bottom left corner of the first image. 
}
\end{figure*}

\begin{figure*}[t!] 
{\centering
\includegraphics[width=\textwidth]{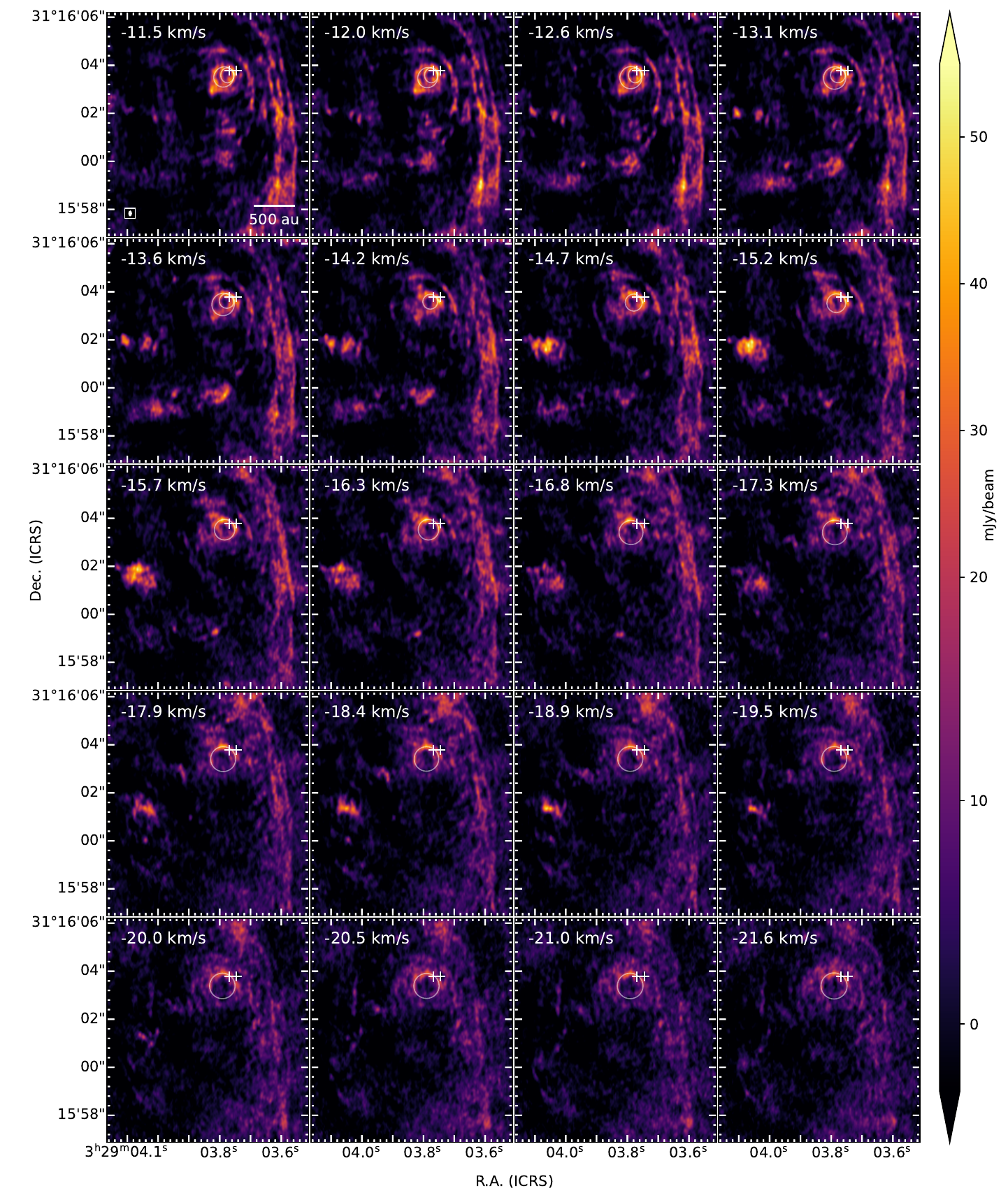}\\
}
{\bf Supplementary Figure 1: } Continued.
\end{figure*}

\begin{figure*}[t!] 
{\centering
\includegraphics[width=\textwidth]{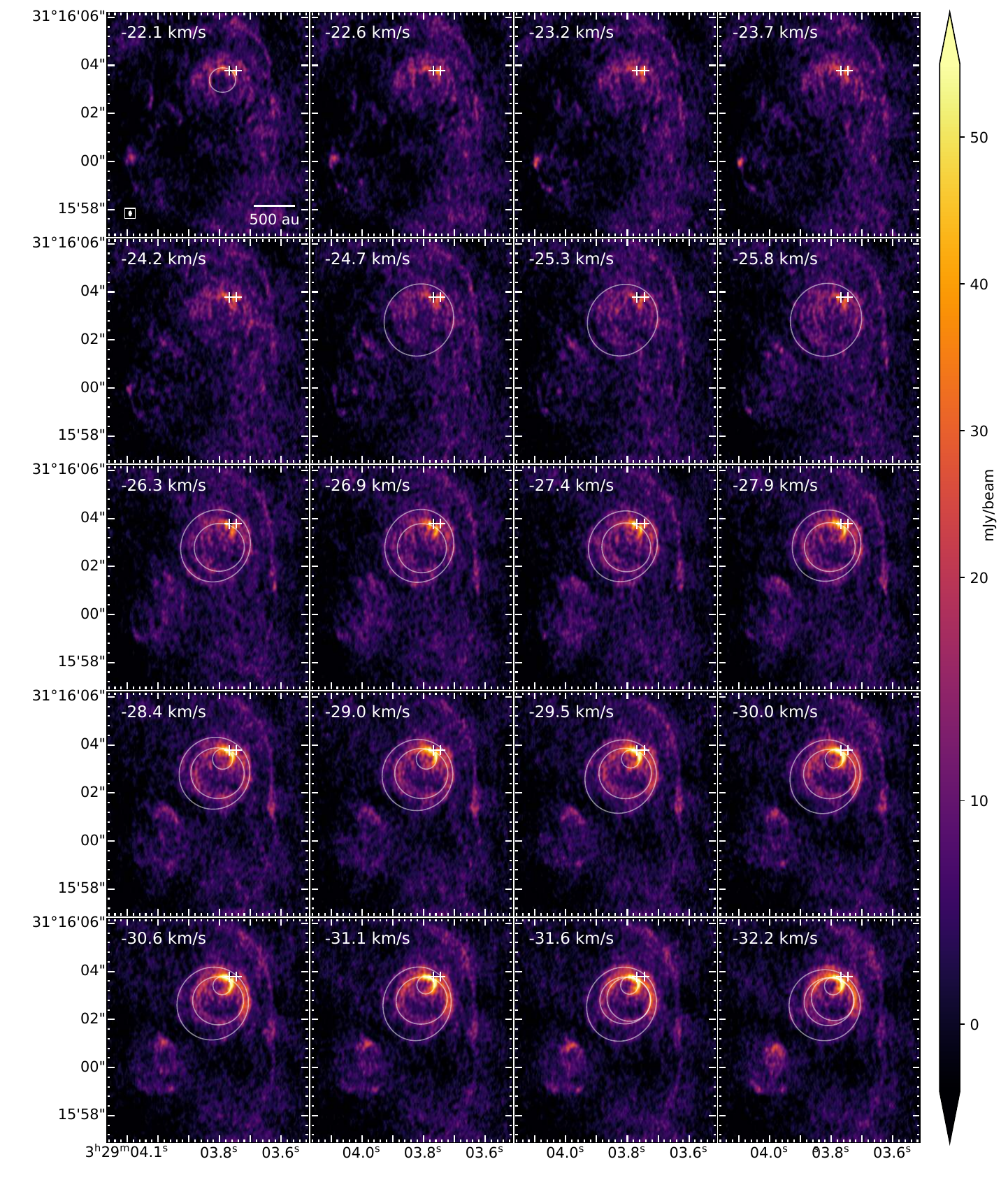}\\
}
{\bf Supplementary Figure 1: } Continued.
\end{figure*}

\begin{figure*}[t!] 
{\centering
\includegraphics[width=\textwidth]{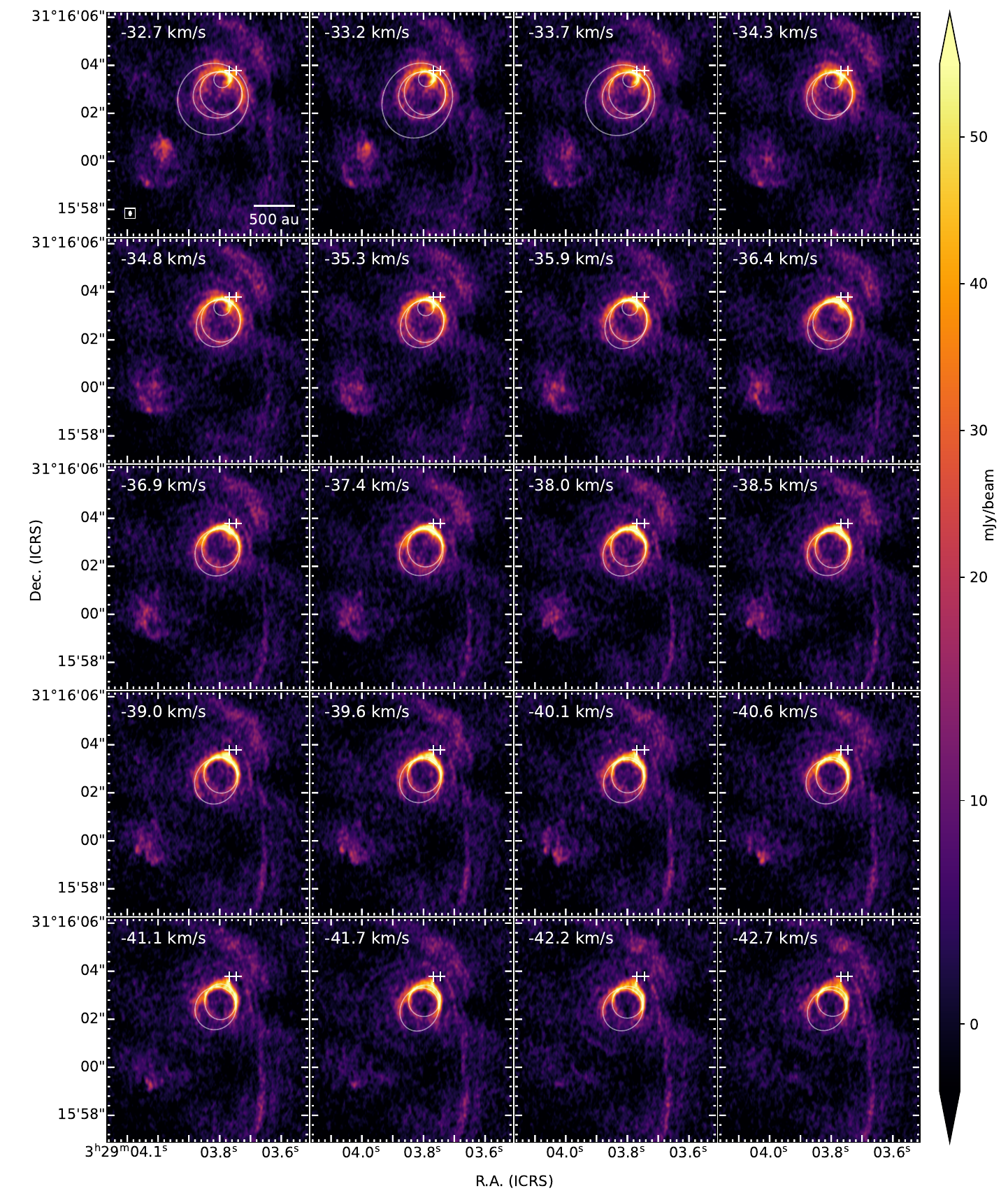}\\
}
{\bf Supplementary Figure 1: } Continued.
\end{figure*}

\begin{figure*}[t!] 
{\centering
\includegraphics[width=\textwidth]{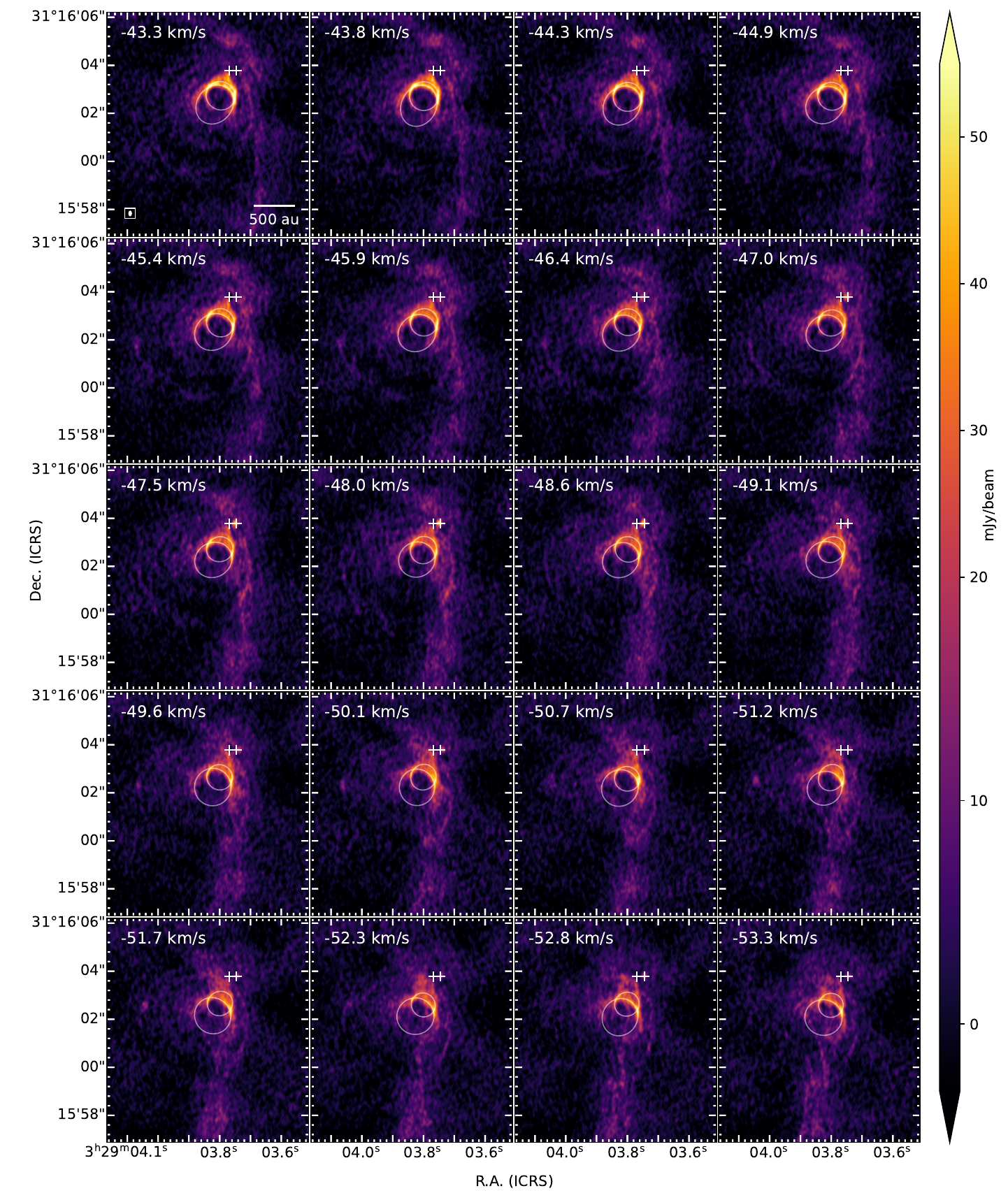}\\
}
{\bf Supplementary Figure 1: } Continued.
\end{figure*}

\begin{figure*}[t!] 
{\centering
\includegraphics[width=\textwidth]{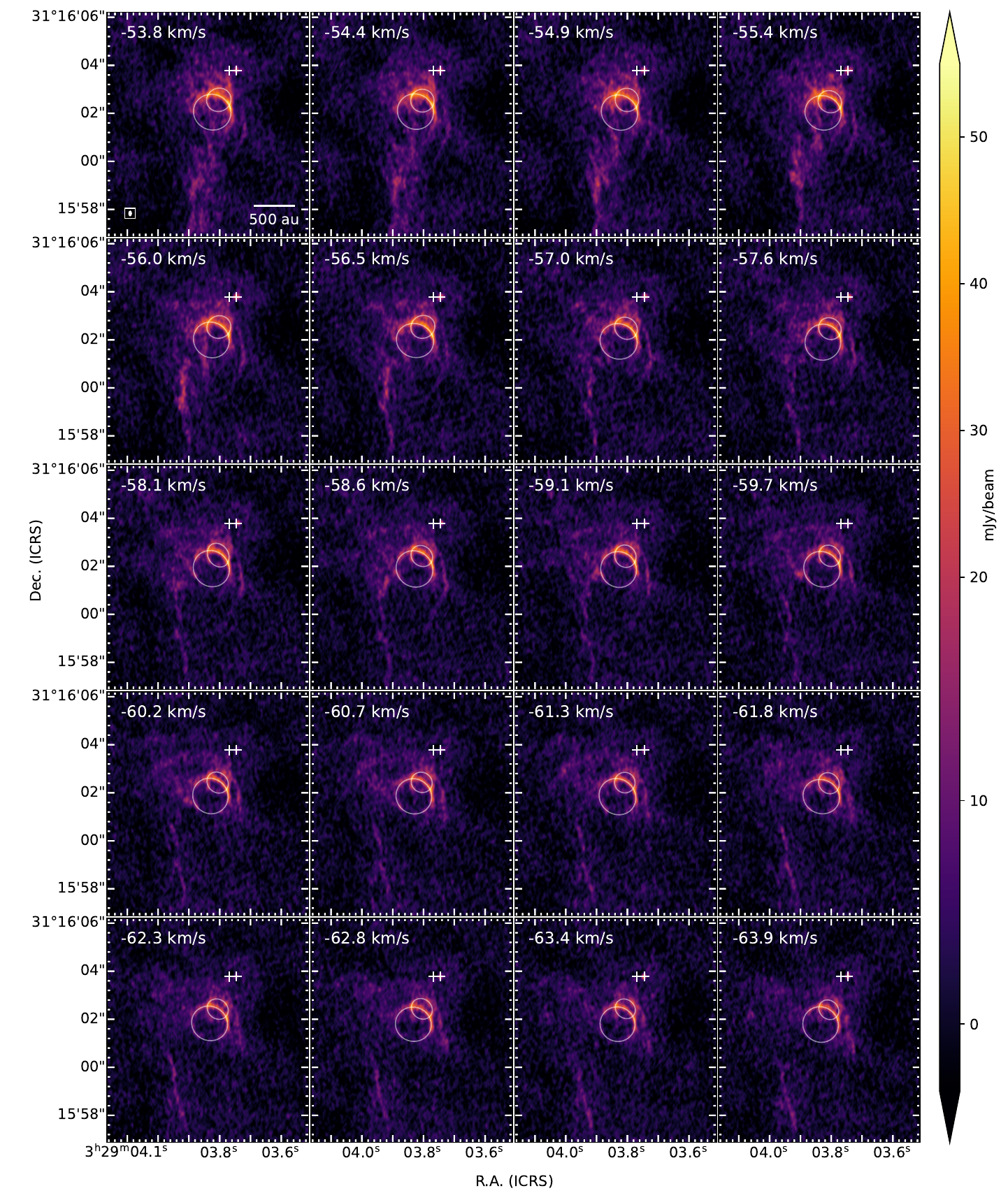}\\
}
{\bf Supplementary Figure 1: } Continued.
\end{figure*}

\begin{figure*}[t!] 
{\centering
\includegraphics[width=\textwidth]{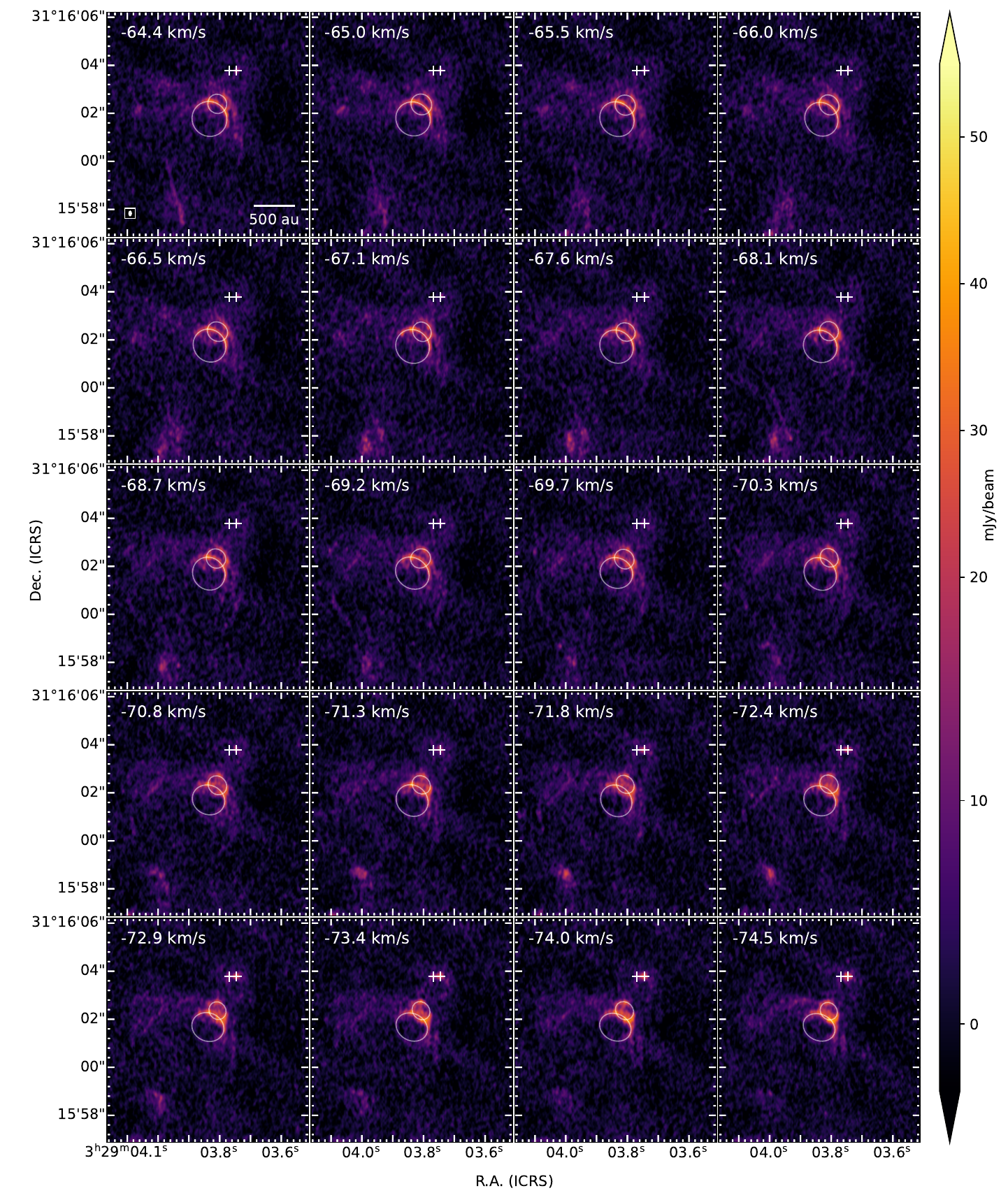}\\
}
{\bf Supplementary Figure 1: } Continued.
\end{figure*}

\begin{figure*}[t!] 
{\centering
\includegraphics[width=\textwidth]{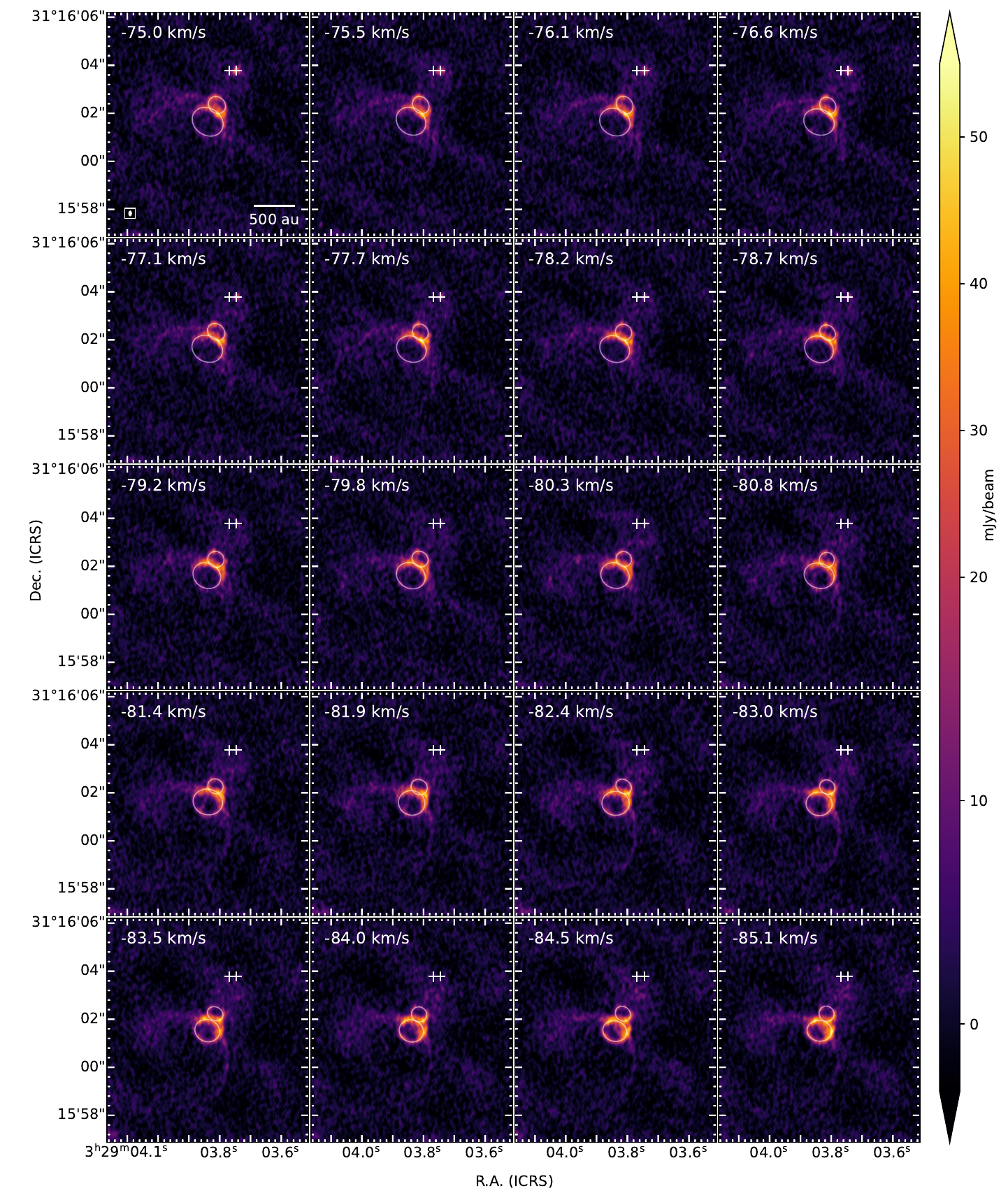}\\
}
{\bf Supplementary Figure 1: } Continued.
\end{figure*}

\begin{figure*}[t!] 
{\centering
\includegraphics[width=\textwidth]{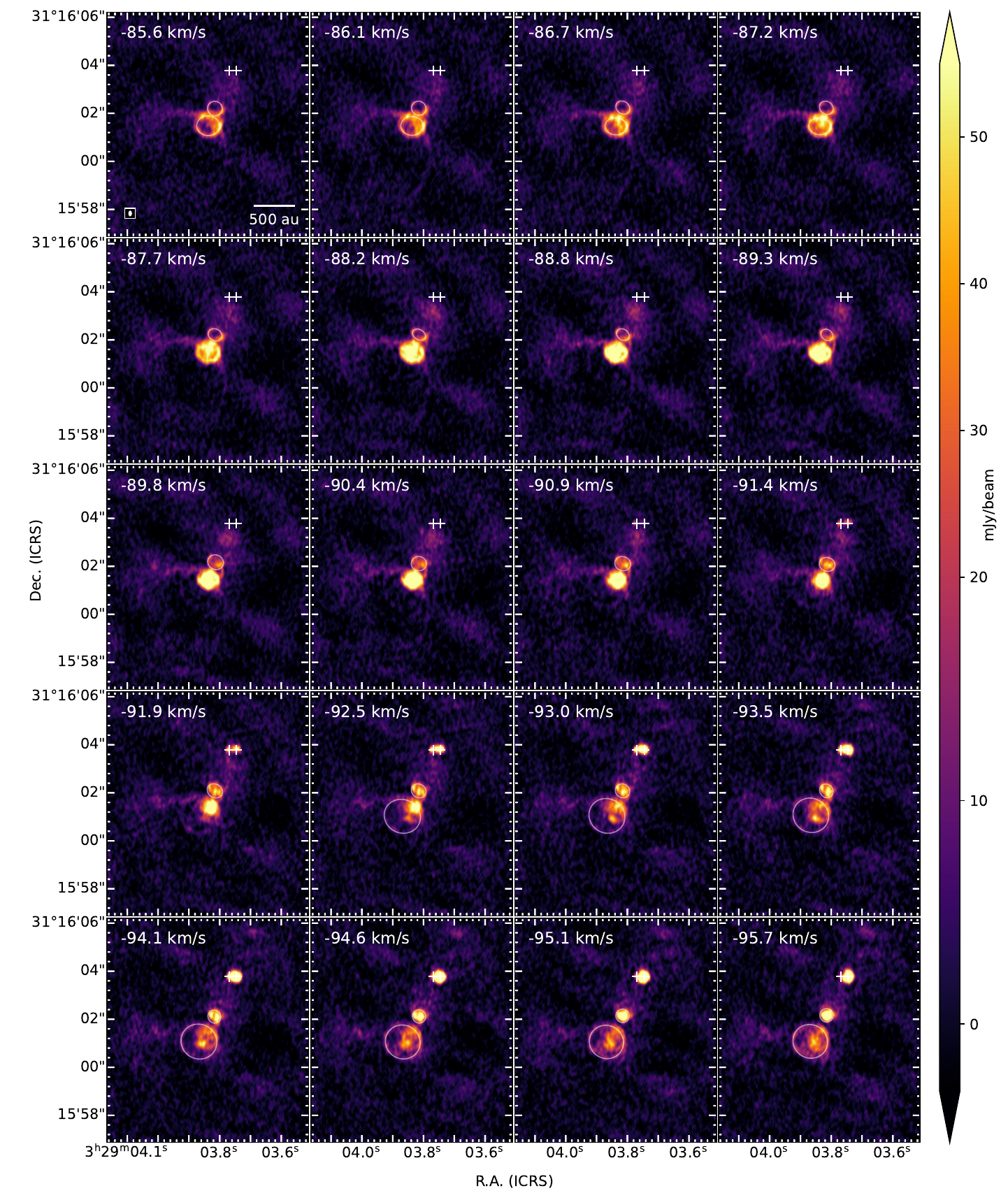}\\
}
{\bf Supplementary Figure 1: } Continued.
\end{figure*}

\begin{figure*}[t!] 
{\centering
\includegraphics[width=\textwidth]{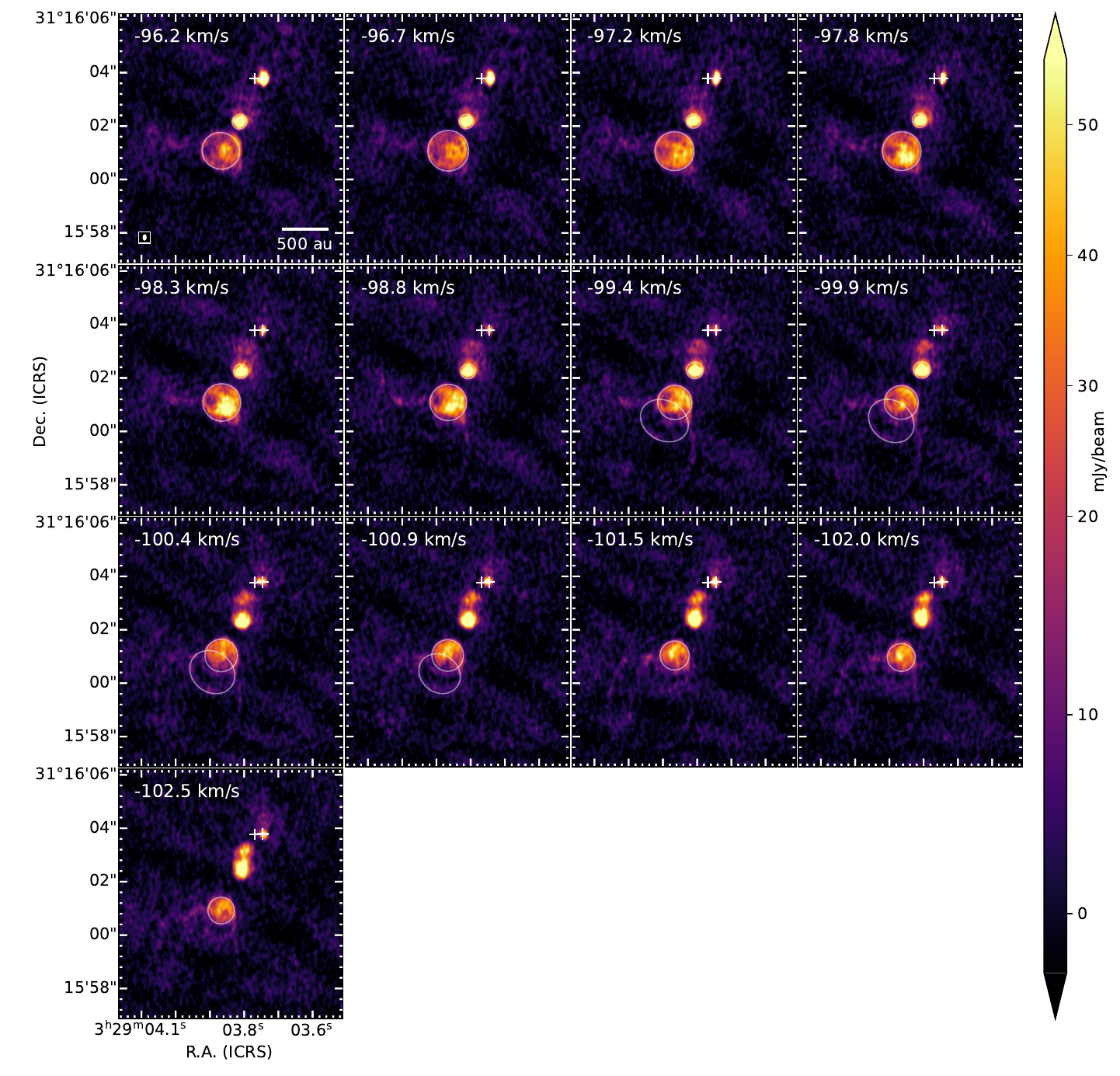}\\
}
{\bf Supplementary Figure 1: } Continued.
\end{figure*}

\begin{figure*}[t!] 
{\centering
\includegraphics[width=\textwidth]{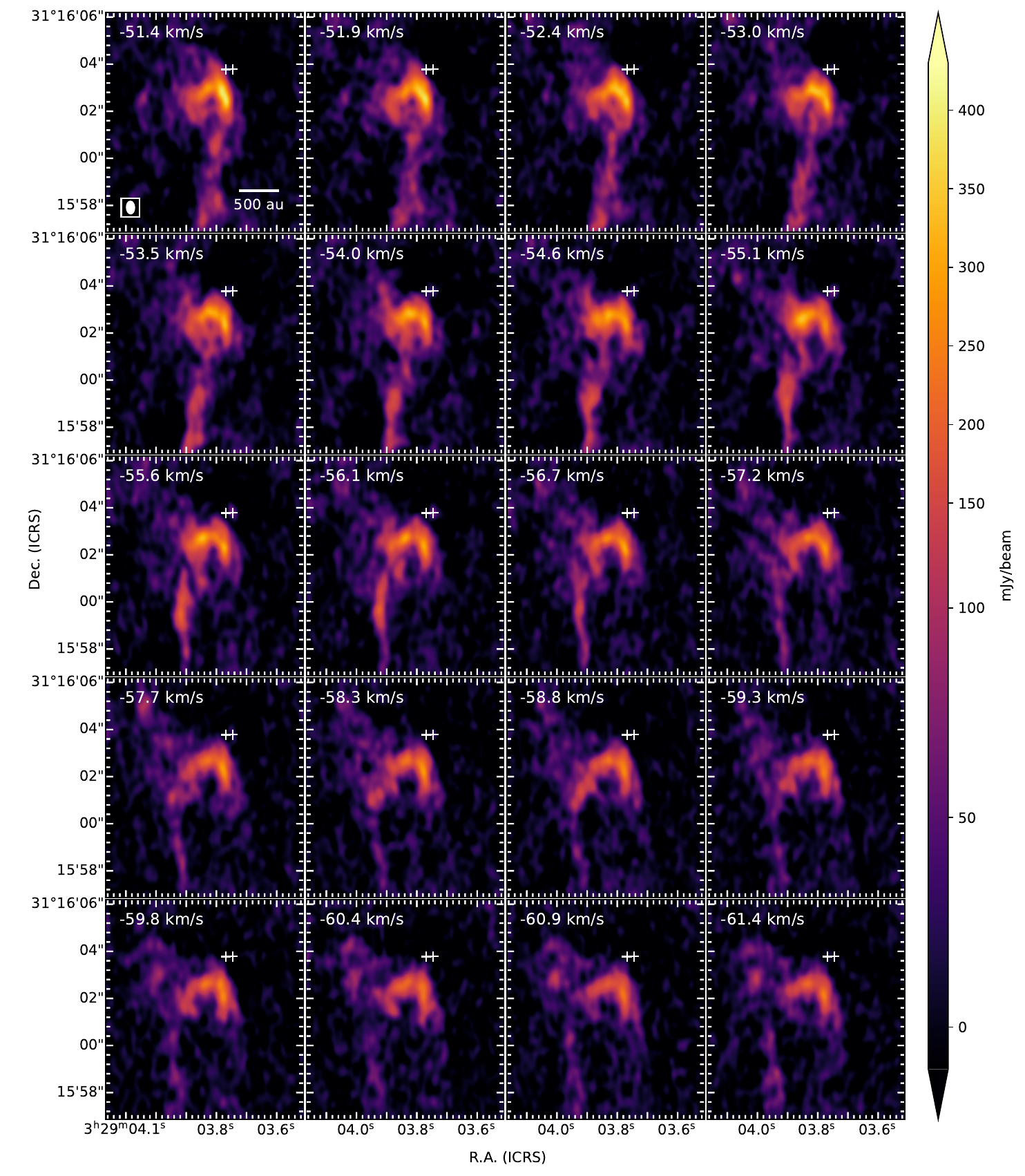}\\
}
{\bf Supplementary Figure 2: Set of CO (${\mathbf J}$=3$\rightarrow$2) spectral channel images of Bullet 1 in SVS~13 observed at low angular resolution.} Channel maps observed with the ALMA 12-m array with a synthesized beam of $0.53'' \times 0.33''$ (PA = $2.7^\circ$). In this set of channel maps the line-of-sight velocity relative to the velocity of VLA~4B ($V_{\rm LSR}$ = +9.3~km~s$^{-1}$; ref. \citen{diaz-rodriguez2022_sup}) ranges from $-$51.4 to $-$129.2 km~s$^{-1}$, and is shown in the top left corner of each image. The positions of the two protostars, VLA~4A (west) and VLA~4B (east), of the SVS~13 binary \cite{anglada2000_sup} are indicated by plus signs. The white ellipses are the elliptical fits to the family of Rings VI. The channel width is 0.53~km~s$^{-1}$ and the r.m.s. of the images (uncorrected by the primary beam response) is 10 mJy~beam$^{-1}$. The beam is plotted as an ellipse in the bottom left corner of the first image. 
\end{figure*}

\begin{figure*}[t!] 
{\centering
\includegraphics[width=\textwidth]{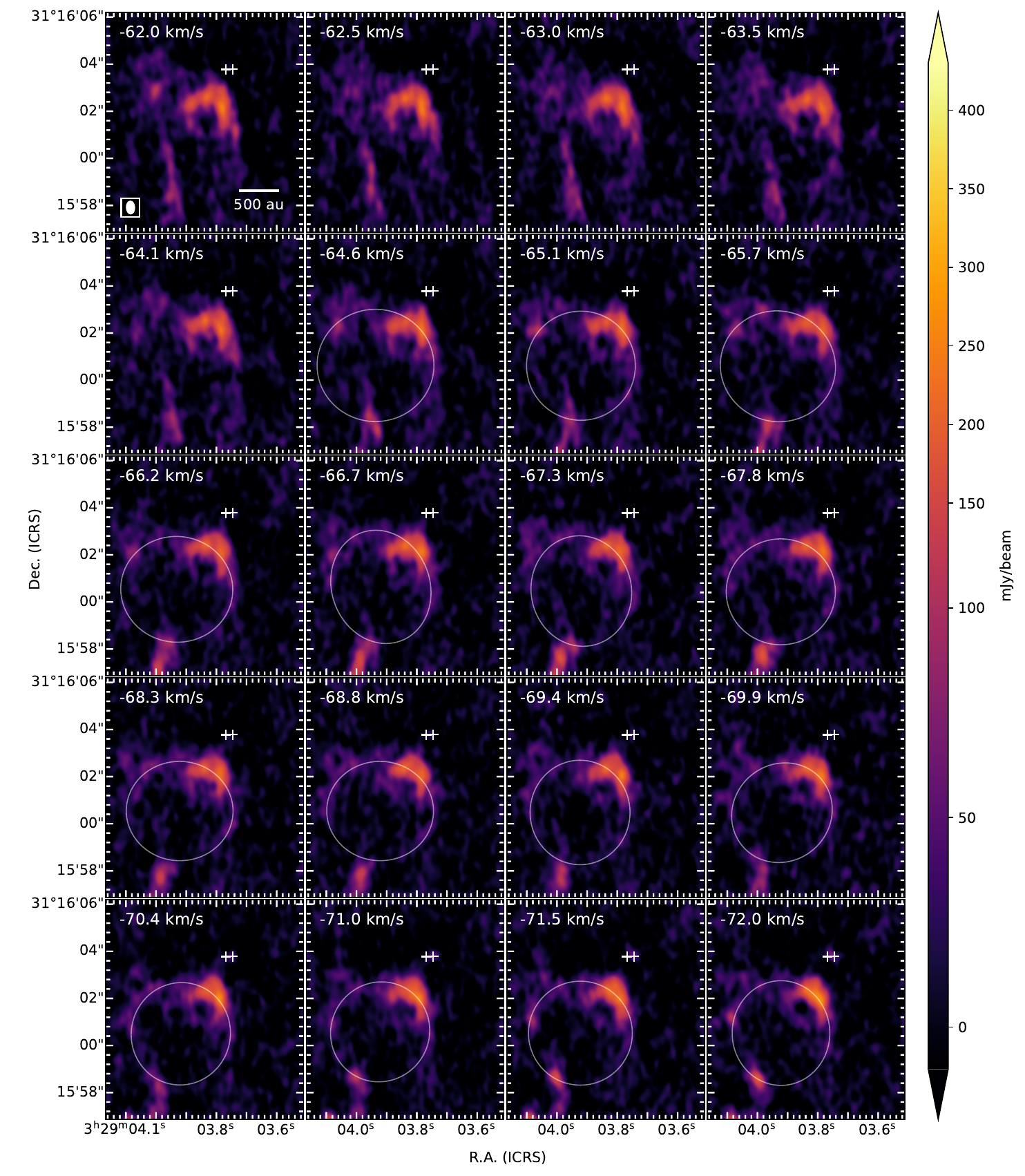}\\
}
{\bf Supplementary Figure 2: } Continued.
\end{figure*}

\begin{figure*}[t!] 
{\centering
\includegraphics[width=\textwidth]{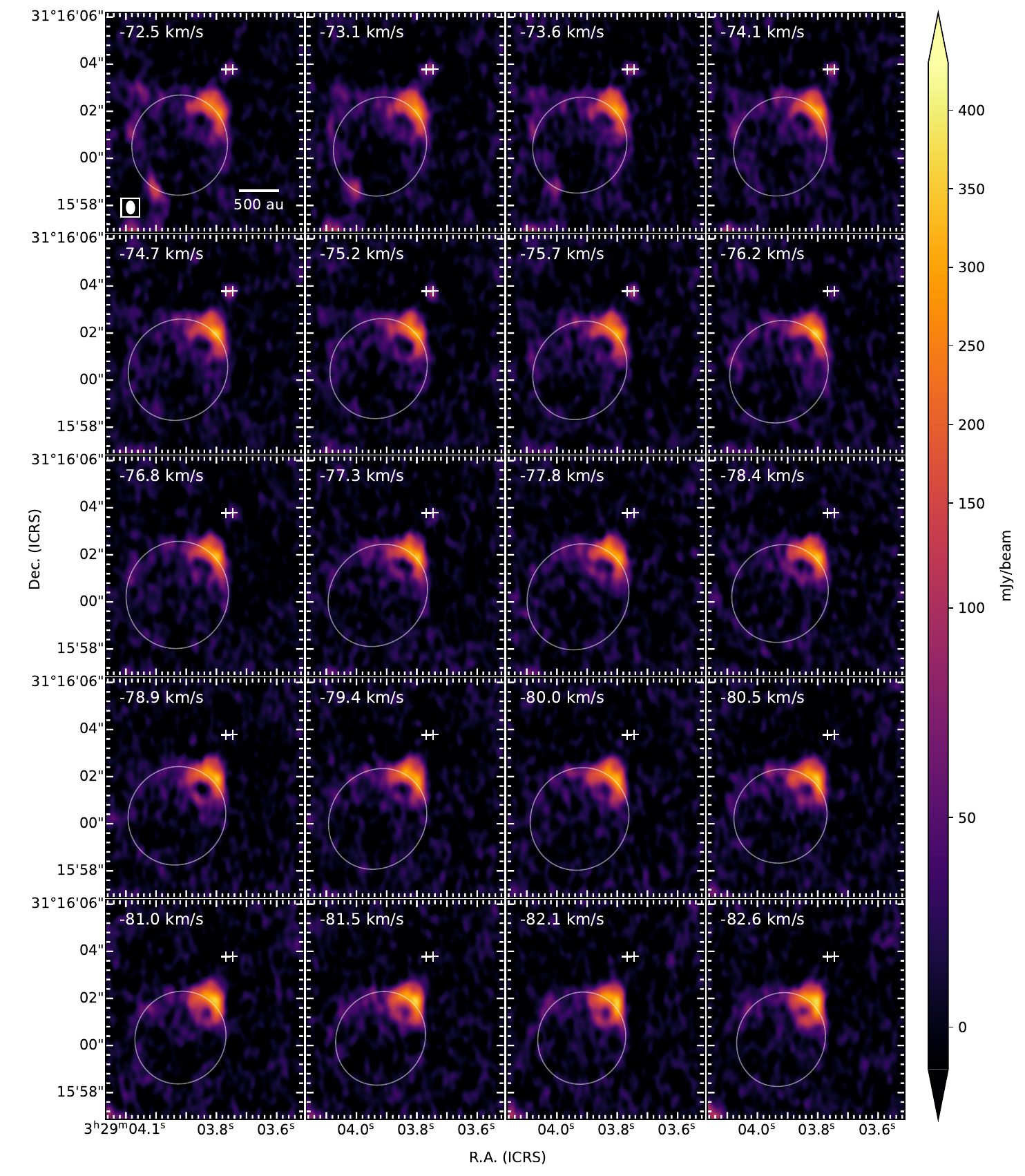}\\
}
{\bf Supplementary Figure 2: } Continued.
\end{figure*}

\begin{figure*}[t!] 
{\centering
\includegraphics[width=\textwidth]{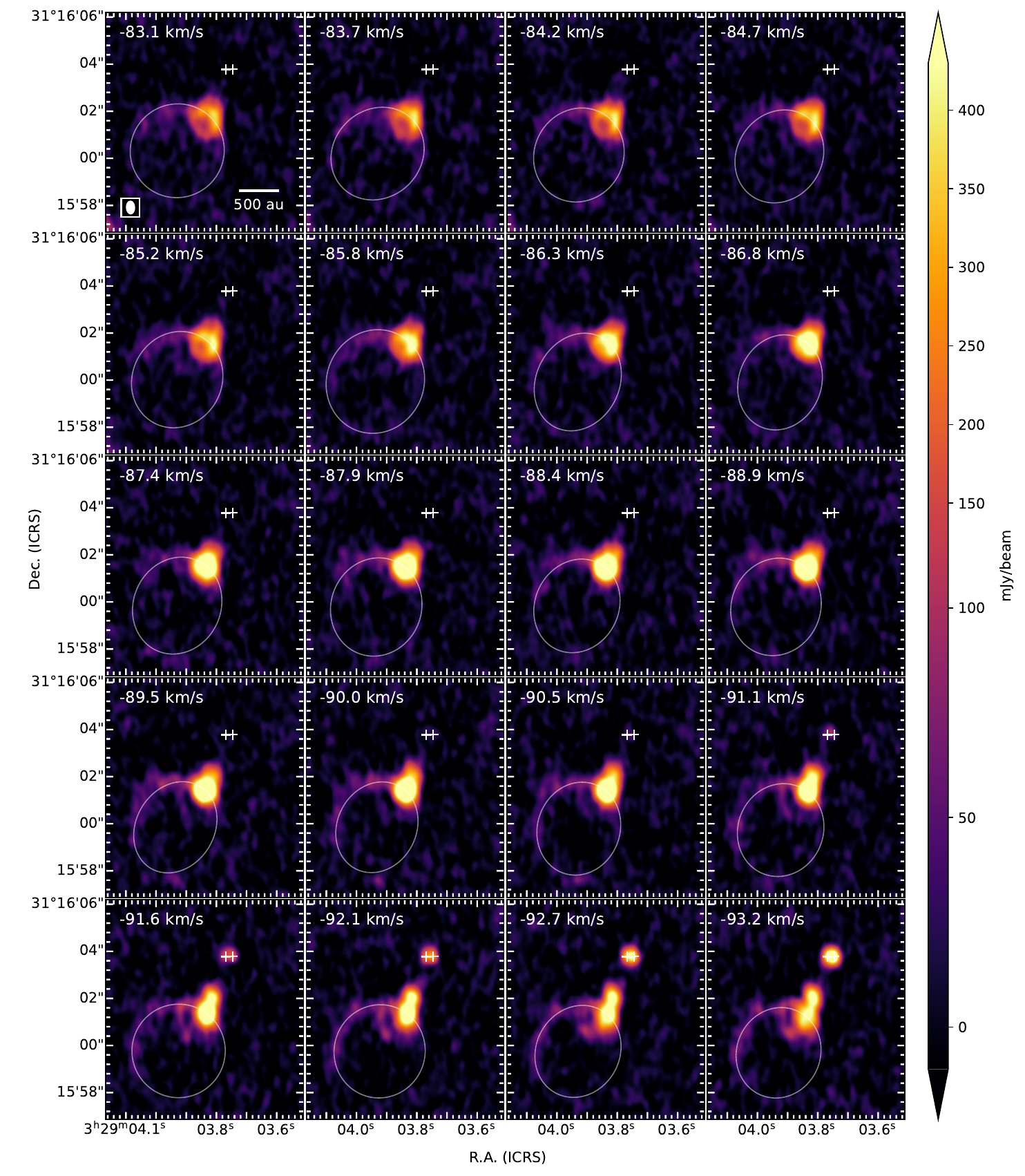}\\
}
{\bf Supplementary Figure 2: } Continued.
\end{figure*}

\begin{figure*}[t!] 
{\centering
\includegraphics[width=\textwidth]{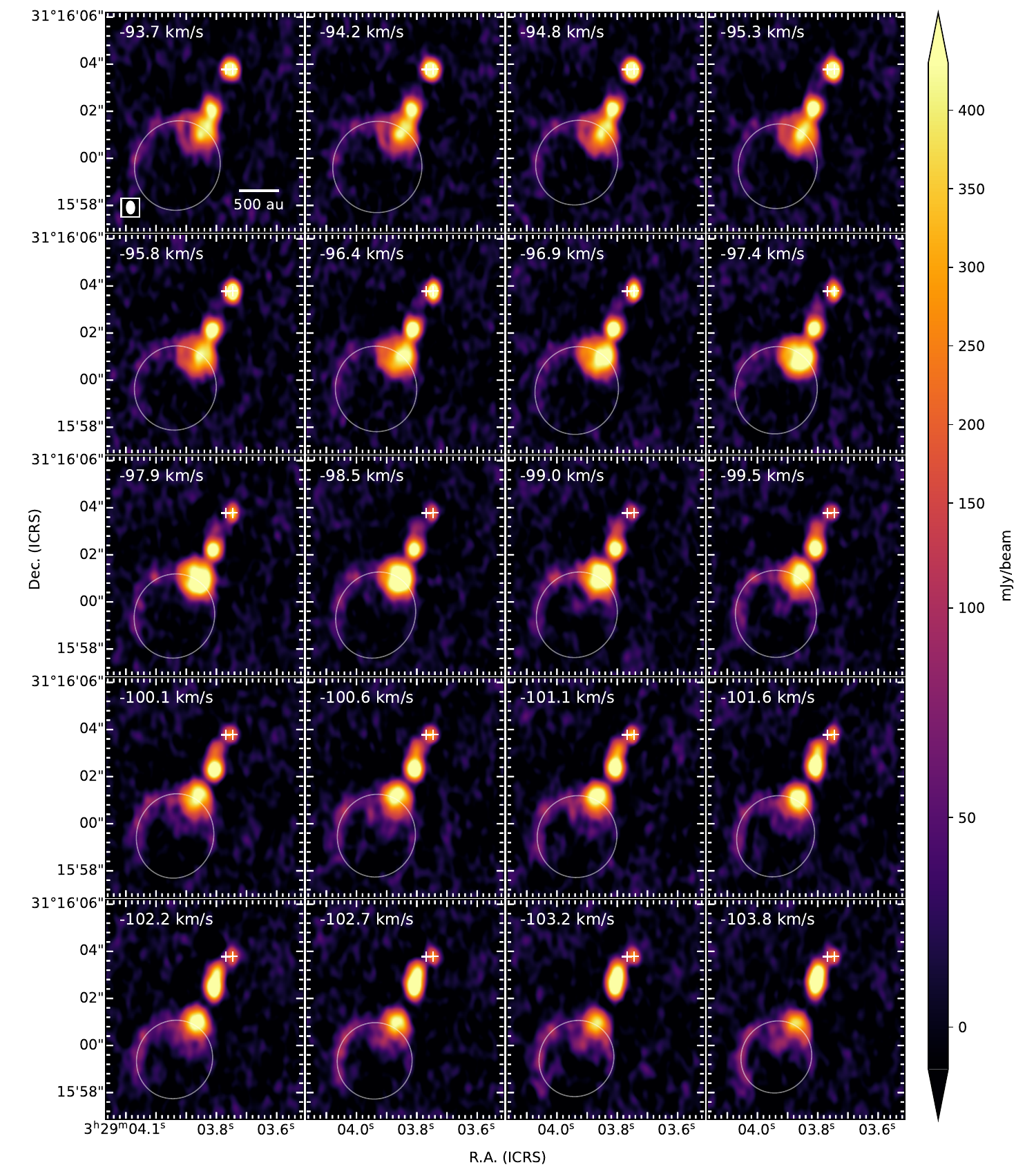}\\
}
{\bf Supplementary Figure 2: } Continued.
\end{figure*}

\begin{figure*}[t!] 
{\centering
\includegraphics[width=\textwidth]{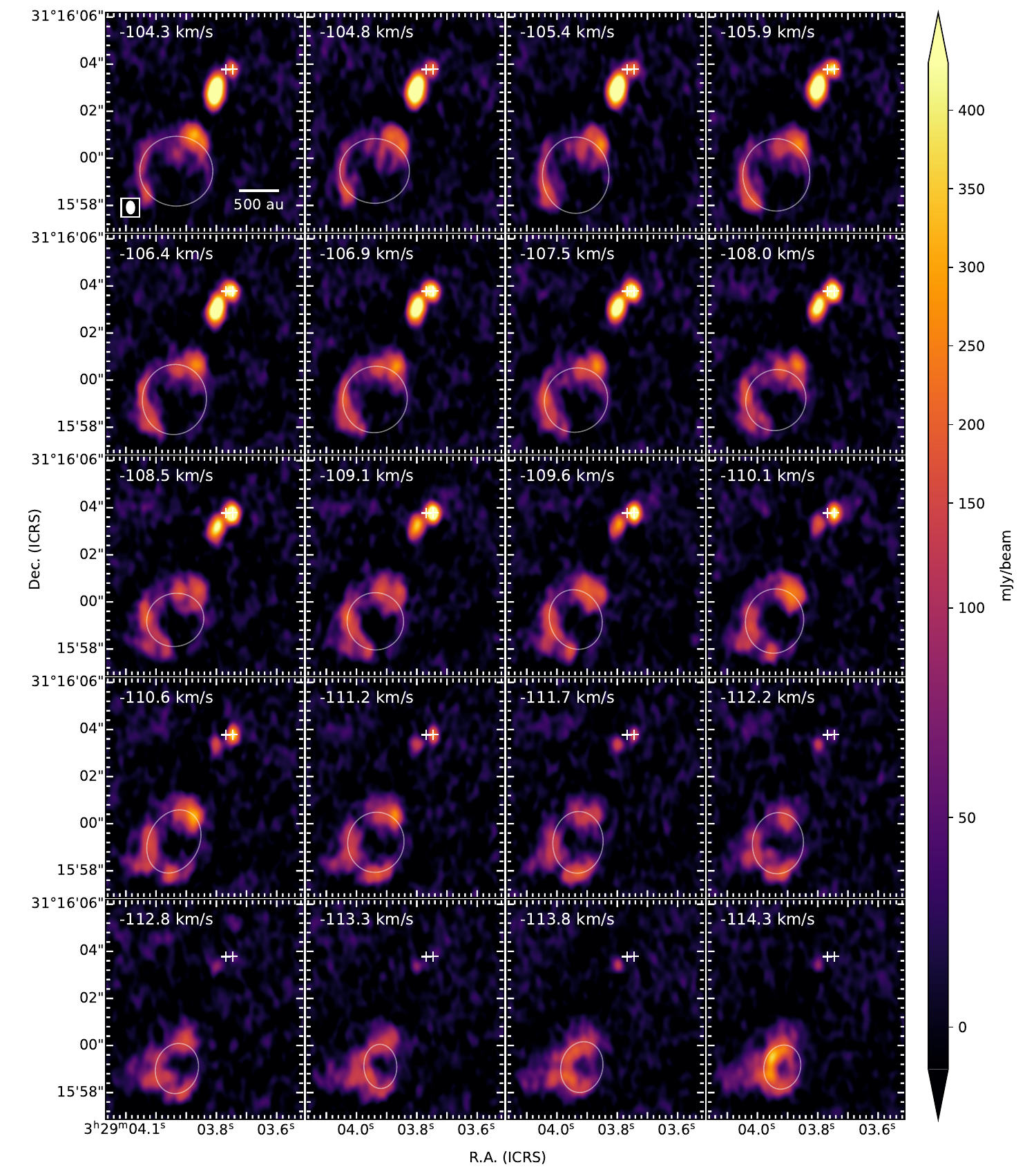}\\
}
{\bf Supplementary Figure 2: } Continued.
\end{figure*}

\begin{figure*}[t!] 
{\centering
\includegraphics[width=\textwidth]{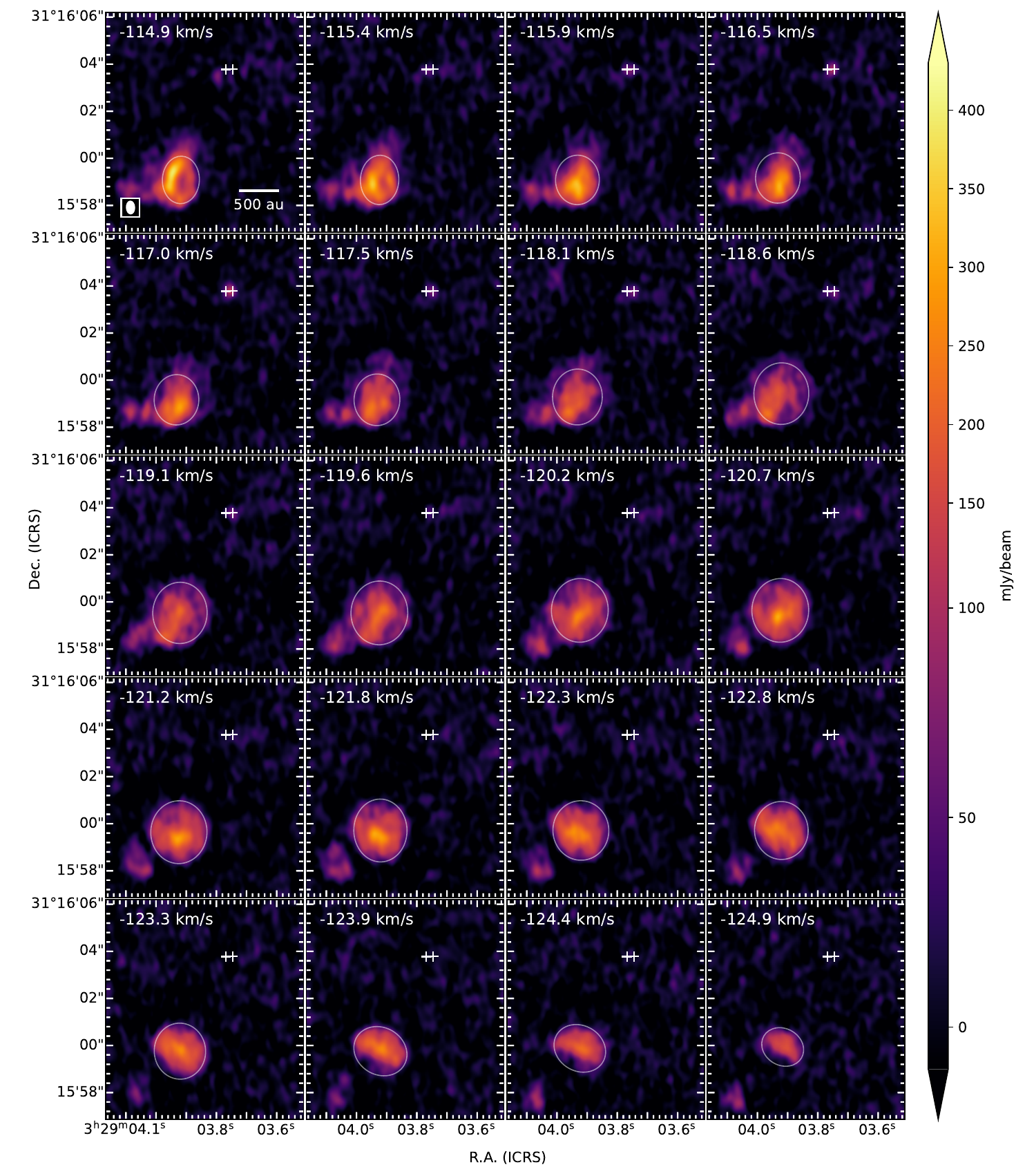}\\
}
{\bf Supplementary Figure 2: } Continued.
\end{figure*}

\begin{figure*}[t!] 
{\centering
\includegraphics[width=\textwidth]{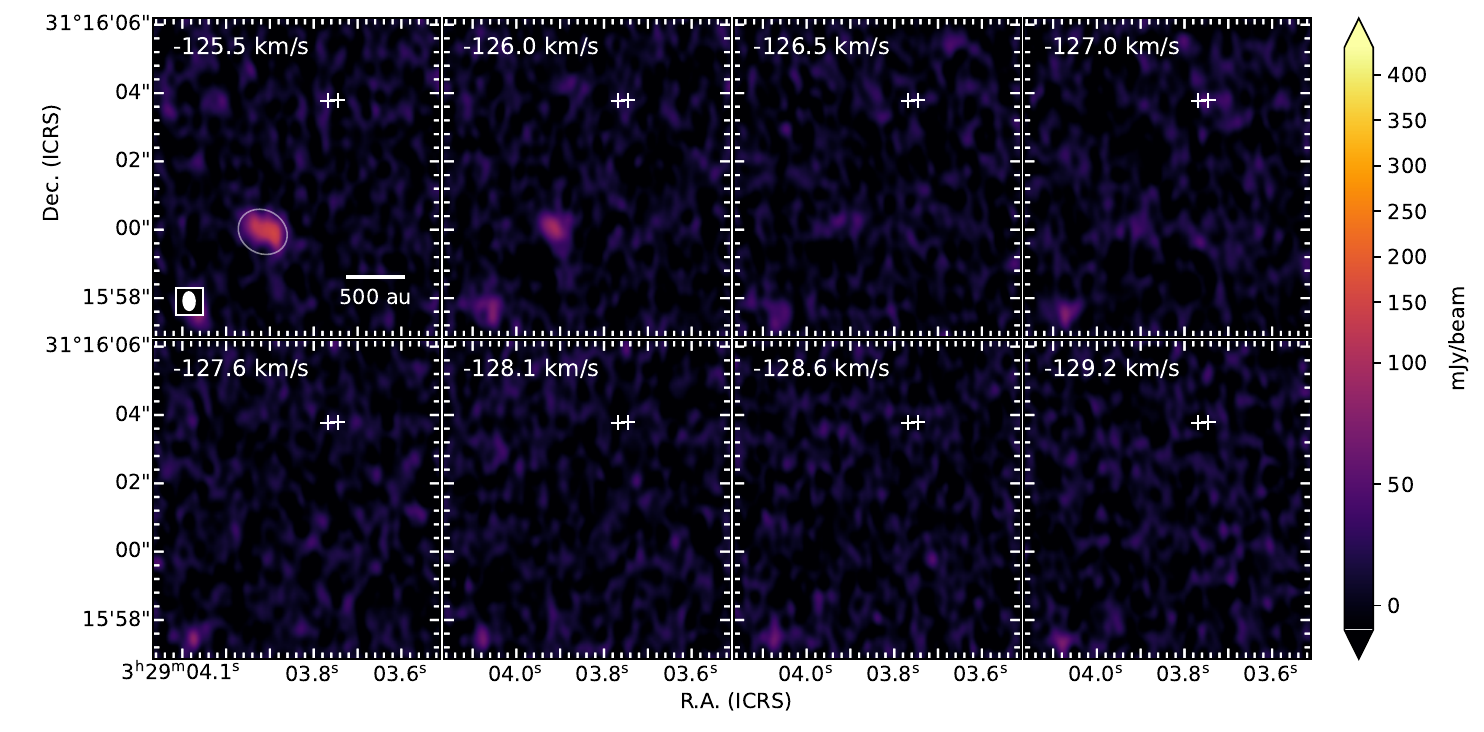}\\
}
{\bf Supplementary Figure 2: } Continued.
\end{figure*}

\begin{figure*}[t!]
\includegraphics[width=\textwidth]{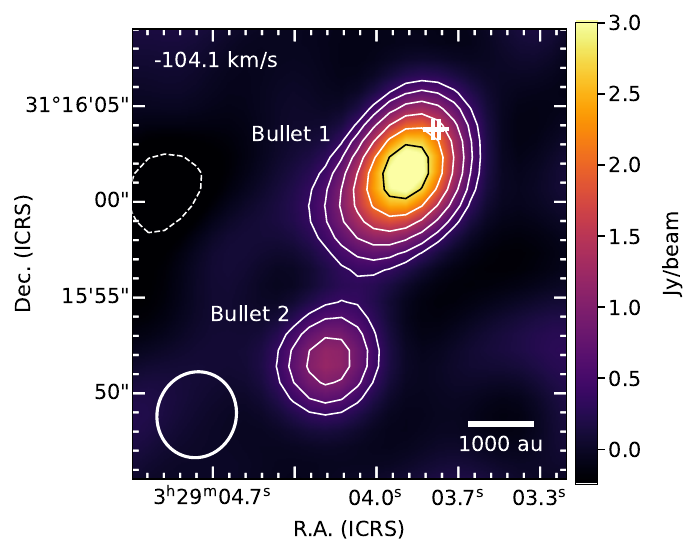} \\
{\bf Supplementary Data Fig.\ 3: Detection of Bullet 2 with ACA.}
Image of the CO($J$=3-2) line emission of the blueshifted bullets 1 and 2 in SVS~13, obtained with the 7-m ACA array, at an LOS velocity of $-$104 km s$^{-1}$ relative to VLA~4B (v$_{\rm LSR}=$ +9.3~km~s$^{-1}$; ref.~\citen{diaz-rodriguez2022_sup}), where the emission of Bullet 2 peaks. The synthesized beam is $4.51''\times 4.13''$ (PA=$-13.90^\circ$), and the channel width is 0.44~km~s$^{-1}$. Contour levels are -3, 3, 5, 8, 12, 17, 23, and 30 times 0.12 mJy~beam$^{-1}$, the r.m.s. of the map. The position of the SVS~13 binary is indicated with white plus signs. The image has not been corrected for the primary beam response. Bullet 2 appears unresolved in the ACA image, with a flux density is 1.9 Jy, after correction for the primary beam response. This second bullet is not detected with the ALMA 12-m array because of a lack of sensitivity at the distances where it is located, due to the smaller size of the 12-m array primary beam. The second bullet was previously detected only at longer wavelengths \cite{bachiller2000_sup,chen2016_sup,lefevre2017_sup}. 
\label{fig:v_theta_bow}
\end{figure*}

\begin{figure*}[t!]
\includegraphics[width=\textwidth]{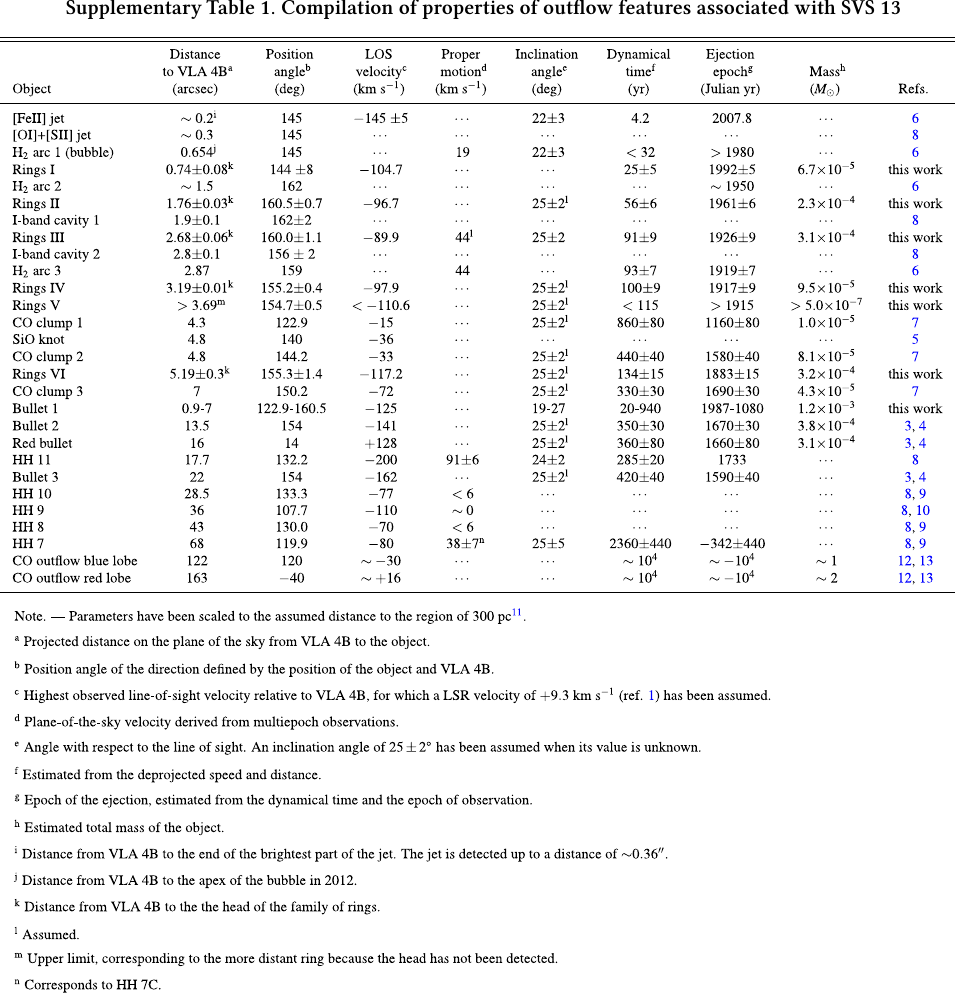} \\
\end{figure*}

\clearpage
\begin{figure*}[t!] 
{\centering
	\includegraphics[width=0.7\textwidth]{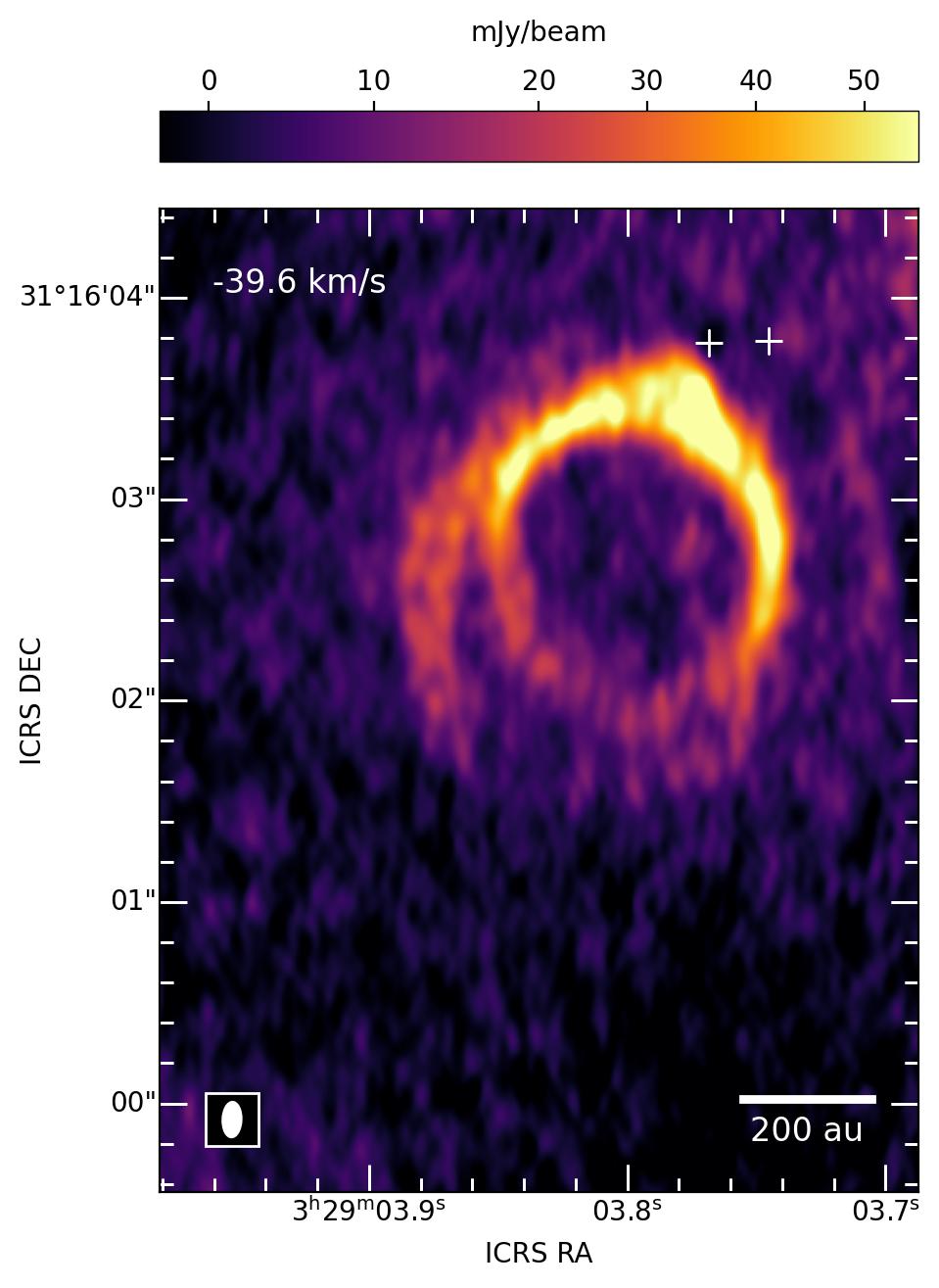}\\
}
{\bf Supplementary Video 1:} 
CO ($J$=3$\rightarrow$2) channel maps of Bullet 1 in SVS~13, in the line-of-sight velocity range (relative to VLA 4B) from $-$0.9 to $-$102.5~km~s$^{-1}$, observed with the ALMA 12-m array with an angular resolution of $0.17'' \times 0.09''$.
The LSR velocity of VLA~4B is $+$9.3~km~s$^{-1}$ (ref. \citen{diaz-rodriguez2022_sup}).  \href{https://static-content.springer.com/esm/art%3A10.1038%2Fs41550-025-02716-2/MediaObjects/41550_2025_2716_MOESM2_ESM.mp4}{Link to supplementary\_video\_1.mp4}
\end{figure*}

\clearpage
\begin{figure*}[t!] 
{\centering
	\includegraphics[width=0.7\textwidth]{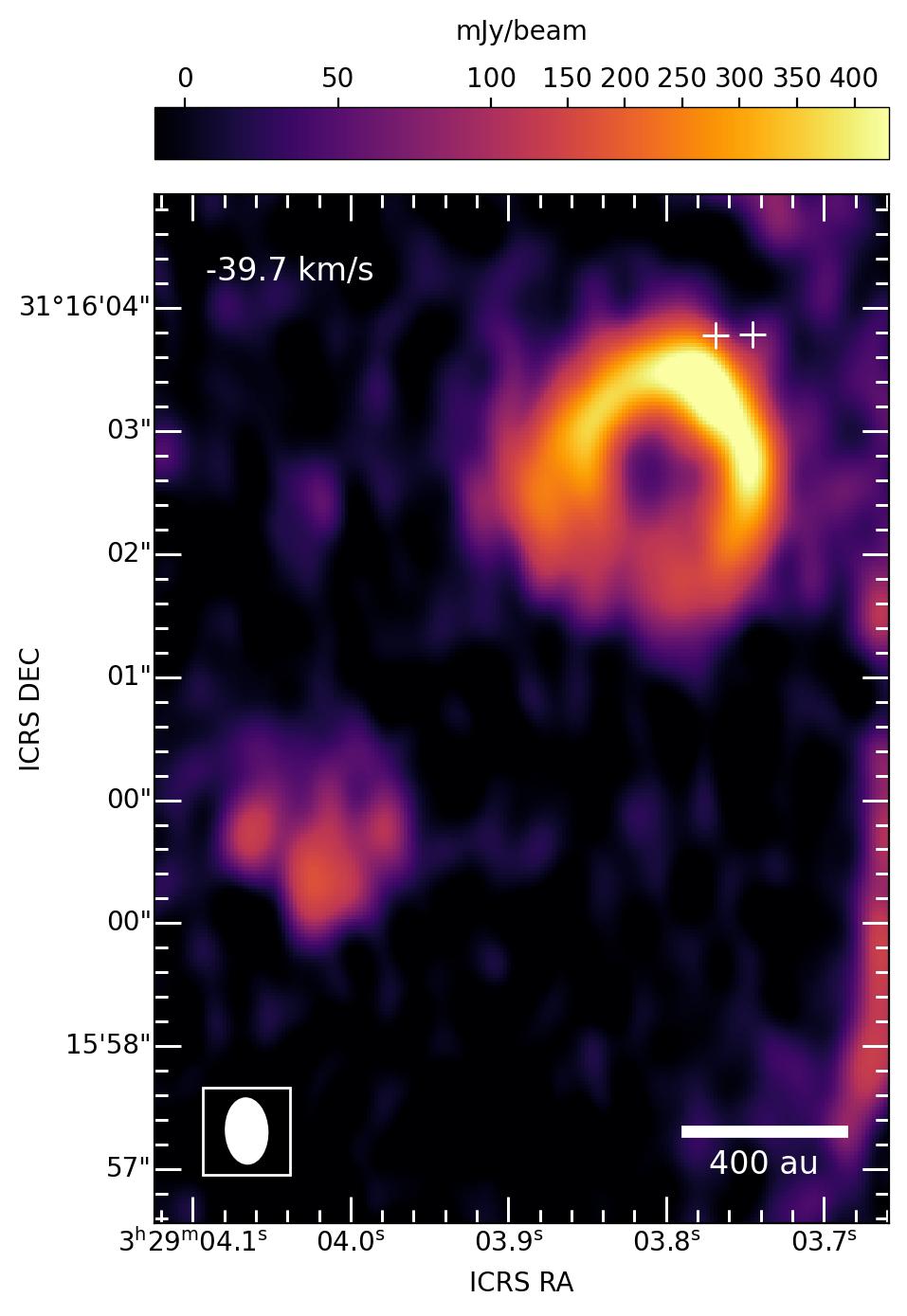}\\
}
{\bf Supplementary Video 2:} CO ($J$=3$\rightarrow$2) channel maps of Bullet 1 in SVS~13, in the line-of-sight velocity range (relative to VLA~4B) from $-$9.0 to $-$129.7 \kms, observed with the ALMA 12-m array with an angular resolution of $0.53'' \times 0.33''$. 
The LSR velocity of VLA~4B is $+$9.3~km~s$^{-1}$ (ref. \citen{diaz-rodriguez2022_sup}). \href{https://static-content.springer.com/esm/art%3A10.1038%2Fs41550-025-02716-2/MediaObjects/41550_2025_2716_MOESM3_ESM.mp4}{Link to supplementary\_video\_2.mp4}
\end{figure*}

\clearpage
\begin{figure*}[t!] 
{\centering
	\includegraphics[width=\textwidth]{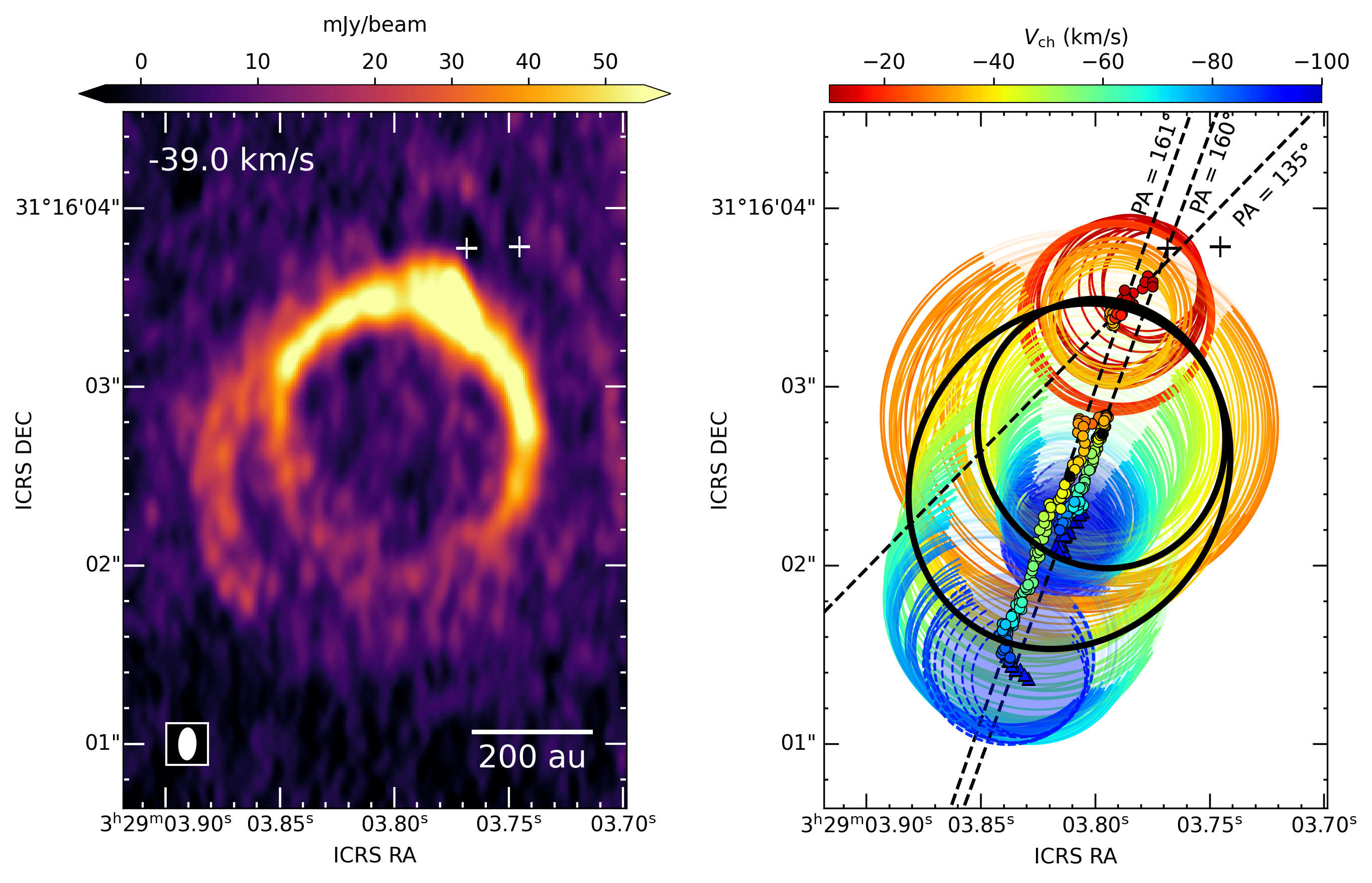}\\
}
{\bf Supplementary Video 3:} Decomposition of the observed CO ($J$=3$\rightarrow$2) channel map emission of Bullet 1 in SVS~13 into elliptical rings. The line-of-sight velocity, relative to VLA~4B ($V_{\rm LSR}$ = +9.3~\kms; ref. \citen{diaz-rodriguez2022_sup}), ranges from $-$0.9 to $-$102.5~km~s$^{-1}$. \href{https://static-content.springer.com/esm/art%3A10.1038%2Fs41550-025-02716-2/MediaObjects/41550_2025_2716_MOESM4_ESM.mp4}{Link to supplementary\_video\_3.mp4}
\end{figure*}

\clearpage
\begin{figure*}[t!] 
{\centering
	\includegraphics[width=0.9\textwidth]{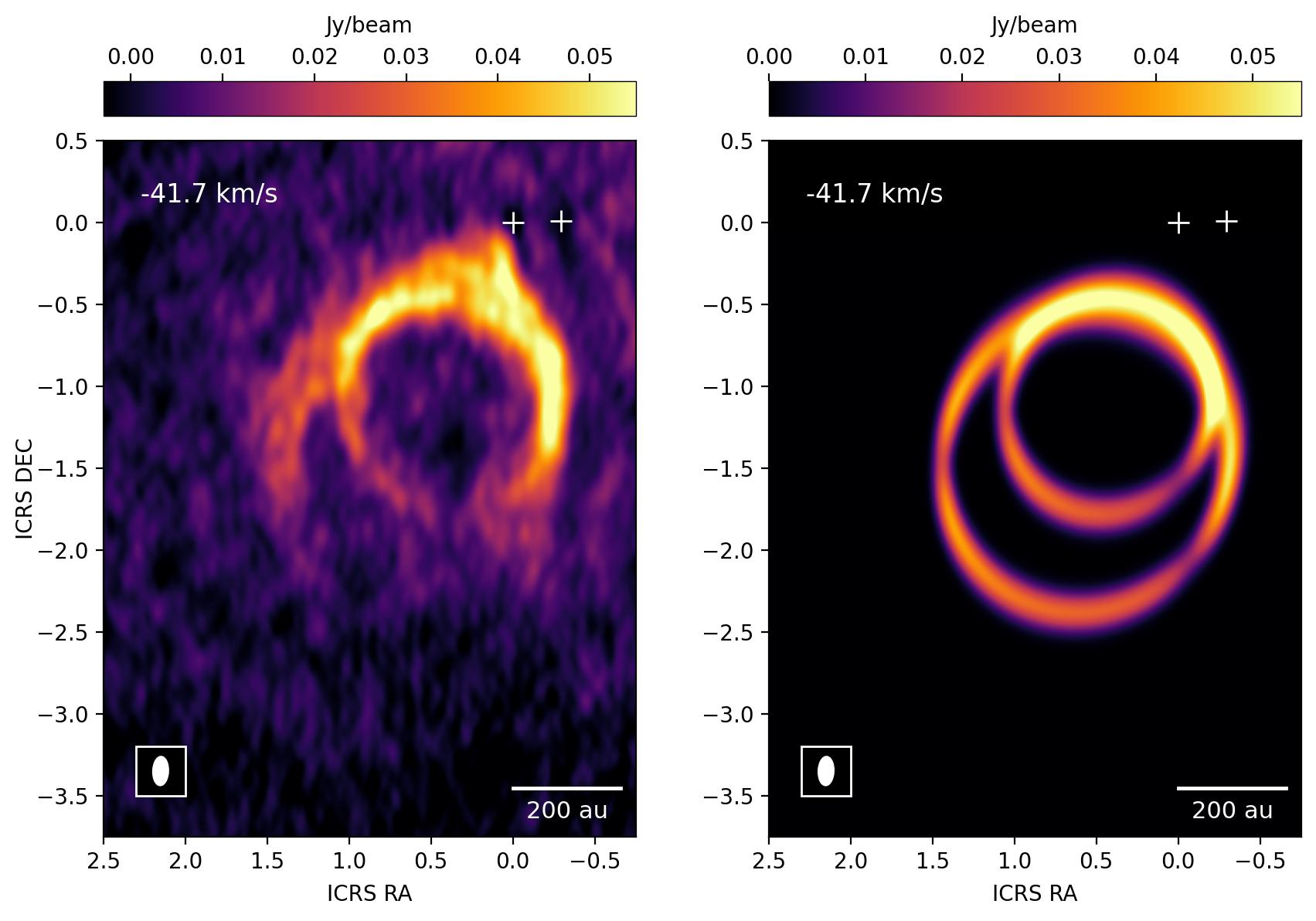}\\
}
{\bf Supplementary Video 4:} Comparison of the observed and bowshock model CO ($J$=3$\rightarrow$2) emission of the families of rings II and III in the line-of-sight velocity range (relative to VLA~4B) from $-$38.3 to $-$102.5~km~s$^{-1}$. \href{https://static-content.springer.com/esm/art%3A10.1038%2Fs41550-025-02716-2/MediaObjects/41550_2025_2716_MOESM5_ESM.mp4}{Link to supplementary\_video\_4.mp4}
\end{figure*}

\end{document}